\pdfoutput=1
\documentclass[11pt]{elsarticle}
\usepackage{graphicx}
\usepackage{amssymb}
\usepackage{textcomp}
\usepackage{amsmath}
\usepackage{bm}
\usepackage{times}
\usepackage{epsfig}
\usepackage{color}
\usepackage{array,multirow}
\usepackage{subfigure,ulem,tabularx,latexsym}
\usepackage{braket}
\usepackage{hyperref}

\newcommand{\trix}[1]{\left(\begin{array}{#1}}
\newcommand{\notrix}{\end{array}\right)}
\newcommand{\comment}[1]{}

\def\beq{\begin{equation}}
\def\eeq{\end{equation}}
\def\bea{\begin{eqnarray}}
\def\eea{\end{eqnarray}}
\def\nn{\nonumber}

\def\nn{\nonumber}

\def\be{\begin{equation}}
\def\ee{\end{equation}}
\def\ba{\begin{eqnarray}}
\def\ea{\end{eqnarray}}

\def\nn{\nonumber}
\def\lf{\left}
\def\rt{\right}

\def\La{\mathcal{L}}
\mathchardef\mhf="2D
\newcommand{\BL}{$B \mhf L $~}
\newcommand{\BLMSSM}{$B \mhf L $ MSSM~}

\begin{document}

\begin{titlepage}
\vspace{-4cm}
\title{\huge {Perturbative} Reheating in \\ Sneutrino-Higgs Cosmology{\LARGE \\[.5cm]}}

\date{}

\author[pnn,uca]{Yong Cai\fnref{fn1}}

\author[pnn]{Rehan Deen\fnref{fn2}}

\author[pnn]{Burt A.~Ovrut\fnref{fn3}}

\author[man]{Austin Purves\fnref{fn4}}

\address[pnn]{Department of Physics and Astronomy, University of Pennsylvania,
Philadelphia, PA 19104--6396, USA \\[0.5cm]}
\address[uca]{School of Physics, University of Chinese Academy of Sciences, Beijing 100049, China\\[0.5cm]}
\address[man]{Department of Physics, Manhattanville College, Purchase, NY 10577, USA }

\fntext[fn1]{caiyong13@mails.ucas.edu.cn}
\fntext[fn2]{rdeen@sas.upenn.edu}
\fntext[fn3]{ovrut@elcapitan.hep.upenn.edu}
\fntext[fn4]{austin.purves@mville.edu}

\begin{abstract}
{\noindent  The theory of {perturbative} reheating in the Sneutrino-Higgs cosmology of the \BLMSSM is presented. It is shown that following an epoch of inflation consistent with all Planck2015 data, the inflaton begins to oscillate around its minimum at zero and to reheat to various species of standard model and supersymmetric matter. The {perturbative} decay rates to this matter are computed, both analytically and numerically. Using these results, the Hubble parameter and the relative energy densities for each matter species, including that of the inflaton, are calculated numerically. The inflaton energy density is demonstrated to vanish at an energy scale of ${\cal{O}}(10^{13})~{\rm GeV}$, signaling the end of the period of reheating. The newly created matter background is shown to be in thermal equilibrium, with a reheating temperature of $\simeq 1.13 \times 10^{13}~{\rm GeV}$. To allow for a \BL breaking scale sufficiently smaller than the reheating scale, we extend the statistical method of determining the soft supersymmetry breaking parameters developed in previous work. The result is that one can determine a large number of phenomenologically realistic initial conditions for which the \BL breaking scale is an order of magnitude or more smaller than the reheating scale.}
\end{abstract}

\maketitle

\end{titlepage}

\newpage

\tableofcontents

\section{Introduction}

The introduction of Higgs inflation in the standard model of particle physics \cite{Bezrukov:2007ep,Bezrukov:2008ut,Bezrukov:2008ej,Barbon:2009ya,Bezrukov:2010jz,Bezrukov:2012sa}, emphasized the important idea that the ``inflaton'' associated with inflationary cosmology might well be--and, perhaps, should be--a fundamental scalar field in the standard model. In the non-supersymmetric case, this could only be the neutral Higgs scalar. A careful analysis of this possibility led to the result that an acceptable theory of inflation could potentially arise in this context--but only at the cost of assuming several unnatural features. For example, the square of the Higgs magnitude must necessarily be coupled to the curvature scalar in the Lagrangian density with an unnaturally large coupling parameter. Generalizing this idea to the supersymmetric standard model (MSSM), and some variants thereof, potentially extends this idea by introducing a large number of scalar fields into the theory, a subset of which might play the role of the inflaton. Although at first seeming promising, these models \cite{Einhorn:2009bh,Ferrara:2010yw,Ibanez:2014kia,Bielleman:2015lka} continued to exhibit various fundamental problems. Not the least of which is the fact that the plethora of scalar fields now greatly complicates that potential energy. In particular, it now becomes very difficult to obtain a potential for the ``inflaton'' that is stable with respect to the other directions in field space. That is, the inflating field usually rolls off into another scalar direction, rapidly losing potential energy and ending the inflationary period far short of the required 60 e-foldings.

Fortunately, there is one specific supersymmetric theory--called the \BLMSSM--that naturally solves this difficulty. The \BLMSSM is a vacuum solution of heterotic M-theory \cite{Horava:1995qa,Witten:1996mz,Horava:1996ma,Lukas:1998yy,Lukas:1997fg,Lukas:1998tt,Donagi:1999gc,Donagi:1999jp,Anderson:2010mh} whose observable sector has precisely the spectrum of the MSSM, where each of the three lepton families contains a right-handed neutrino chiral supermultiplet. The theory differs from the MSSM in that the gauge group is enhanced from the $SU(3)_{C} \times SU(2)_{L} \times U(1)_{Y}$ group of the standard model to $SU(3)_{C} \times SU(2)_{L} \times U(1)_{Y} \times U(1)_{B-L}$. That is, the global \BL symmetry of the MSSM is gauged. This theory is known to remain anomaly free and is the simplest possible extension of the MSSM. This theory was introduced in \cite{Braun:2005nv}, and its properties studied in a series of papers \cite{Barger:2008wn,Ambroso:2009jd,Donagi:2004ia,Braun:2005ux,Braun:2005bw,Ovrut:2012wg,Marshall:2014kea,Marshall:2014cwa,Ovrut:2014rba,Ovrut:2015uea,Deen:2016vyh}. Importantly, it was shown in \cite{Ovrut:2015uea} that for a very wide and diverse range of initial conditions, the \BLMSSM is completely consistent with all phenomenological data; that is, the \BL symmetry is appropriately radiatively broken, electroweak symmetry is also radiatively broken  with the correct values of the $Z^{0}$ and $W^{\pm}$ masses, all sparticles exceed the present experimental bounds and, remarkably, the prediction for the mass of the neutral Higgs boson is within 2$\sigma$ of the present ATLAS bounds from CERN \cite{Aad:2012tfa,Aad:2015zhl,Patrignani:2016xqp}. Thus, if a successful  inflationary theory can be developed in this context, it would simultaneously be consistent with all low energy phenomenology.

Remarkably, a completely successful theory of inflationary cosmology {\it can} be developed within this context. The reason for this is relatively straightforward. The potential energy of the scalar field has three contributions; the D-term potential $V_{D}=\frac{1}{2} \sum_{a}D_{a}^{2}$, the F-term potential $V_{F}$ and contributions from the soft supersymmetry breaking terms $V_{soft}$. As will be discussed below, the F-term potential essentially vanishes after the beginning of inflation. What is remarkable about the \BLMSSM is that,  for a precise combination--denoted by $\phi_{1}$--of the right-handed sneutrino, the left-handed sneutrino and the neutral up Higgs field, all $D_{a}$ contributions also vanish, rendering $V_{D}=0$. This fact, which is due to the existence the right-handed sneutrino, lowers the entire D-flat potential into a ``valley'' that is stable against roll-off into any other scalar direction. It follows that the inflaton stably inflates along the 
$V_{soft}$ potential--completely consistent with all Plank2015 cosmological data. This inflationary theory, which we have named ``Sneutrino-Higgs inflation" was developed in detail in \cite{Deen:2016zfr}. In fact, this theory is very similar to the so-called ``no scale'' supergravity theory developed in a different context in \cite{Cremmer:1983bf,Ellis:2003sq,Ellis:2013xoa,Ellis:2016ipm}. 
What we believe is important about Sneutrino-Higgs inflation in the \BLMSSM is that it is, as stated above, entirely consistent with all low energy particle physics phenomenology.
That is, not only is it an accurate inflationary theory, it is within the context of realistic low-energy particle physics as well. 

With this in mind, in the present paper we move beyond the purely inflationary epoch and study the precise theory of perturbative reheating of the Sneutrino-Higgs theory. 
The subject of reheating in inflationary models has been studied extensively; see for example \cite{PhysRevLett.48.1437,Abbott:1982hn,Traschen:1990sw,Kofman:1994rk,Khlebnikov:1996mc,Kofman:1997yn,Chung:1998rq,Bassett:2005xm,Allahverdi:2010xz,Amin:2014eta,Allahverdi:2011aj,Ellis:2015pla,Ellis:2015jpg,Ellis:2016spb}. As discussed in those papers, reheating can occur after the inflationary epoch via preheating (that is, parametric resonance), via perturbative decay of the inflaton or as a combination of both of these processes. Which of these two processes dominates reheating is highly model dependent. Of all the theories discussed in the preceding references, only the models in \cite{Ellis:2015pla,Ellis:2015jpg,Ellis:2016spb} are closely related to the Sneutrino-Higgs inflation in the \BLMSSM discussed in our work. Specifically, the theories in  \cite{Ellis:2015pla,Ellis:2015jpg,Ellis:2016spb} involve Higgs and sneutrino inflatons coupled to so-called ``no-scale'' $N=1$ supergravity--very similar to Sneutrino-Higgs inflation in the \BLMSSM. Within this context, the first two papers found that perturbative reheating is the dominant process for both decay to standard model particles \cite{Ellis:2015pla} as well as to the gravitino \cite{Ellis:2015jpg}. This was due to the fact that the inflaton in these papers was purely a right-handed sneutrino and, hence, the coupling to matter is via small Yukawa and gravitational interactions. In \cite{Ellis:2016spb}, however, the authors analyzed the case of a pure Higgs inflaton, which couples to matter via gauge interactions. They conclude that non-perturbative production may or may not be a significant component of reheating, depending on the specific parameters of the model. However, it is possible that preheating could be the more efficient reheating mechanism. Be that as it may, in the present work we will compute only the perturbative decays of the Sneutrino-Higgs inflaton, since this is expected to play a role in the reheating process. The possible effects of preheating, which may dominate over perturbative decays, will be explored elsewhere.
As we will show, in the Sneutrino-Higgs theory, this reheating epoch is completely amenable to exact computation. The end result is that this theory also reheats in a technically determined manner and appears to be completely acceptable physically.

Specifically, we will do the following. In Section 2, we give a brief review of the theory of Sneutrino-Higgs inflation presented in \cite{Deen:2016zfr}. We outline the basic formalism and present the specific parameters that lead to a successful theory of inflation--completely consistent with the Planck2015 data \cite{Ade:2015lrj}. Using the statistical method for choosing the soft supersymmetry breaking parameters introduced in \cite{Ovrut:2015uea}, we show that there is a large and diverse set of phenomenologically valid initial conditions that are consistent with the appropriate cosmological parameters. We conclude this section by presenting a new result--specifically, we compute the range of values of the \BL breaking scale in this context and plot this as a histogram against the number of valid points at a given \BL scale. We find that the smallest \BL scale that one can attain in this formalism is $\simeq 2 \times 10^{12}~{\rm GeV}$--and this for only a very small number of valid initial points. Although this is marginally acceptable, we would like to modify our formalism in such a way as to attain smaller values for the \BL scale. The reason is that we expect the reheating temperature to be of ${\cal{O}}(10^{13}~ {\rm GeV})$ and would, for simplicity, prefer that \BL breaking, and, hence, the breaking of baryon number, lepton number and $R$-parity, to occur at a much smaller scale. This allows one to separate the reheating epoch from the period of baryo- and lepto-genesis \cite{Trodden:2004mj}. This is carried out in detail in Appendix A, where we introduce an extension of our previous formalism that allows one to lower the \BL scale arbitrarily, while remaining completely phenomenologically realistic. In this Appendix, we give an explicit example where the \BL scale is reduced to a range from a low of $10^{6}~{\rm GeV}$ to well above $10^{13}~{\rm GeV}$--all this occurring for phenomenologically acceptable initial conditions. We also give an explicit formula for the degree of fine-tuning required to set the \BL scale to any given value. Such a wide range is not required for cosmology. Hence, Section 3 is devoted to explaining this new formalism, and then using it to alter the \BL scale to a more reduced range of values--namely from $10^{10}~{\rm GeV}$ to $10^{12}~{\rm GeV}$. We find, in this context, that the greatest number of phenomenologically valid initial points occurs for a \BL scale of $10^{11}~{\rm GeV}$ and, hence, when specificity is required later in our analysis, we will choose this as the value of \BL symmetry breaking.

The remainder of the paper concentrates on the post-inflationary epoch and reheating. In Section 4, we give a detailed discussion of the classical behavior of both the inflaton and the Hubble parameter--ignoring for the time being the quantum mechanical decays of the inflaton into matter. We begin by calculating these quantities numerically and show that at a given time  after the end of inflation, denoted by $t_{osc}$, the inflaton begins to oscillate around its minimum at zero. Then, using  an iterative method, we analytically compute  both the inflaton and Hubble parameter to a high degree of approximation. We show that this analytic solution becomes well-defined after a time, $t_{MD} > t_{osc}$, which marks the beginning of the period of matter-domination. The numerical and analytic results are plotted simultaneously and shown to very closely approximate each other for $t > t_{MD}$. In Section 5, using these results, we compute the decay of the inflaton into matter and, hence, reheating. We show that there are four major decay channels--up-type standard model particles, charginos, neutralinos and gauge bosons--and analytically calculate the decay rates $\Gamma$ of the inflaton into each of these species. We present the details of this analysis both in Section 5 and in Appendix C. In Section 6, we analyze the full set of differential equations for the inflaton and Hubble parameter--now sourced by the decay rates and associated energy density of the matter species into which the inflaton decays--as well as the differential equations for the evolution of the matter itself. These equations are solved numerically for $t>t_{MD}$. The time-dependent behavior of the individual decay rates, as well as the behavior of the Hubble parameter, are plotted. Using these results, the fraction of the individual energy densities, 
$\frac{\rho_{i}}{\rho_{total}}$, including the inflaton density, are computed and plotted. The time of reheating, $t_{R}$, is defined to be the time at which the inflaton density vanishes--indicating that its energy has then been entirely transferred to matter. Finally, in Section 7 we show that by $t_{R}$ all the newly created matter species are in thermal equilibrium, and calculate the associated reheating temperature.

\section{Sneutrino-Higgs Inflation}

In this section, we will briefly review the theory of supersymmetric Sneutrino-Higgs inflation, first presented in \cite{Deen:2016zfr}, that arises within the context of the \BLMSSM \cite{Braun:2005nv, Barger:2008wn,Ambroso:2009jd,Everett:2009vy,FileviezPerez:2012mj,Ovrut:2012wg,Marshall:2014kea,Marshall:2014cwa,Ovrut:2014rba,Ovrut:2015uea,Deen:2016vyh} coupled to $N=1$ supergravity. 

\subsection{~Supergravitational \BL MSSM} 

In the flat superspace limit, the particle content of the \BLMSSM is precisely that of the $N=1$ supersymmetric MSSM; that is, three families of quark and lepton chiral supermultiplets, each family with a right-handed neutrino chiral multiplet, and a pair of Higgs, Higgs conjugate chiral superfields. The gauge group, however, is extended from the standard model gauge group to
\begin{equation}
G=SU(3)_{C} \times SU(2)_{L} \times U(1)_{Y} \times U(1)_{B-L} \ .
\label{1}
\end{equation}
The additional Abelian gauge group $U(1)_{B-L}$ is anomaly free and arises from the fact that the \BLMSSM is a low energy vacuum of heterotic M-theory \cite{Horava:1995qa,Lukas:1997fg,Donagi:1999gc,Ambroso:2009jd} compactified to four-dimensions on a specific Calabi-Yau threefold \cite{Braun:2005zv} with a given holomorphic vector bundle \cite{Braun:2005nv}. As discussed in \cite{Ovrut:2015uea}, it is convenient (and physically equivalent) to use the group $U(1)_{3R}$, with generator  $T_{3R}=Y-\frac{1}{2}(B-L)$ in place of $U(1)_{Y}$, since this eliminates kinetic mixing of the associated field strength with that of \BL .\footnote{Similarly, it is useful to rescale the generator of the \BL group so as to replace $U(1)_{B-L}$ with $U(1)_{BL^{\prime}}$, as demonstrated in \cite{Ovrut:2015uea}. However, for clarity, in the remainder of this paper we will continue to denote the rescaled gauge group as $U(1)_{B-L}$; that is, without the prime.}

The formalism for coupling of a generic four-dimensional compactification of M-theory to $N=1$ supergravity was presented in \cite{Witten:1985xb}, and can easily be applied to the \BLMSSM. We note that the Planck mass used in this formalism is the so-called ``reduced'' Planck mass given by $M_{P}=2.435 \times 10^{18}~{\rm GeV}$. For the bosonic sector of the theory, this was carried out in \cite{Lukas:1997fg}. Suffice it here to say that, to order $\kappa^{2/3}$ in heterotic M-theory, by appropriately redefining the matter scalars $C_{i}$ and setting the Planck mass $M_{P}=1$, the matter scalar K\"ahler potential and each of the gauge kinetic functions in the Lagrangian become
\begin{equation}
K= -3~{\rm ln}(1 - \sum_{i}{ \frac{|C_{i}|^{2}}{3}})~,  \quad  f_{a}=1~~~{\rm for}~ a=3,2,3R, B-L
\label{2}
\end{equation}
respectively. Recalling that the matter scalar kinetic energy terms in the Lagrangian are given by
\begin{equation}
-K_{i\bar{j}}\partial_{\mu}C^{i}\partial^{\mu}C^{\bar{j}} \ ,
\label{3}
\end{equation}
it follows that for small values of $C_{i}$ these kinetic energy terms are canonically normalized. Importantly, however, this is no longer true when one or more of these fields are of the order of the Planck scale. Note that throughout this paper, we will often, for notational simplicity, set $M_{P}=1$. However, there are sections where the physical content is more transparent when expressed in mass dimensions of {\rm GeV}. Which notation is being used will be explicitly indicated.

In addition to the kinetic terms, the \BLMSSM Lagrangian contains a holomorphic superpotential given by
\begin{eqnarray}
	W=Y_u Q H_u u^c_R - Y_d Q H_d d^c_R -Y_e L H_d e^c_R +Y_\nu L H_u \nu^c_R +\mu H_u H_d \ ,
\label{4}
\end{eqnarray}
where both generational and gauge indices have been suppressed. As discussed in \cite{Ovrut:2015uea}, the Yukawa couplings can be taken to be flavor diagonal and real for the purposes this analysis.
The $\mu$ parameter can also be chosen to be real without loss of generality. This superpotential leads to the F-term contributions to the potential energy for matter scalars in the Lagrangian, given by the usual expression
\begin{equation}
V_{F}= e^{K}\big( K^{i \bar{j}}D_{i}W \overline{D_{\bar{j}}W}-3|W|^{2}\big) \ .
\label{5}
\end{equation}
Similarly, eliminating the auxiliary fields in the vector superfields associated with the $SU(3)_{C}, SU(2)_{L}, U(1)_{3R}$ and $U(1)_{B-L}$ gauge groups leads to the D-term contributions to the matter scalar potential energy given by
\begin{equation}
V_{D}=\frac{1}{2} \sum_{a}D_{a}^{2} \ ,
\label{6}
\end{equation}
where the  $D_{a}$, $a=3,2,3R,B-L$ functions are 
\begin{equation}
D^{r}_{a} = -g_a \frac{\partial K}{\partial C_i} [T_{(a)}^r]_i\,^j C_j 
= \frac{g_a }{\left(1 - \tfrac{1}{3}\sum_i |C_i|^2\right)} \mathcal{D}^{r}_{(a)}~,  \qquad  
\mathcal{D}_{(a)}^{r} = -\overline{C}^i [T_{(a)}^r]_i\,^j C_j
\label{7}
\end{equation}
and $T_{(a)}^{r}$, $r=1,\dots,{\rm dim}~G_{a}$  are the generators of the group $G_{a}$. Both $V_{F}$ and $V_{D}$ were computed for all matter scalars in \cite{Deen:2016zfr}. We refer the reader there for details. Finally, it is assumed in \cite{Ovrut:2015uea} that supersymmetry is spontaneously broken in a hidden sector. This leads to ``soft'' supersymmetry breaking terms in the observable low-energy Lagrangian. These contain a final contribution to the matter scalar potential energy given by
\begin{eqnarray}
V_{soft} &=& m_{\tilde Q}^2|\tilde Q|^2+m_{\tilde u^c_R}^2|\tilde u^c_R|^2+m_{\tilde d^c_R}^2|\tilde d^c_R|^2+m_{\tilde L}^2|\tilde L|^2
	+m_{\tilde \nu^c_R}^2|\tilde \nu^c_R|^2+m_{\tilde e^c_R}^2|\tilde e^c_R|^2  +m_{H_u}^2|H_u|^2 \nn \\
&+&m_{H_d}^2|H_d|^2  + \left(a_u \tilde Q H_u \tilde u^c_R - a_d \tilde Q H_d \tilde d^c_R - a_e \tilde L H_d \tilde e^c_R
		+ a_\nu \tilde L H_u \tilde \nu^c_R + b H_u H_d + h.c.\right) \label{8} \ .
\end{eqnarray}
We will make the usual assumption that each of the dimensionless couplings for the cubic terms is proportional to the associated Yukawa coupling. Since it will become important in the analysis of reheating given later in this paper, we point out that, in addition to the scalar terms in \eqref{8}, the soft supersymmetry breaking terms also include mass terms for each of the gauginos given by
\begin{equation}
\mathcal -{\cal{L}}_{soft}^{(gaugino)}  = \left(\frac{1}{2} M_3 \tilde g^2+ \frac{1}{2} M_2 \tilde W^2+ \frac{1}{2} M_R \tilde W_R^2+\frac{1}{2} M_{BL} \tilde {B}^2 + h.c.\right) \ .
\label{9}
\end{equation}
The fermions $\tilde g, \tilde W, \tilde W_R$ and $\tilde {B}$ in \eqref{8} are the $SU(3)_{C}, SU(2)_{L}, U(1)_{3R}$ and
$U(1)_{B-L}$ 
gauginos respectively.

Finally, we point out that the coupling of the \BLMSSM to $N=1$ supergravity using the M-theory formalism given in \cite{Lukas:1997fg}, leads to a canonical Lagrangian for the metric tensor $g_{\mu\nu}$.  That is, we find that the pure gravitational action is simply given by
\begin{equation}
-\frac{1}{2} \int_{M_{4}} \sqrt{-g}R \ .
\label{10}
\end{equation}
It follows that in the Sneutrino-Higgs inflation theory, we do not require any ``non-canonical'' coupling of matter to the curvature tensor $R$.

\subsection{~The Inflaton and the Inflationary Potential}

The inflationary potential in \cite{Deen:2016zfr} was constructed as follows. In order to help guarantee the stability of the potential energy, we first searched for a direction in scalar field space for which all of the above D-terms vanish. We were naturally lead to the field space configuration
\begin{equation}
H_u^0 = \tilde{\nu}_{R,3}^c = \tilde{\nu}_{L,3} \ ,
\label{11}
\end{equation}
with all other fields set to zero. We note that only in a model such as the \BLMSSM with right-handed neutrino superfields would such a D-flat direction arise. To proceed, we defined three new fields $\phi_{i}, i=1,2,3$ by
\bea
H_u^0 &=&\tfrac{1}{\sqrt{3}}\left(\phi _1- \phi _2- \phi _3 \right), \nonumber \\
\tilde{\nu} _{3,L} &=& \tfrac{1}{\sqrt{3}}\phi _1+\left(\tfrac{1}{2}+\tfrac{1}{2 \sqrt{3}}\right) \phi _2+\left(\tfrac{1}{2 \sqrt{3}}-\tfrac{1}{2}\right) \phi _3, \nonumber \\
\tilde{\nu} _{3,R}^c &=& \tfrac{1}{\sqrt{3}}\phi _1+\left(\tfrac{1}{2 \sqrt{3}}-\tfrac{1}{2}\right) \phi _2+\left(\tfrac{1}{2}+\tfrac{1}{2 \sqrt{3}}\right) \phi _3.
\label{12}
\eea
The field $\phi_1$ corresponds to the D-flat field direction while $\phi_2$ and $\phi_3$ are two orthogonal directions. We note that 
\begin{equation}
\phi_1 = \tfrac{1}{\sqrt{3}} \left( H_u^0 + \tilde{\nu} _{L,3} + \tilde{\nu} _{R,3}^c\right)
\label{13}
\end{equation}
and the associated quadratic soft mass squared is given by
\begin{equation}
m^{2}=\frac{1}{3}(m^{2}_{H_{u} }+ m_{\tilde{L}_3}^{2}+m_{{\tilde{\nu}}_{R,3}^c}^{2}) \ .
\label{14}
\end{equation}
Setting all fields to zero with the exception of $\phi_{1}$, the $V_{D}$ potential vanishes and the Lagrangian becomes
\begin{equation}
\mathcal{L} = -\frac{1}{\left(1 - \tfrac{1}{3}|\phi_1|^2\right)^2}\partial_\mu{\overline \phi_1} \partial^\mu{\phi_1}  - V_F(\phi_1) - V_{soft} (\phi_1)~, 
\label{15}
\end{equation}
where
\begin{equation}
V_F(\phi_1) = \frac{3 |\phi _1|^2 \left(\mu ^2+ Y_{\nu3}^2|\phi _1|^2 \right)}{\left(3 - |\phi _1|^2\right)^2}~, \qquad
V_{soft}(\phi_1) = m^2 |\phi_1|^2 \ .
\label{16}
\end{equation}
Here $Y_{\nu3}$ is the third-family sneutrino Yukawa coupling and $\mu$ is the usual supersymmetric Higgs parameter. In \cite{Deen:2016zfr} it was verified in detail that, modulo several acceptable caveats, the field $\phi_{1}$ lies in a valley of the total potential energy. That is, the potential energy gets larger as one moves away from the pure $\phi_{1}$ direction by making any other scalar field non-zero.

We henceforth, for simplicity, considered only the real part of $\phi_{1}$, which we continue to denote by the same symbol.  Since the value of $\phi_{1}$ can be of Planck scale during inflation, it follows from \eqref{2} and \eqref{3} that its kinetic energy is not canonically normalized in this regime. To canonically normalize the kinetic term, we make a field  redefinition to a real scalar $\psi$ given by
\begin{equation}
\phi_1 = \sqrt{3} \tanh\left(\frac{\psi}{\sqrt{6}} \right) .
\label{17}
\end{equation}
In terms of the new field $\psi$, Lagrangian \eqref{15} now becomes
\begin{equation}
\mathcal{L} = -\frac{1}{2}\partial_\mu \psi \partial^\mu \psi - V_F(\psi)  - V_{soft}(\psi) \ ,
\label{18} 
\end{equation}
where $V_{F}(\psi)$ is obtained from the first term in  \eqref{16} using \eqref{17} and
\begin{equation}
V_{soft}(\psi)= 3 m^2 \tanh^2\left( \frac{\psi}{\sqrt{6}}\right) \ .
\label{19}
\end{equation}
We emphasize again that throughout this paper we will, unless otherwise specified, set the reduced Planck mass to unity. Hence, all terms in Lagrangian density \eqref{18} are dimensionless--including the coordinates $t$ and $x_{i}, i=1,2,3$.

\subsection{~Inflation and the Primordial Parameters}

To analyze inflation, one chooses the metric to be in the Friedman-Robinson-Walker form
\begin{equation}
ds^{2}=-dt^{2}+a^{2}(t)d{\vec{x}}^{2}
\label{20}
\end{equation}
and defines the Hubble parameter as $H={\dot{a}}/a$. Then the complete set of dynamical equations associated with the canonically normalized gravity action \eqref{10} and the normalized $\psi$ Lagrangian \eqref{18} is given by
\bea
&&3H^{2}=\frac{1}{2}{\dot{\psi}}^{2}+V(\psi) \label{21}  \\
&&\dot{H}=-\frac{1}{2}{\dot{\psi}}^{2} \ , \label{22} \\
&&\ddot{\psi} + 3H\dot{\psi} +V_{,\psi}=0 \ , \label{23}  
\eea
where the potential $V$ is given by
\begin{equation}
V(\psi)= V_F(\psi)  + V_{soft}(\psi) \ .
\label{24}
\end{equation}

\begin{figure}[htbp]
   \centering
	\includegraphics[scale=0.4]{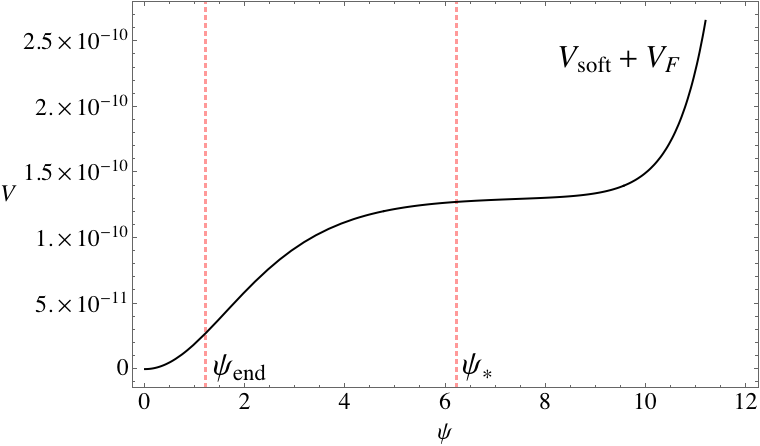}
   \caption{The black line is a graph of the potential $V_{soft}+V_{F}$ for the parameters  $m = 6.49 \times 10^{-6}$, $Y_{\nu3} \sim 10^{-12}$ and $\mu= 4.93 \times 10^{-4}$. For these values of the parameters, the vertical red dashed lines mark $\psi_{end} \simeq 1.21$ and $\psi_{*}  \simeq 6.25$ respectively. We have set $M_{P}=1$.}
   \label{plot2}
\end{figure}

\begin{figure}[htbp]
\centering
\includegraphics[scale=0.1,width=0.85\textwidth]{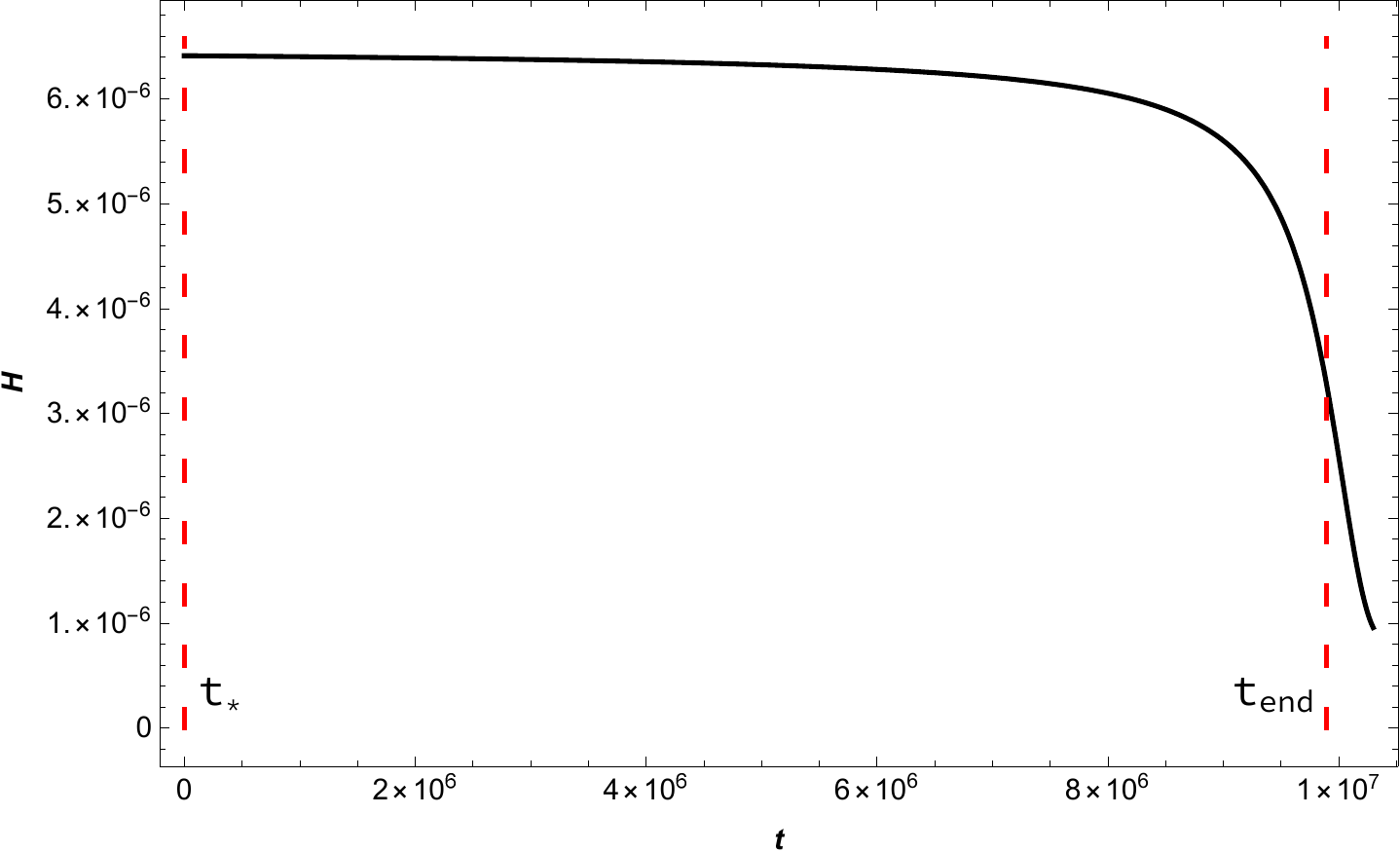}
\caption{The black line displays the evolution of the Hubble parameter $H$. Inflation begins at $t_*=0$ and ends at $t_{end}\simeq 9.89\times 10^6$. We have set $M_P=1$. }
\label{figHinf}
\end{figure}

A careful discussion of the primordial parameters involved in inflationary cosmology, as well as their exact relationship to the potential energy function $V(\psi)$ given in \eqref{24}, was presented in \cite{Deen:2016zfr}. The results were the following. Choosing the parameters
\begin{equation}
m= 6.49 \times 10^{-6} ~, \quad Y_{\nu 3} \sim 10^{-12}~, \quad \mu= 4.93 \times 10^{-4} 
\label{25}
\end{equation}
it was shown that the inflaton produced 60 e-foldings of inflation between  the end of inflation at $\psi_{end}$ and a starting value of $\psi_{*}$ where
\begin{equation}
\psi_{end} \simeq 1.21~, \quad \psi_* \simeq 6.25 \ .
\label{30}
\end{equation}
It followed from this that the spectral index $n_{s}$, the tensor-to-scalar ratio $r$ and the energy scale of inflation associated with the above parameters are given by
\begin{equation}
n_s \simeq 0.967~, \qquad r \simeq 0.00326~, \qquad V_{*}^{1/4}= 3.27 \times 10^{-3} 
\label{31}
\end{equation}
respectively. These results are completely consistent with all of the Planck2015 data \cite{Ade:2015lrj}. For the explicit parameters in \eqref{25}, the graph of the complete potential energy associated with the inflationary epoch is shown in Figure \ref{plot2}.
Similarly, for the same choice of parameters, the plot of the Hubble parameter $H$ between the time $t_{end} \simeq 9.89\times 10^6$ at the end of inflation and the time $t_{*}=0$ when inflation begins is shown in Figure \ref{figHinf}.
It is physically enlightening to restore mass units for both the mass of the inflaton as well as the $\mu$ coefficient. Then the results in \eqref{25} are given in by
\begin{equation}
m=1.58 \times 10^{13}~{\rm GeV}~,  \quad Y_{\nu 3} \sim 10^{-12}~, \quad \mu=1.20 \times 10^{10}~{\rm GeV}
\label{wk1}
\end{equation}
respectively. It follows that the scale of supersymmetry breaking--which is associated with $m$--is very large, whereas the $\mu$-term is considerably smaller. Both scales play an important role in the analysis of Sneutrino-Higgs inflation and reheating.

The parameters presented in \eqref{25},\eqref{wk1} are specific representatives of a wider class of parameters, all of which satisfy the Planck2015 data. Generically, one can choose any parameter $m \sim 10^{13} ~\mathrm{GeV}$ and $Y_{\nu 3} \sim 10^{-12}$. The allowed value of the parameter $\mu$, however, is much wider. As was discussed in \cite{Deen:2016zfr}, its value has an upper bound given by $\mu= 1.20 \times 10^{10} ~\mathrm{GeV}$ presented in \eqref{25},\eqref{wk1}. However, any smaller value is allowed as long as it is consistent with the breaking of electroweak symmetry. For these values of $\mu$, it was shown in \cite{Deen:2016zfr} that $V_{F} \ll V_{soft}$ both during and after the inflationary epoch. That is, $V_{F}$ is only substantial prior to inflation.

\subsection{~The Search for Valid Low Energy Points}

\begin{figure}
\begin{center}
\includegraphics[scale=0.55]{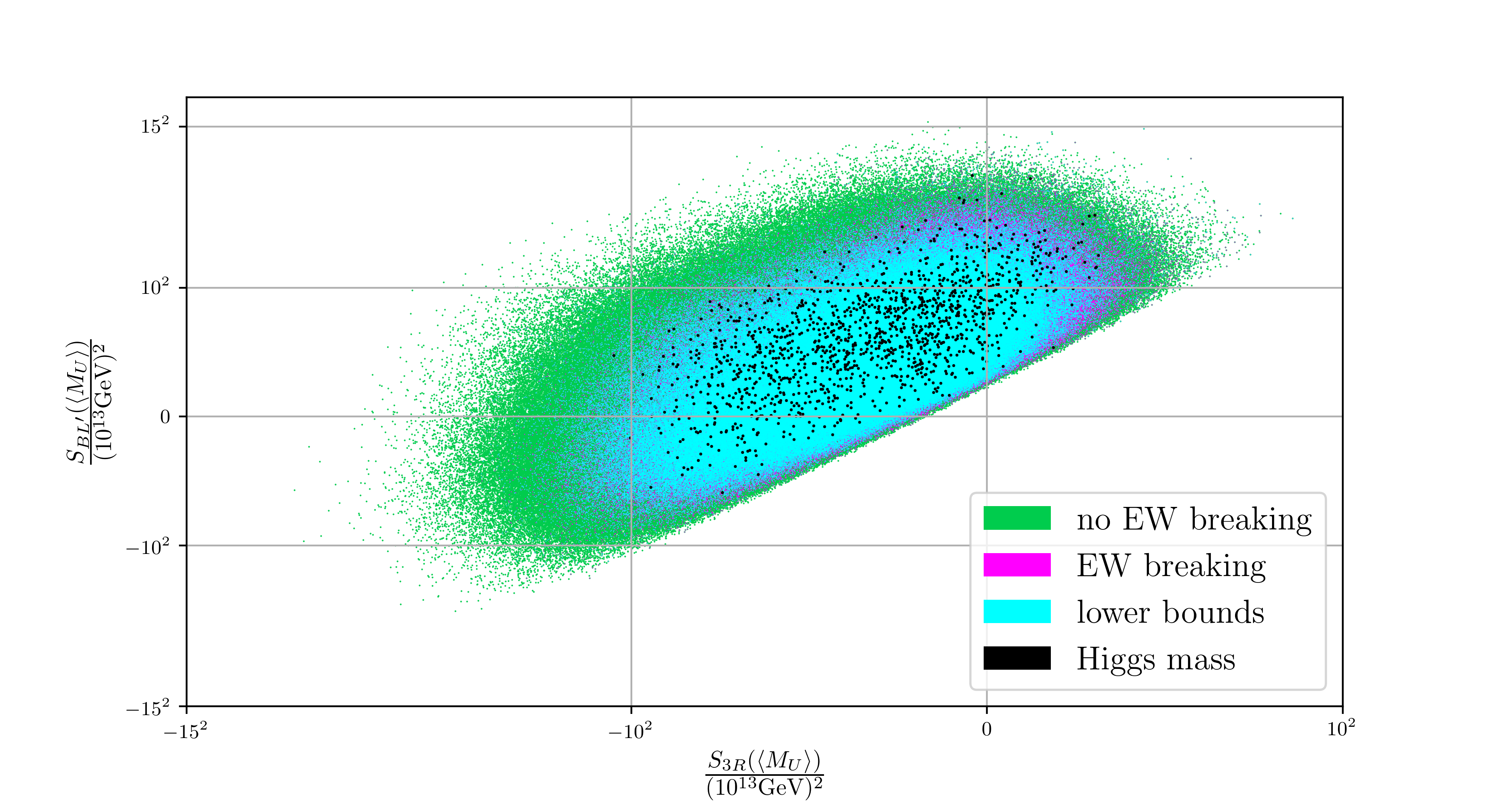}
\caption{Results from generating 50 million sets of initial data where $m_{H_u}^2$ is fixed by the cosmological constraint. We find that 4,209,300 points break \BL but not electroweak symmetry, and 860,084 points appropriately break both \BL and electroweak symmetry. Of the latter, 545,753 points are consistent with current LHC bounds on sparticle searches. Finally, we have 1406  points which satisfy all these conditions and are within the $2\sigma$ window of the measured Higgs mass. The black points are enlarged for legibility.The axes are two dominant parameters of the renormalization group equations and are defined in \cite{Ovrut:2015uea}.}
\label{fig:blackpoints2}
\end{center}
\end{figure}

In \cite{Deen:2016zfr}, we used the formalism presented in \cite{Ovrut:2015uea} to statistically search the space of initial soft supersymmetry breaking parameters for those points which 1) satisfy the Planck2015 data  by using the parameters presented in \eqref{wk1} while 2) simultaneously being consistent with all present low energy phenomenological data--that is, appropriate \BL and EW breaking, all lower bounds on SUSY sparticles and the experimentally measured lightest neutral Higgs mass. We refer the reader to \cite{Ovrut:2015uea} for details of this formalism. Since the relevant phenomenological data is usually presented in GeV, we will work in this unit for this entire analysis, including the discussion of lowering the \BL scale presented in the following section. Suffice it here to say that initial dimensional soft SUSY breaking parameters are analyzed by randomly scattering all of them, with the exception of $m_{H_{u}}$, 
in the interval $[m/f,fm]$, where $m= 1.58 \times 10^{13}~ \mathrm{GeV}$ is the cosmologically consistent mass presented in \eqref{wk1} and $f=3.3$. The parameter $m_{H_{u}}$ is then fixed, for each initial throw, by demanding that it satisfy the cosmological constraint given by \eqref{14} and \eqref{wk1}. In this paper, to be consistent with the analysis below, we extend the results given in \cite{Deen:2016vyh} by statistically throwing the initial parameters 50 million times instead of the 10 million times presented previously. The results satisfying both requirements 1) and 2) are shown as the ``valid'' black  points in Figure \ref{fig:blackpoints2}.

\begin{figure}[h!]
\begin{center}
\includegraphics[scale=0.3]{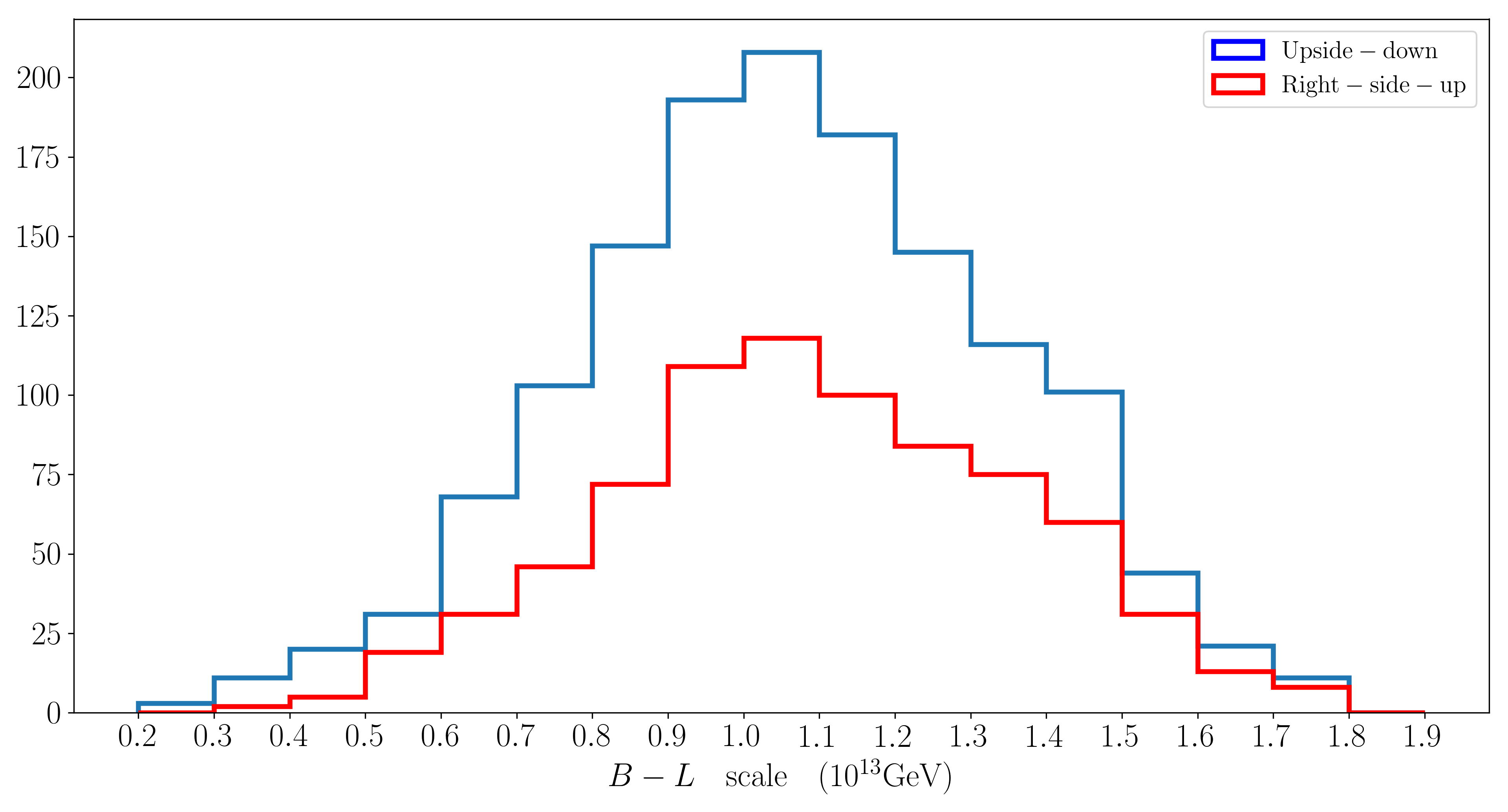}
\caption{Plot of the $U(1)_{B-L}$ breaking scale for the 1406 valid black points shown in Figure \ref{fig:blackpoints2}. The \BL and supersymmetry breaking scales are computed using the formalism presented in \cite{Ovrut:2015uea}. We indicate what fraction of each bin consists of those cases in which $M_{BL} > M_{SUSY}$ (right-side-up), and in which $M_{BL}<M_{SUSY}$ (upside-down). For example, between 1.0 and 1.1 $ \times 10^{13}~{\text{GeV}}$ the number of right-side-up valid points is $\approx 120$ whereas the number of upside-down valid points is $\approx 210-120=90$.}
\label{fig:BLscale}
\end{center}
\end{figure}

It is of interest to use the formalism presented in \cite{Ovrut:2015uea} to compute the \BL breaking scale associated with each of the 1406 valid black points presented in Figure \ref{fig:blackpoints2}. This quantity was not  
analyzed in any of the previous papers on the \BLMSSM and Sneutrino-Higgs inflation. However, as we will see below, knowledge of the \BL breaking scale is important in discussing reheating. Therefore, we have computed the \BL scales for all valid black points and present a statistical graph of the results in Figure \ref{fig:BLscale}. For completeness, we have indicated the percentage of points for which the \BL scale $M_{BL}$ exceeds or is smaller than the supersymmetry breaking scale $M_{SUSY}$ defined in \cite{Ovrut:2015uea}. This is referred to as  ``right-side-up'' and ``upside-down'' \BL breaking respectively. 
It is important to note that the smallest \BL scale associated with the valid black points is approximately $2 \times 10^{12}~\mathrm{GeV}$. As we will see in the following, appropriate reheating will occur most naturally for values of $M_{BL} \lesssim 10^{12}~\mathrm{GeV}$. It is important, therefore, to see if one can modify the initial statistical input of the soft breaking parameters so that one obtains physically realistic valid black points for which $M_{BL}$ is substantially smaller than $10^{12}~\mathrm{GeV}$.
The answer is affirmative, as will be shown in the following section.

\section{Lowering the \BL scale}

We would like to arbitrarily lower the scale of \BL breaking for a given set of initial data so that the RGE evolution of this data is consistent with all phenomenological constraints; that is, 1) the electroweak scale is radiatively broken with the correct Z and W boson masses, 2) all sparticle masses exceed their present experimental lower bounds and 3) the Higgs mass is given to within 2$\sigma$ of its measured value of $125~\mathrm{GeV}$. To accomplish this, we will no longer compute the scale of \BL breaking as an ``output" of the initial class of data discussed in \cite{Ovrut:2015uea} and used in the previous section.  Rather, we will input the \BL scale as an arbitrary parameter as part of the initial data. 

To do this, we recall from \cite{Ovrut:2015uea} that the $U(1)_{B-L}$ symmetry breaks when the right-handed sneutrino $\tilde{\nu}_3^c$ obtains a non-vanishing vacuum expectation value. This occurs when the parameter $m^2_{\tilde{\nu}_3^c}$ turns negative as one runs down in energy-momentum from the unification scale $\langle M_{U} \rangle$ defined in \cite{Deen:2016vyh}. Since natural reheating will require a lower value of $M_{BL}$, we will assume that our \BL scale will be less than the scale of SUSY breaking; that is, we will only  consider the ``upside-down'' hierarchy where $M_{BL} < M_{SUSY}$. The \BL scale $M_{BL}$ is defined in \cite{Ovrut:2015uea} via the recursive relation
\begin{eqnarray}
M_{Z_R}(M_{BL}) = M_{BL} \,,
\label{eq:BL1}
\end{eqnarray}
where $M_{Z_R}$ is the mass of the $Z^\prime$ boson, which receives a mass when the right-handed sneutrino develops a vacuum expectation. We also have the relation 
\begin{eqnarray}
M_{Z_R} \simeq \sqrt{2} |m_{\tilde{\nu}_3^c}| \ .
\label{eq:BL2}
\end{eqnarray}
It follows that 
\begin{eqnarray}
M^2_{BL} \simeq - 2 m_{\tilde{\nu}_3^c}^2 (M_{BL})\, ,
\label{eq:BL3}
\end{eqnarray}
where $m_{\tilde{\nu}_3^c}^2 <0$ when \BL is broken.
From this relation, we see that fixing $M_{BL}$ at a particular value demands that $m_{\tilde{\nu}_3^c}^2 (M_{BL})$ also be fixed. If we continue to randomly generate $m_{\tilde{\nu}_3^c}$ at the unification scale $\langle M_U \rangle$, as was done in all previous analyses, it is exceedingly improbable that, upon running the sneutrino mass down, we will arrive at our desired value of $m_{\tilde{\nu}_3^c}^2 (M_{BL})$ and, hence, of $M_{BL}$. This problem is concretely expressed by the results shown in Figure \ref{fig:BLscale}.

We therefore change our approach from previous work, and no longer randomly generate the sneutrino mass $m_{\tilde{\nu}_3^c}^2$ at the unification scale. Instead, we specify the desired \BL breaking scale and use \eqref{eq:BL3} and the relevant RGEs to {\it determine the required soft sneutrino mass parameter at $\langle M_U \rangle$} . Once this process is accomplished, we then have a complete set of initial data against which the phenomenological constraints 1), 2) and 3) above can be verified. To carry this out in detail relies heavily on a generalization of the formalism for the renormalization group equations of the \BLMSSM previously given in \cite{Ovrut:2015uea}. This is technically non-trivial and, hence, we present the mathematical details in Appendix A of this paper. In this section, we will simply use the results obtained in that Appendix.

\begin{figure}
\begin{center}
\includegraphics[scale=0.5]{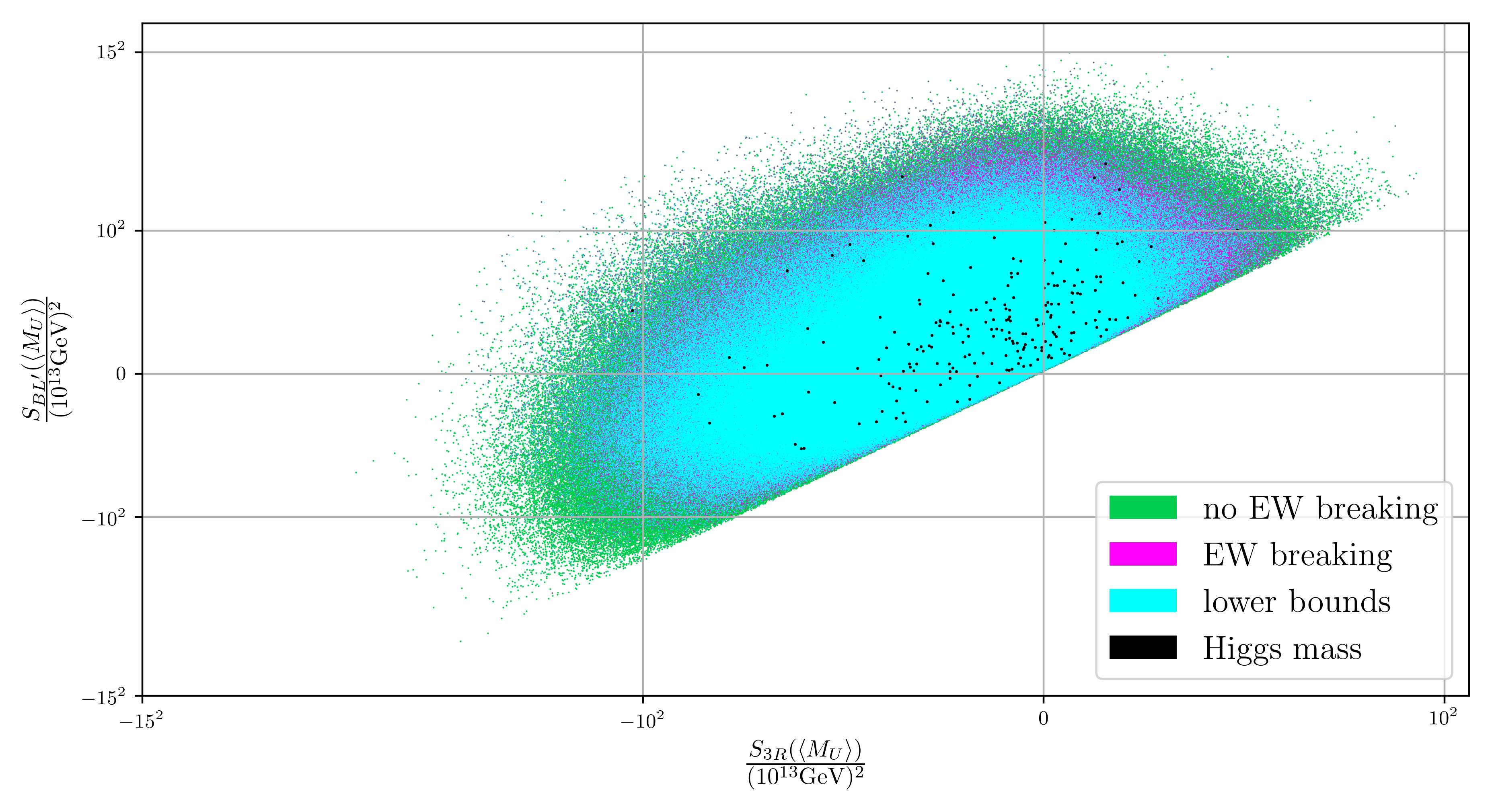}
\caption{Results from generating 50 million sets of initial data where the \BL scale is chosen from a log-uniform distribution between $10^{10}$ GeV and $10^{12}$ GeV. We find that 5,949,281 points break \BL but not electroweak symmetry, and
1,937,174 points break \BL and electroweak symmetry. Of the latter
1,283,484 points are consistent with current LHC bounds on sparticle searches. Finally, we have 215 points which satisfy all these conditions and are within the $2\sigma$ window of the measured Higgs mass.}
\label{fig:blackpoints3}
\end{center}
\end{figure}

\begin{figure}[h!]
\begin{center}
\includegraphics[scale=0.35]{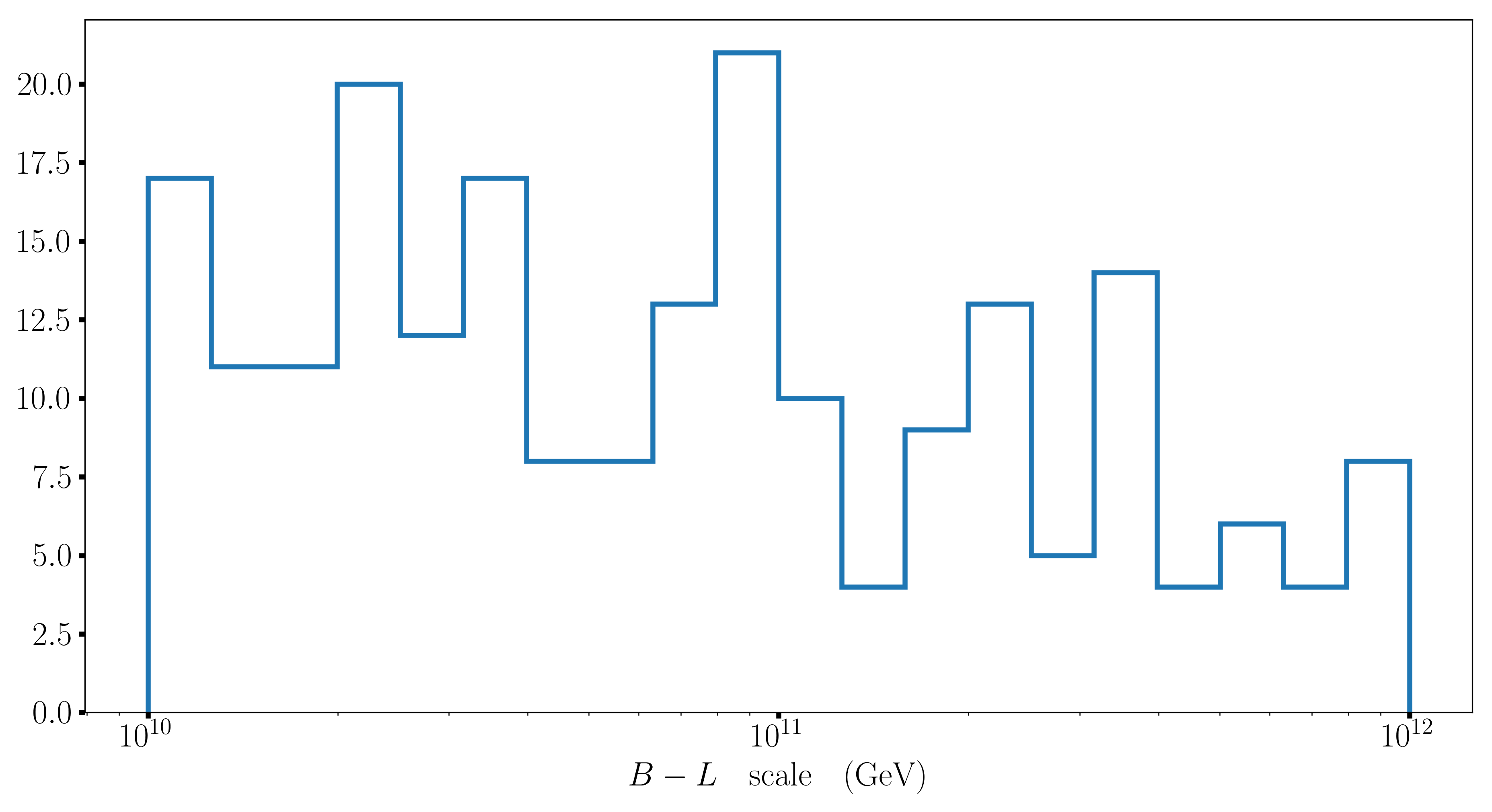}
\caption{Distribution of the \BL breaking scale for the 215 black points displayed in Figure \ref{fig:blackpoints3}. The vertical axis labels the number of valid black points.}
\label{fig:BLScale2}
\end{center}
\end{figure}

As discussed in subsection A.2, one can choose a range over which one wants to input the \BL scale and then, using the formalism presented there, determine the phenomenologically acceptable valid black points whose \BL scales lie in that range. In the Appendix, we carried this out over a very wide range of \BL scales, specifically
from $10^6$ {\rm GeV} to $10^{14}$ {\rm GeV}. However, for the discussion of reheating in this paper, such a wide range for $M_{BL}$ is not required. Instead, we will limit our discussion in the text to \BL scales in the range  $10^{10} ~{\rm GeV} \leq M_{BL} \leq 10^{12} ~{\rm GeV}$.
Let us implement the procedure outlined in the Appendix, now, however, for this restricted range of \BL scales.  We will generate 50 million initial throws of the soft masses with the inputted scale of $U(1)_{B-L}$ breaking randomly generated from a log-uniform distribution between $10^{10}$ GeV and $10^{12}$ GeV. Carrying out our checks, we find that this ultimately leads to 215 sets of initial data which satisfy all phenomenological constraints. These physically valid black points are shown in Figure \ref{fig:blackpoints3}.
The distribution of the \BL breaking scale for the black points in Figure \ref{fig:blackpoints3}
is given in Figure \ref{fig:BLScale2}. 

\begin{figure}
\begin{center}
\includegraphics[scale=0.4]{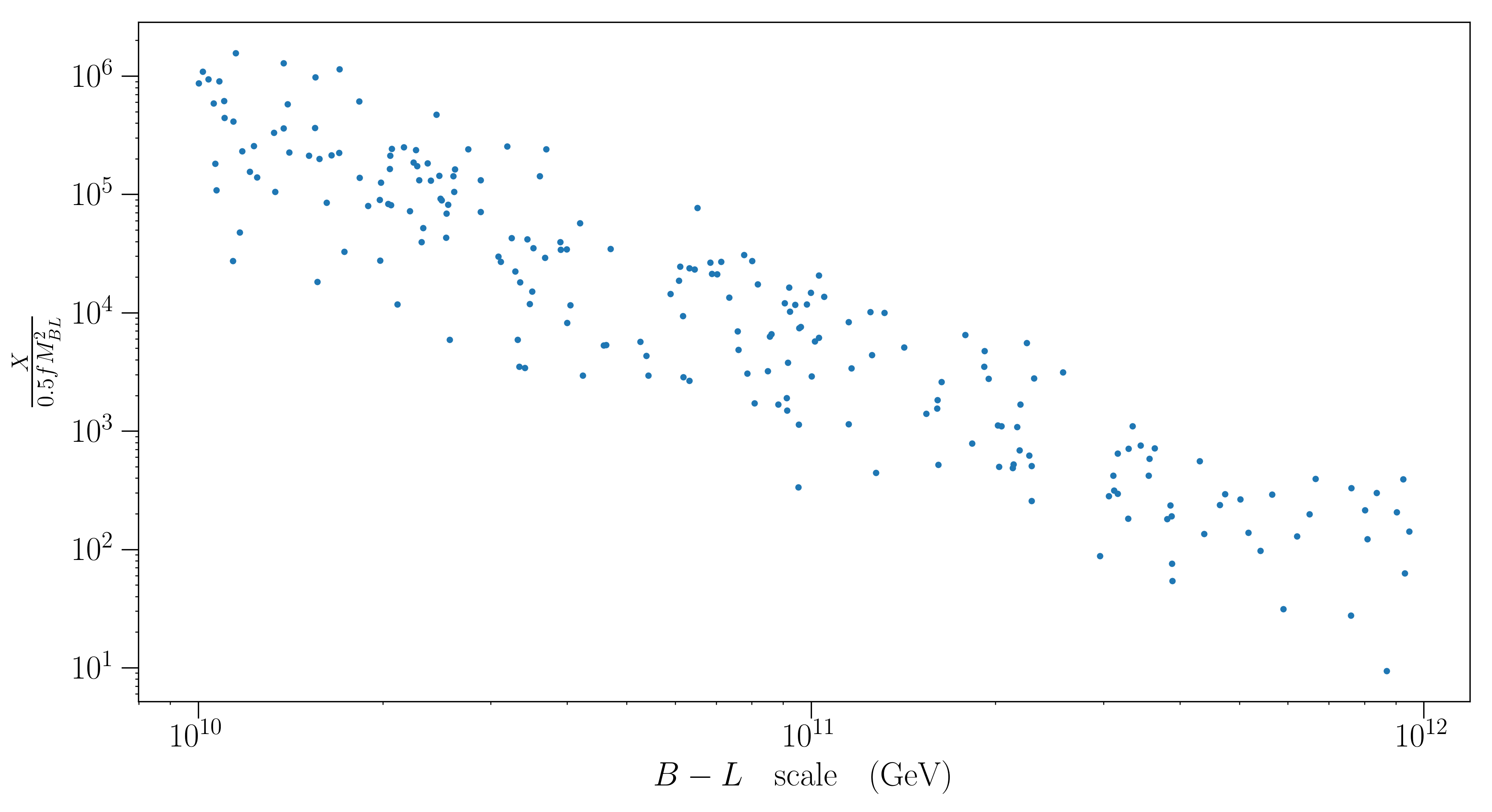}
\end{center}
\caption{Log-log plot of $\frac{X}{\frac{1}{2} f M_{BL}^2}$ against the \BL scale, for the valid black points shown in Figure \ref{fig:blackpoints3} from the scan of 50 million sets of initial conditions. The quantity $\frac{X}{\frac{1}{2} f M_{BL}^2}$ expresses the degree of fine-tuning required to achieve the associated value of the \BL scale. The expression for $X$ is presented in Appendix A.3.}
\label{fig:fine-tuning-log}
\end{figure}

Finally, using the formalism discussed in subsection A.3 of the Appendix, we compute the amount of fine-tuning required to lower the \BL scale into the $10^{10}$ GeV to $10^{12}$ GeV range. The results for the 215 valid black points are shown in Figure \ref{fig:fine-tuning-log}. Note that the degree of fine-tuning is of ${\cal{O}}(10^{4}-10^{6})$ for \BL scale of order $10^{10}$ GeV and of ${\cal{O}}(10^2-10^{3})$ for \BL scale of order $10^{12}$ GeV. We note in passing that all the black points in Figure \ref{fig:blackpoints3} are in the so-called ``upside-down" hierarchy, with $M_{BL} < M_{SUSY}$.

\section{Post-Inflationary Epoch: Classical Behavior of $\psi$ and $H$}

In this Section, we will begin our discussion of the post-inflationary epoch, assuming, for the time being, that the inflaton does {\it not} decay to normal matter.  Within this context, we will calculate the classical behavior of the inflaton $\psi$ and the Hubble parameter $H$ numerically, and then present analytic solutions for both quantities which closely approximate the numerical results.  Having achieved this, we will, in the next section, begin our discussion of the details of the inflaton decay to normal matter and reheating. For this analysis, it is far more convenient to work in units where $M_{P}=1$.

\begin{figure}[htbp]
        \subfigure[~~]{\includegraphics[width=.9\textwidth]{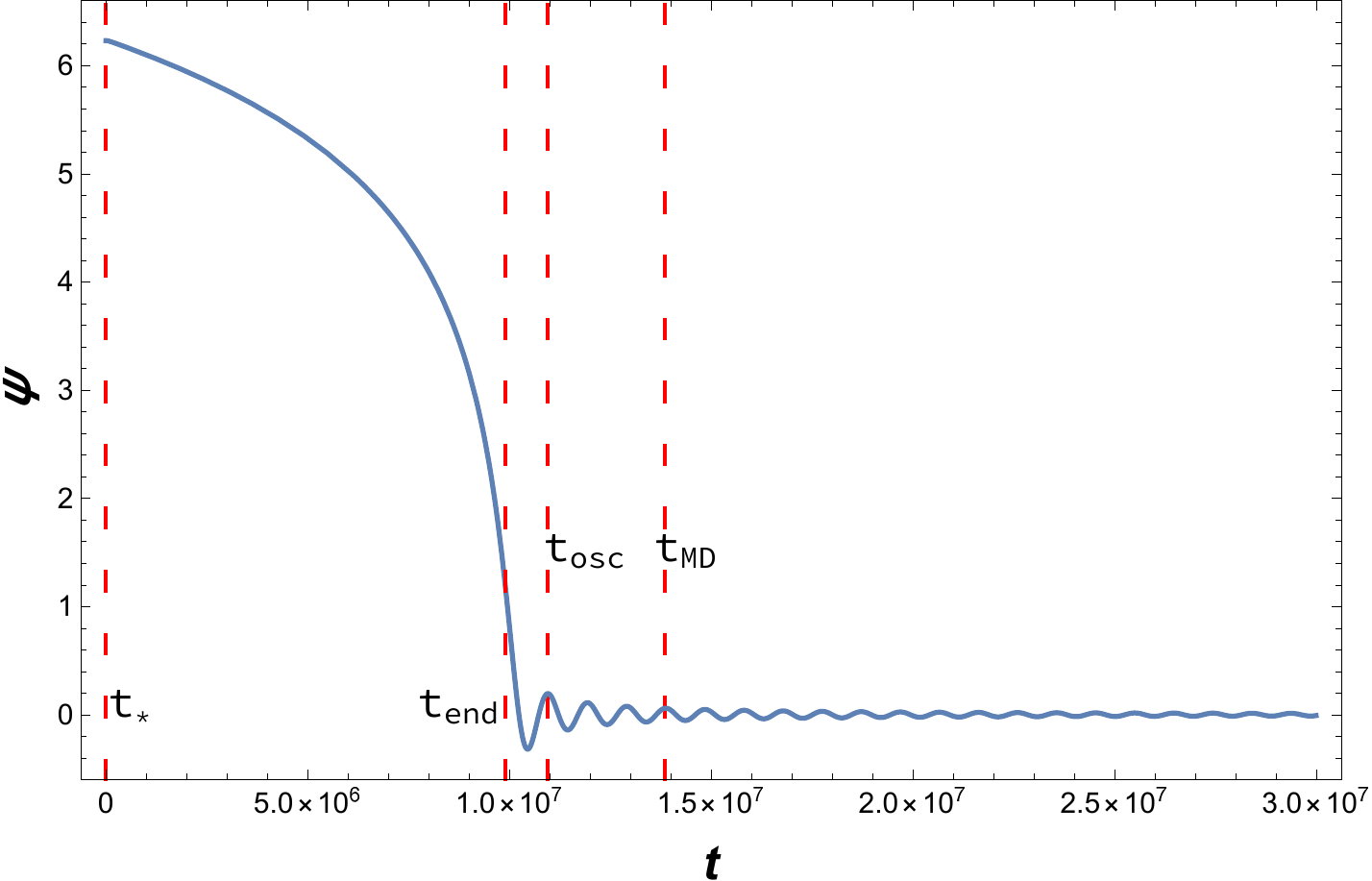} }
        \subfigure[~~]{\includegraphics[width=.9\textwidth]{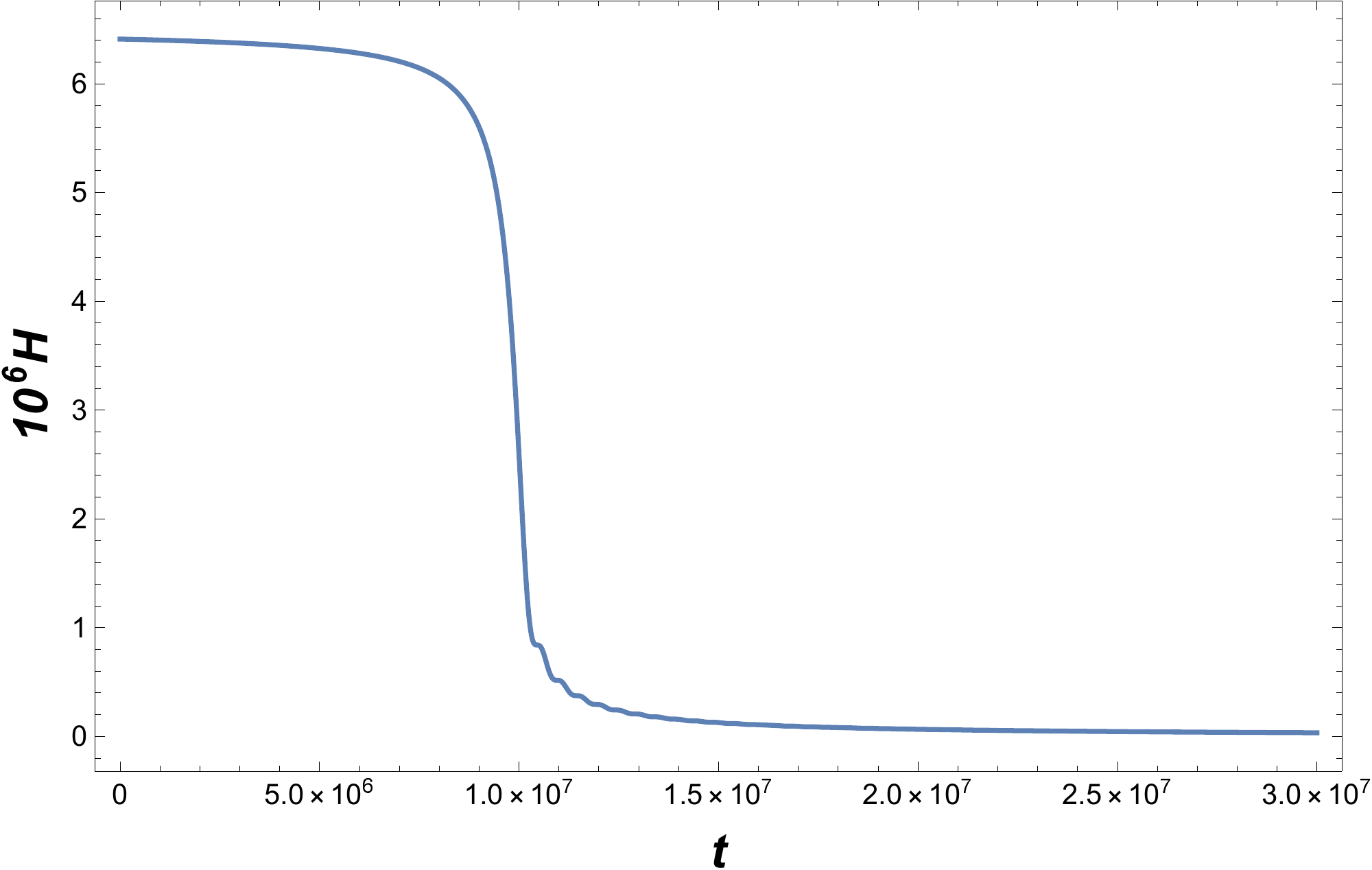} }
	\caption{The numerical solutions for $\psi(t)$ and $H(t)$, where we have set $M_P=1$. Note that $t_*=0$ and $t_{end}\simeq 9.89\times 10^6$ mark the beginning and end of the inflationary period.  The times $t>t_{end}$ correspond to the post inflationary epoch. As defined in the text,  $t_{osc}\simeq 1.096\times 10^7$ marks the time at which the potential energy is well approximated by $V=\frac{1}{2} m{\psi^{2}}$ and $t_{MD}\simeq 1.387\times 10^7$  is the time at which our analytic solutions for $\psi$ and $H$ become valid.} \label{figbg01}
\end{figure}

In the inflationary and post-inflation epoch, the equations of motion for $\psi$ and the Hubble parameter $H$ are specified by equations \eqref{21}, \eqref{22} and \eqref{23}. The values for the parameters $m$, $Y_{\nu3}$ and $\mu$ will be chosen to be those given in \eqref{wk1} to ensure that the inflating epoch is consistent with all phenomenological data. As discussed above, for this choice of the $Y_{\nu3}$ and $\mu$ the F-term potential satisfies $V_{F} \ll V_{soft}$ in both the  inflationary and post-inflation regimes. Therefore, in this section, we can, to a high degree of accuracy, simply take the potential energy to be $V=V_{soft}$. Then the relevant equations of motion are given by
\bea
&&3H^{2}=\frac{1}{2}{\dot{\psi}}^{2}+V(\psi) \label{p1}  \\
&&\dot{H}=-\frac{1}{2}{\dot{\psi}}^{2} \ , \label{p2} \\
&&\ddot{\psi} + 3H\dot{\psi} +V_{,\psi}=0 \ , \label{p3}  
\eea
where 
\begin{equation}
V(\psi)= 3 m^2 \tanh^2\left( \frac{\psi}{\sqrt{6}}\right) \ .
\label{p4}
\end{equation}
These equations can be solved numerically for both $\psi$ and $H$ as functions of time. The results for $\psi(t)$ and $H(t) $ starting 1) at the beginning of inflation at $t_{*}=0$, 2) running through the inflationary epoch to $t_{end} \simeq 9.89 \times 10^{6}$, and then 3) continuing into the post-inflation epoch with for $t >t_{end}$, are shown in Figure \ref{figbg01} (a) and (b) respectively.

As is clear from Figure \ref{figbg01}(a), shortly after the end of the inflationary period, the inflaton will begin to oscillate around the minimum of its potential at $\psi=0$.  
Taylor expanding  $V(\psi)$ in \eqref{p4} around the origin, one obtains
\be V(\psi)= {1\over 2} m^2\psi^2\lf[1-\lf({\psi\over 3}\rt)^2+\lf(\frac{17\psi^4}{1620}\rt)+ \dots \rt]\,.\label{potentialexpand}
\ee
When $\psi\ll 3$, $V\approx {1\over 2} m^2\psi^2$ and the mass of inflaton is
$m_\psi=\sqrt{V_{\psi\psi}}\approx m$. Noting that $m=6.5 \times 10^{-6}$, it follows from Figure \ref{figbg01}(b) that $m\gg H$ everywhere in the post-inflationary period. Hence, $\psi$ will oscillate coherently around the minimum of $V$ with a frequency $\omega=m_\psi$, although with decreasing amplitude. From Figure \ref{figbg01}(a), we can numerically show that the height of the first oscillatory peak corresponds to $\lf({\psi\over 3}\rt)^2 \simeq 4.2 \times 10^{-3}$, which easily satisfies the above criterion that $\psi\ll 3$. Henceforth, for specificity, we consider this first peak as the beginning of the oscillatory phase and will denote the corresponding time, which we numerically compute, to be $t_{osc} \simeq 1.096\times 10^7$. This time is indicated by a dashed red line in Figure \ref{figbg01}(a). For all $t > t_{osc}$ we will, henceforth, take $V(\psi)= {1\over 2} m^2\psi^2$.
During the oscillatory  period, that is, when, $t>t_{osc}$, equations \eqref{p1}, \eqref{p2} and \eqref{p3}  then become
\ba &\,&3H^2={1\over 2}\dot{\psi}^2+{1\over 2} m^2\psi^2\,,\label{Fried02}\\
&\,& \dot{H}=-{1\over2}\dot{\psi}^2\,,\label{Fried03}\\
&\,& \ddot{\psi}+3H\dot{\psi}+ m^2\psi=0\,.\label{eqpsi02}
\ea

We now want to find an approximate analytic solution to these equations for both $H$ and $\psi$. To do this, we first neglect the effect of $3H\dot{\psi}$ in (\ref{eqpsi02}). That is, we will take the lowest order approximation to $H$ to be $H_0=0$. Then the solution of $\psi$ to this order is nothing but an harmonic oscillator. That is
\be \psi(t) \approx \psi_0(t)=A_0 \sin[m (t-c_1)]\,,\label{psi0}
\ee
where $A_0$ and $c_1$ are constants. Using (\ref{psi0}), we have
\ba {1\over2}\dot{\psi}_0^2&=&{1\over2}A_0^2 m^2\cos^2[m (t-c_1)]\,,\\
V(\psi_0)&=&{1\over2}A_0^2 m^2\sin^2[m (t-c_1)]\,.
\ea
Plug these expressions into (\ref{Fried02}) and (\ref{Fried03}), yields
\ba 3H_1^2&=&{1\over2}A_0^2 m^2\,,\\
\dot{H}_1&=&-{1\over2}A_0^2 m^2\cos^2[m (t-c_1)]\,,
\ea
where $H_{1}$ is the first order approximation to $H$. It then follows that
\be {d\over dt}\lf({1\over H_1} \rt)=-{\dot{H_1}\over H_1^2}=3\cos^2[m (t-c_1)]\,.
\ee
Integrating this from $t_{osc}$ to $t (> t_{osc})$ yields 
\ba 
&&{1\over H_1(t)}-{1\over H_1(t_{osc})}=3\int_{t_{osc}}^t \cos^2[m (t'-c_1)] dt'  \nn\\
&&={3\over2}\int_{t_{osc}}^t 1+ \cos[2m (t'-c_1)] dt'\nn\\
&&={3\over2}(t-t_{osc})+{3\over 4m}\lf\{\sin[2m (t-c_1)]-\sin[2m (t_{osc}-c_1)] \rt\}\,.
\ea
Clearly, for times $t$ where $t-t_{osc}\gg 1/m$, we have 
\be H\approx H_1(t) = {2\over 3(t-c_2)}\label{H1-01}
\ee
where 
\be  c_2=t_{osc}-{2\over 3 H(t_{osc})}\,.
\ee
By numerically evaluating $H(t_{osc})$ using the results displayed in Figure \ref{figbg01}(b), we find that $c_{2} \simeq 9.67 \times 10^{6}$.

Having found an approximate analytic expression for $H(t)$ beyond leading order, we now want to find the next order correction to $\psi_{0}$ given in \eqref{psi0}. Due to the non-vanishing expansion of the Universe given by $H\approx H_1$, the amplitude of the oscillations of $\psi$ will necessarily  be damped. However, the frequency of the $\psi$ oscillations will hardly be effected as long as $m\gg H$ which, as mentioned previously, will be true for all $t>t_{osc}$ and, hence, for $t-t_{osc}\gg 1/m$. Thus we can set
\be \psi(t)\approx \psi_1(t)=A_1(t) \sin[m (t-c_1)]\,,\label{psi1}
\ee
where $A_1(t)$ can be obtained by
inserting expressions (\ref{H1-01}) and (\ref{psi1}) into equation (\ref{eqpsi02}). This gives
\be \lf(\ddot{A}_1+{2\over t-c_2}\dot{A}_1\rt)\sin[m (t-c_1)]+\lf(2\dot{A}_1 m+{2\over t-c_2} A_1 m\rt)\cos[m (t-c_1)]=0
\ee
and, hence,
\ba \ddot{A}_1+{2\over t-c_2}\dot{A}_1&=&0\,,\\
\dot{A}_1 +{1\over t-c_2} A_1 &=&0\,.
\ea
The solution is
\be A_1(t)={{\cal A}_1\over t-c_2}\,,\label{A_1}
\ee
where ${\cal A}_1$ is a constant. Putting  (\ref{H1-01}), (\ref{psi1}) and (\ref{A_1}) into (\ref{Fried02}), we find that
\be \frac{4}{3}={\cal A}_1^2\lf(\frac{\sin ^2\left[m \left(t-c_1\right)\right]}{2 \left(t-c_2\right){}^2}-\frac{m \sin
	\left[m \left(t-c_1\right)\right] \cos \left[m
	\left(t-c_1\right)\right]}{t-c_2}+\frac{m^2}{2} \rt)\,.
\ee 
When  $t-c_2\gg 1/m$, which is automatically satisfied when $t-t_{osc}\gg 1/m$, we simply have
\be {\cal A}_1=\sqrt{8\over3}{1\over m}\,.
\ee
Therefore, the next order analytic solution for $\psi(t)$, valid in the region where  $t-t_{osc}\gg 1/m$, is given by
\begin{equation}
\psi(t)\approx \psi_1(t)=\sqrt{8\over3}{1\over m(t-c_2)} \sin[m (t-c_1)]  \\
\label{p6}
\end{equation}
where $c_{2} \simeq 9.67 \times 10^{6}$ was evaluated above. By matching \eqref{p6}  with the oscillations in the numerical solution of $\psi$, see Figure \ref{figbg01}(a), we can find that $c_1 \simeq 9.78\times10^6$.
For specificity, we note from the numerical calculation that the time associated with the fourth oscillatory peak in Figure \ref{figbg01}(a) is given by $t \simeq 1.387\times 10^7$ and satisfies $t-t_{osc} > 6\pi/m \gg 1/m$. Hence, to a high degree of approximation, the analytic solutions for $H$ and $\psi$ are both valid for any time larger than the time of the fourth oscillation peak. Since, as we will show below, this corresponds to the period of matter domination, we henceforth denote this time as $t_{MD}$ and indicated it by a red line in Figure \ref{figbg01}(a). 
To summarize, when $t>t_{MD} \simeq1.387\times 10^7$
\ba &\,&H(t)\approx H_1(t) = {2\over 3(t-c_2)}\label{H1-02}\,,
\\
&\,&\psi(t)\approx \psi_1(t)=\sqrt{8\over3}{1\over m(t-c_2)} \sin[m (t-c_1)]\,,\label{psi1-02}
\ea
where $c_2 = t_{osc}-{2\over 3 H(t_{osc})} \simeq9.67\times 10^6$ and $c_1 \simeq 9.78\times10^6$. 

\begin{figure}[htbp]
	\subfigure[~~]{\includegraphics[width=.9\textwidth]{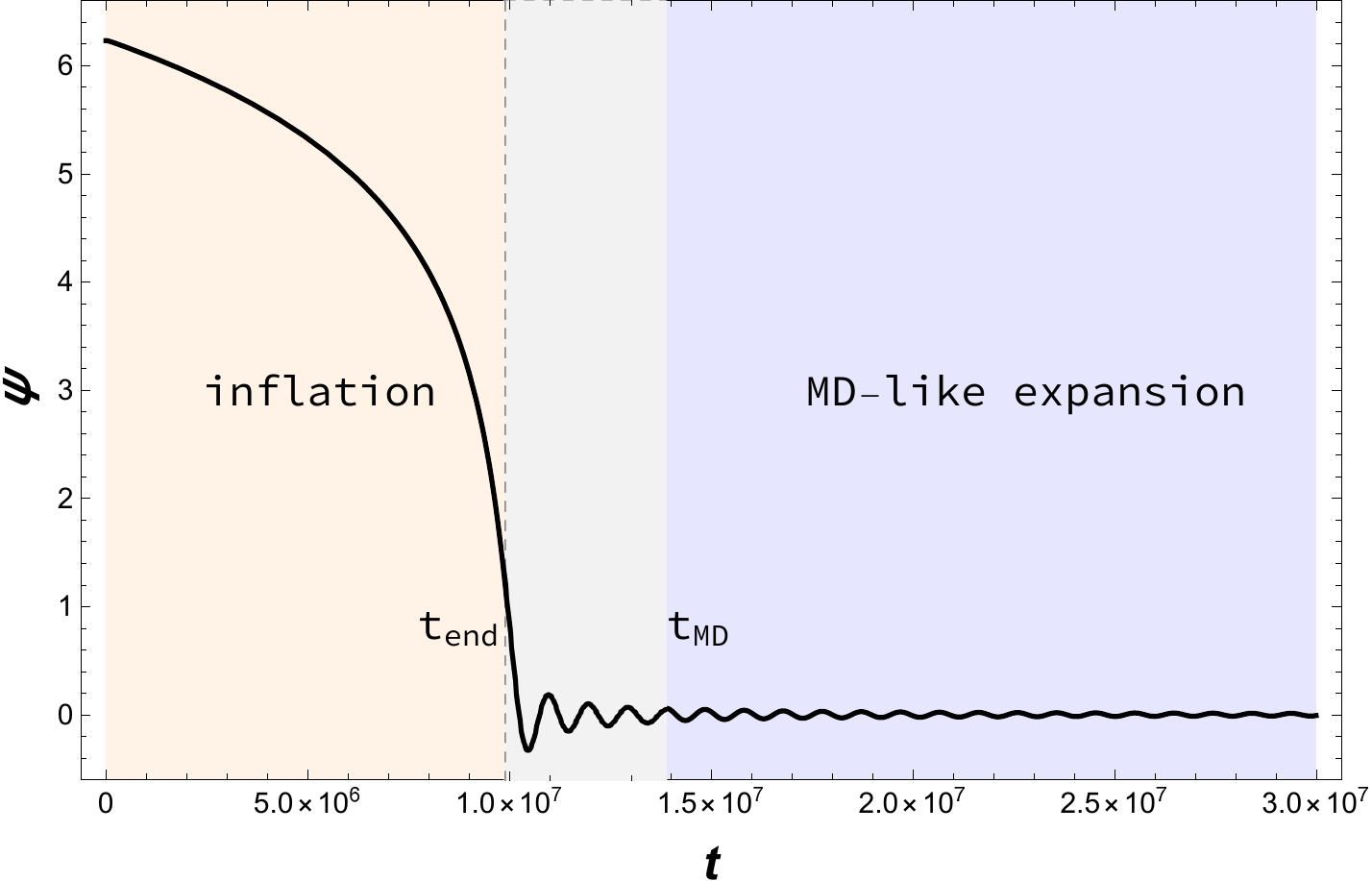} }
	\caption{The inflationary, transition and matter dominated regimes are shown in yellow, grey and blue respectively. We have used $t_{MD} \simeq1.387\times 10^7$ and set $M_P=1$.} \label{figpsinodecay}
\end{figure}

\begin{figure}[htbp]
	\subfigure[~~$t>t_{MD}$]{\includegraphics[width=.5\textwidth]{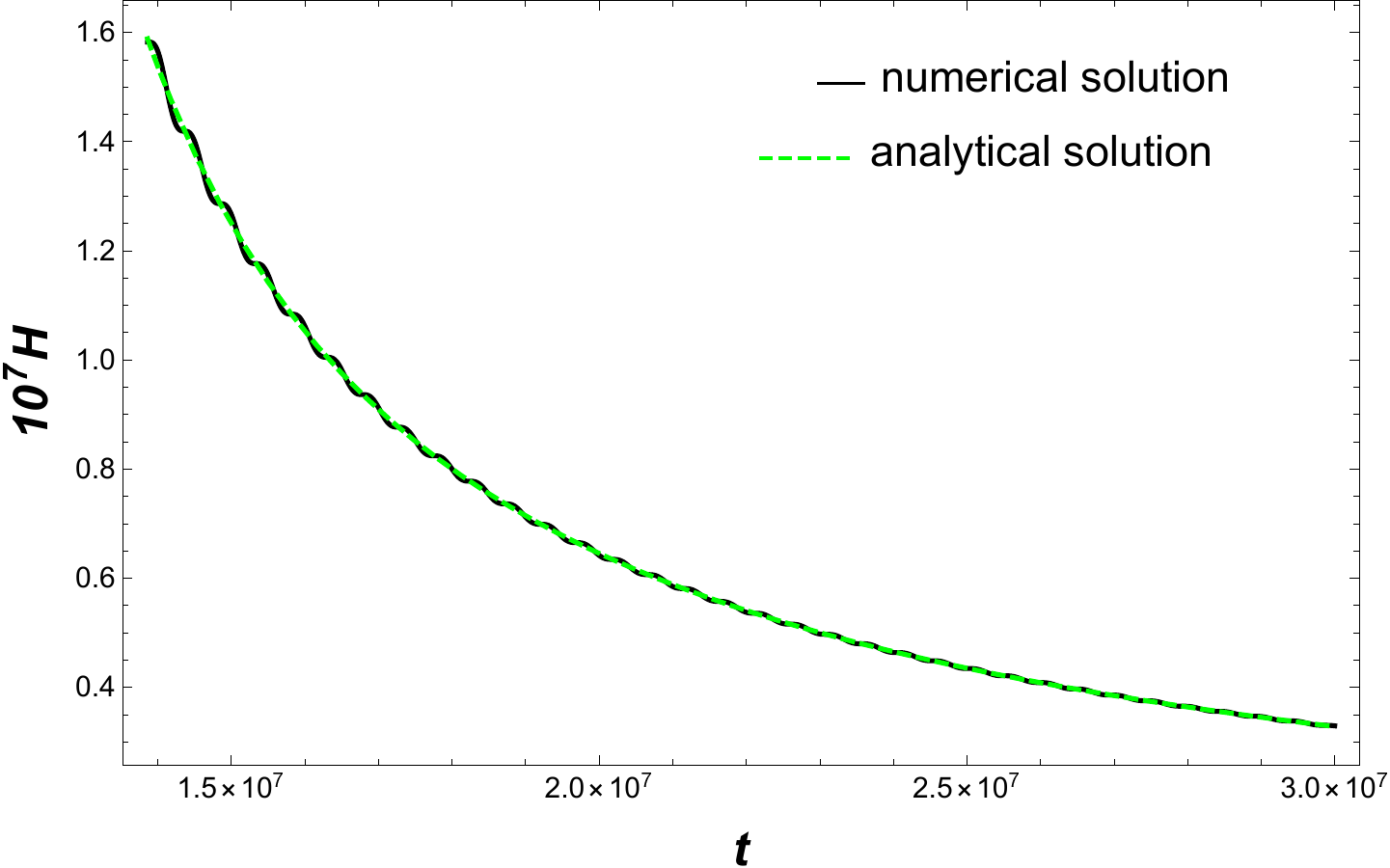} }
	\subfigure[~~$t>t_{MD}$]{\includegraphics[width=.5\textwidth]{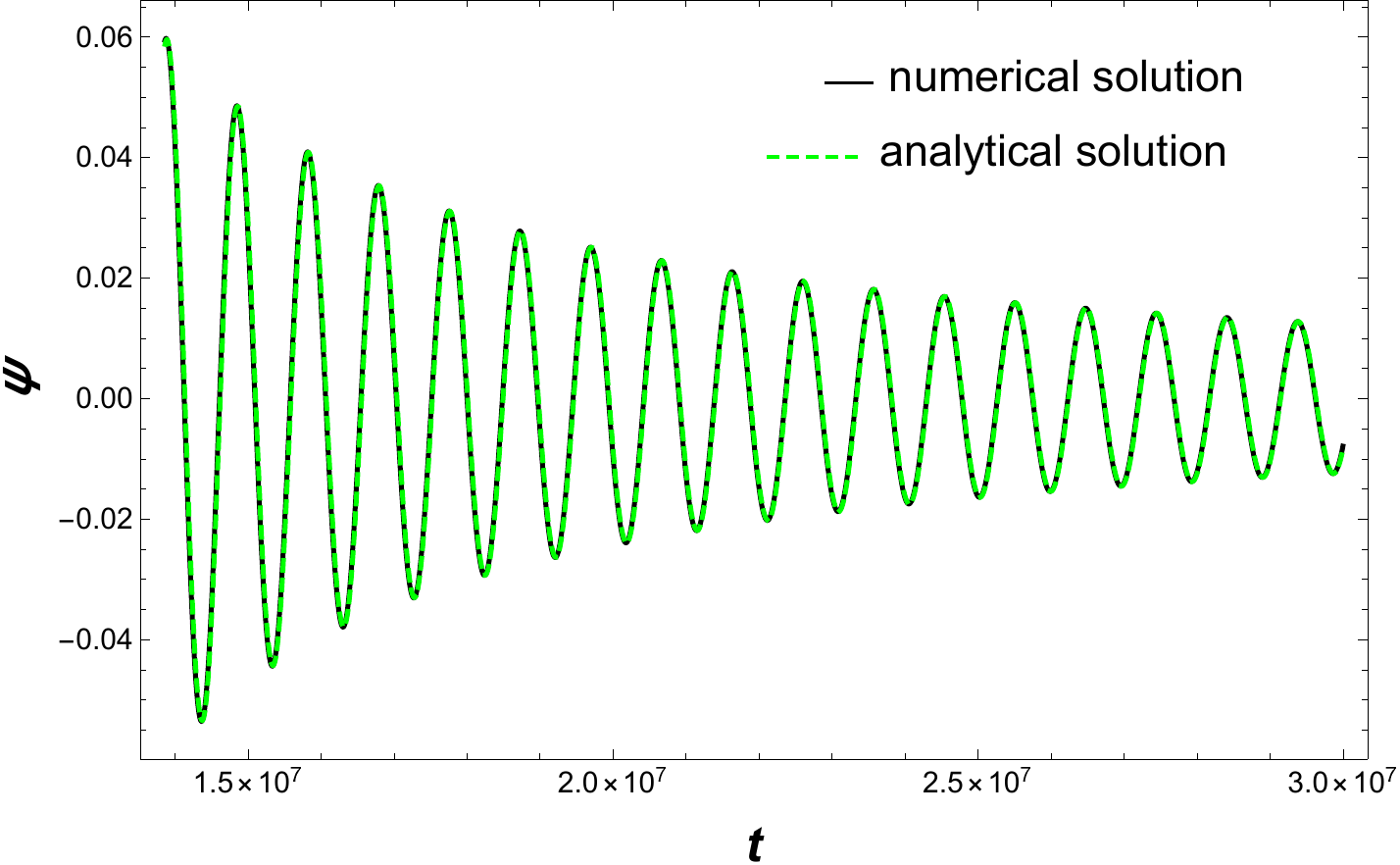} }
	\caption{In (a), the black solid curve and the green dashed curve are the numerical solution and analytical solution (\ref{H1-02}) of $H(t)$, respectively. 
	In (b), the black solid curve and the green dashed curve are the numerical solution and analytical solution (\ref{psi1-02}) of $\psi(t)$, respectively.
	We have used $t_{MD}\simeq 1.387\times 10^7$ and set $M_P=1$.} \label{figpsinodecay2}
\end{figure}

The numerical values of $H$ and $(\psi(t) / 3)^2$ at $t_*$,  $t_{end}$, $t_{osc}$ and $t_{MD}$ are displayed in Table \ref{Table-at-t}.
\begin{table*}[htbp!]
	\begin{center}
		\begin{tabular}{|c|c|c|}
			\hline
			~~ \, ~~& ~~$H(t)$~~ & ~~$\lf({\psi(t) \over 3}\rt)^2$~~~\\
			\hline
			$t_*=0$   & $6.41\times10^{-6}$ & 4.31  \\
			\hline 
			$t_{end}\approx 9.89\times 10^6$   & $3.29\times10^{-6}$ & $0.16$   \\
			\hline 
			$t_{osc}\approx 1.096\times 10^7$   & $5.16\times10^{-7}$ & $4.22\times 10^{-3}$   \\
			\hline 
			~$t_{MD}\approx 1.387\times 10^7$~   & ~~$1.58\times10^{-7}$~~& ~~$3.93\times10^{-4}$~~   \\
			\hline 
			\end{tabular}
		\caption {The values for $H$ and $\lf({\psi\over 3}\rt)^2$ at the beginning and end of inflation, and at the beginning of both the oscillatory and matter dominated regimes respectively. We have set $M_p=1$.}\label{Table-at-t}
	\end{center}
\end{table*}
The regimes of inflation and matter domination are shown as the yellow and blue regions of Figure \ref{figpsinodecay} respectively. 
The duration of the intermediate phase, that is, the gray area in Figure \ref{figpsinodecay}, is given by $\Delta t \simeq t_{MD}-t_{end} \simeq 3.97\times 10^6$. As will be shown below, this is negligible compared with the duration of the reheating period. For that reason, this ``transition'' regime will, henceforth, be ignored. %
Finally, as a check on our approximate analytic solution for $H(t)$ in \eqref{H1-02} and for  $\psi(t)$ in \eqref{psi1-02}, we compare them in Figure \ref{figpsinodecay2} (a) and (b) respectively against the exact numerical solutions for $H$ and $\psi$ in the region $t>t_{MD}$. It is clear that our analytic solution is a very accurate approximation.

\section{Post Inflationary Epoch:  Perturbative Decay of $\psi$ to Matter}

In the previous Section, we ignored the quantum mechanical decay of the inflaton into various species of matter, focussing instead on its purely classical behavior and the associated classical behavior of the Hubble parameter. However, the $\psi$ {\it does} decay into various species of matter, thus reheating the Universe. In this section we commence our discussion of these decays. 

\subsection{Dynamics of $\psi$ and $H$ During Particle Decay}

Different decay processes can be occurring simultaneously, although they may have started at different times.
In general, taking account of the decay of the inflaton, the $\psi$ and $H$ equations \eqref{Fried02}, \eqref{Fried03} and \eqref{eqpsi02} are modified to become
\ba &\,&3H^2={1\over 2}\dot{\psi}^2+V(\psi)+\sum_i\rho_i\,, \label{f1}\\
&\,& \dot{H}=-{1\over2}\dot{\psi}^2-{1\over2}\sum_i\lf(\rho_i+p_i\rt)\,, \label{f2}\\
&\,& \ddot{\psi}+\lf(3H+\sum_i \Gamma_{d,i}\rt)\dot{\psi}+V'(\psi)=0\,,\label{f3}\\
&\,&\dot{\rho}_{i}+3\lf(1+\omega_i(t)\rt)H\rho_{i}-\Gamma_{d,i}\dot{\psi}^2=0\,, \label{f4}
\ea 
where $\Gamma_{d,i}$ is the decay rate of $\psi$ into the $i$-th matter species, and $\rho_i$ and $p_i$ are the energy density and pressure respectively of the $i$-th species in the decay products. The quantities $\rho_i$ and $p_i$ are related by the relation $p_i=\omega_i(t) \rho_i$, where $\omega_i=0$ and $1/3$ respectively for matter and radiation. The initial conditions for $\psi$ and $H$ are set by their classical values at the end of the inflationary epoch, and can be determined from the results in the previous Section. Additionally, we have $\rho_i = 0$ until the time at which the $i$-th decay process commences; that is, when $\Gamma_{d,i}$ becomes non-zero. For convenience, we define the fraction of energy density of 
the $i$-th species as
\be \Omega_i(t)={\rho_i(t)\over \rho_{total}}\,,
\label{ff1}
\ee
where, as follows from \eqref{f1}, the total energy density of the Universe is given by $\rho_{total}=3H^2$. The fraction of energy density of the inflaton can be defined by $\Omega_{\psi}=\rho_{\psi}/\rho_{total}$ with $\rho_{\psi}={1\over 2}\dot{\psi}^2+V(\psi)$.

In our theory, the inflaton is defined in \eqref{13} to be
\begin{equation}
\phi_1 = \tfrac{1}{\sqrt{3}} \left( H_u^0 + \tilde{\nu} _{L,3} + \tilde{\nu}^c _{R,3}\right)
\label{f5}
\end{equation}
with the associated quadratic soft mass squared given in \eqref{14} by
\begin{equation}
m^{2}=\frac{1}{3}(m^{2}_{H_{u}}+ m_{\tilde{L}_3}^{2}+m_{\tilde{\nu}_{R,3}^c}^{2}) \ .
\label{f6}
\end{equation}
The value of $m$ was fixed as 
\begin{equation}
m= 6.49 \times 10^{-6}
\label{f7}
\end{equation}
to be consistent with the results of Planck2015 \cite{Ade:2015lrj}. The relationship of $H_u^0$,  $\tilde{\nu} _{L,3}$ and $\tilde{\nu} _{R,3}^c$ to $\phi_{1}$ was presented in \eqref{12}. Setting $\phi_{2}$ and $\phi_{3}$ to zero in those expressions--since their values vanish in the D-flat potential energy valley of the inflaton--gives
\bea
H_u^0= \tilde{\nu} _{L,3}= \tilde{\nu}_{R,3}^c = \frac{1}{\sqrt{3}}~\phi _1 \ .
\label{f7a}
\eea
Hence, each of these three fields can each be replaced by $\phi_{1}$ in the Lagrangian density.
However, as discussed above, $\phi_{1}$ has a non-trivial K\"ahler potential and, hence, non-canonical kinetic energy. By performing the field redefinition in \eqref{17}, that is
\begin{equation}
\phi_1 = \frac{1}{\sqrt{3}} \tanh\left(\frac{\psi}{\sqrt{6}} \right) \ ,
\label{f8}
\end{equation}
we find that the $\psi$ field is canonically normalized. We therefore used $\psi$ in all our previous analysis. To analyze inflaton decay it is, therefore, essential that we re-express $\phi_{1}$ in terms of $\psi$ in the Lagrangian. Happily, expression \eqref{f8} can be simplified in the post-inflationary regime. Taylor expanding \eqref{f8} around $\psi=0$, we find
\begin{equation}
\phi_{1}= \frac{\psi}{\sqrt{2}} + {\cal{O}}(({\psi})^{2})
\label{f9}
\end{equation}
As can be seen from Figure \ref{figpsinodecay}(b), in the matter dominated period $t>t_{MD}$ we find $\psi \ll 1$. Hence, to a high degree of approximation, one can simply set 
\begin{equation}
\phi_{1}= \frac{\psi}{\sqrt{2}}
\label{f10}
\end{equation}
in the Lagrangian. We do this in the following analysis.
It follows that the decay of $H_u^0$, $\tilde{\nu}_{R,3}^c$ and $\tilde{\nu}_{L,3}$ can be viewed as the decay of the canonical scalar field $\psi$.
As will be shown in detail, $\psi$ is coupled with different classes of particles; including the standard model particles, charginos, nuetralinos, gauge bosons and scalar particles. 

A specific decay process can occur only when the total mass of the decay products is smaller than the mass of $\psi$.
Since shortly after inflation, namely, when $t>t_{osc}$, $\psi$ will oscillate around the minimum of its potential $V(\psi)={1\over 2}m^2\psi^2$, the mass of the inflaton is 
\begin{equation}
m_\psi=\sqrt{V''}=m. 
\label{f11}
\end{equation}
The mass of potential decay products can have two origins; namely, from soft mass terms in the Lagrangian or from the non-zero expectation value of the inflaton $\psi$. In fact, as well as inducing mass terms, the expectation value of the inflaton will also give rise to mixing terms between different fields. For example, the coupling $\frac{1}{\sqrt{2}}g_2 \psi \tilde{W}^- \tilde{\psi}_u^+$, which arises from a super-covariant derivative term, gives rise to mixing between the $\tilde{W}^-$ gaugino and the Higgsino $\psi_u^+$. This will be discussed in more detail below. 

Given the essential role of the inflaton ``expectation value'' in determining the masses and couplings of its decay products, it is essential that this be carefully defined. 
Since $\psi$ oscillates around $0$ when $t>t_{osc}$, its ``naive'' expectation value will vanish. However, it is clear that this is not the physical expectation value effecting the inflaton decay. Rather, we will use the root mean squared value of $\psi$, that is, $\sqrt{\langle\psi^2\rangle}$, where $\langle\psi^2\rangle$ can be defined by 
\be \langle \psi^2\rangle={1\over 2\delta}\int^{t+\delta}_{t-\delta}\psi^2(\tilde{t})d\tilde{t}
\label{f12}
\ee
with $\delta$ being the period of the oscillations of $\psi$. As long as the total decay rate $\sum_i \Gamma_{d,i}\ll m$, which as shown below will always be satisfied, then the frequency of the oscillations of $\psi$ can accurately be taken to be $\delta={2\pi/m_\psi}$. Having defined this root mean squared VEV for the inflaton, we will henceforth expand $\psi$ as 
\begin{equation}
\psi=\sqrt{\langle{\psi^{2}} \rangle} + \delta \psi \ ,
\label{rz1}
\end{equation}
where $\delta \psi$ is a small fluctuation. As we will see below, using this expansion in the Lagrangian density will have two important ramifications; first, it will produce time dependent mass terms proportional to $\sqrt{\langle{\psi^{2}} \rangle}$ for each particle species and second, it will lead to a coupling of the inflaton fluctuation to matter--thus inducing  quantum mechanical reheating of the universe. We will, for simplicity, often abuse notation and denote the fluctuation $\delta \psi$ simply as $\psi$. The correct meaning of the symbol will always be clear from the context.

\subsection{Decay classes}

In this subsection, we present a detailed analysis of the different types of matter into which the inflaton can decay. These are
\begin{enumerate} 
\item up type standard model particles $(\psi\rightarrow t\bar{t}, c\bar{c}, u\bar{u})$; 
\item charginos ($\psi\rightarrow \tilde{C}_1^+ \tilde{C}_1^-$, $\psi_{u}^{-} \tau_{R}$);
\item neutralino ($\psi\rightarrow \tilde{N}_2\tilde{N}_2$).
\item gauge bosons ($\psi\rightarrow W_0^\mu W_{0\mu}, W_{R\mu}W^\mu_R, W^{-\mu}W^+_\mu, B^\mu B_{\mu}$).
\end{enumerate}
These classes are distinguished by whether the decay products are fermions or bosons, and whether their masses must be determined by diagonalizing a mass matrix which arises due to mixing terms from the inflaton VEV.

\subsubsection{Up-Type Standard Model Fermions}

Note from \eqref{13} that the inflaton contains $H_{u}^{0}$ as a component field. It follows that the inflaton is able to decay via the Yukawa interactions directly into up-type standard model fermions. Since, as we will discuss below, the up-type leptons, that is, the neutrinos, can also mix with Higgsinos, we will treat these separately. Here, we consider only up-type quarks, since they cannot mix with other fermions. If we denote {by $F$ any of the $u$, $c$ and $t$ quarks then}
\be 
{\cal L}\supset y_{HF}HF\bar{F}
=y_{\psi F}\psi F\bar{F}\, ,
\ee
where $y_{HF}$ is the usual Yukawa parameter for coupling to the Higgs and, using \eqref{f7a} and \eqref{f10}, the Yukawa parameter for coupling to $\psi$ is
\be 
y_{\psi F}={y_{H F}\over\sqrt{6}}\, .
\ee
The values for $y_{HF}$ depend on the energy scale at which they are evaluated, and can be determined at any given scale using the renormalization group analysis presented in \cite{Ovrut:2015uea}. As discussed in subsection {\it 6.4}, an appropriate scale in the interior of the reheating interval is $5.8\times 10^{13}~{\rm GeV}$.
The values for $y_{HF}$ at this energy are found to be
\begin{equation}
y_{Hu}=6.47 \times10^{-6}~~,~y_{Hc}=3.77 \times 10^{-3}~~,~y_{Ht}=6.07 \times 10^{-1} \ .
\label{yc1} 
\end{equation}
That is, the  inflaton can decay, in the order of the coupling strength, as $\psi\rightarrow t\bar{t}$, $\psi\rightarrow c\bar{c}$ and $\psi\rightarrow u\bar{u}$. 

\begin{figure}[htbp]
	\includegraphics[scale=0.6]{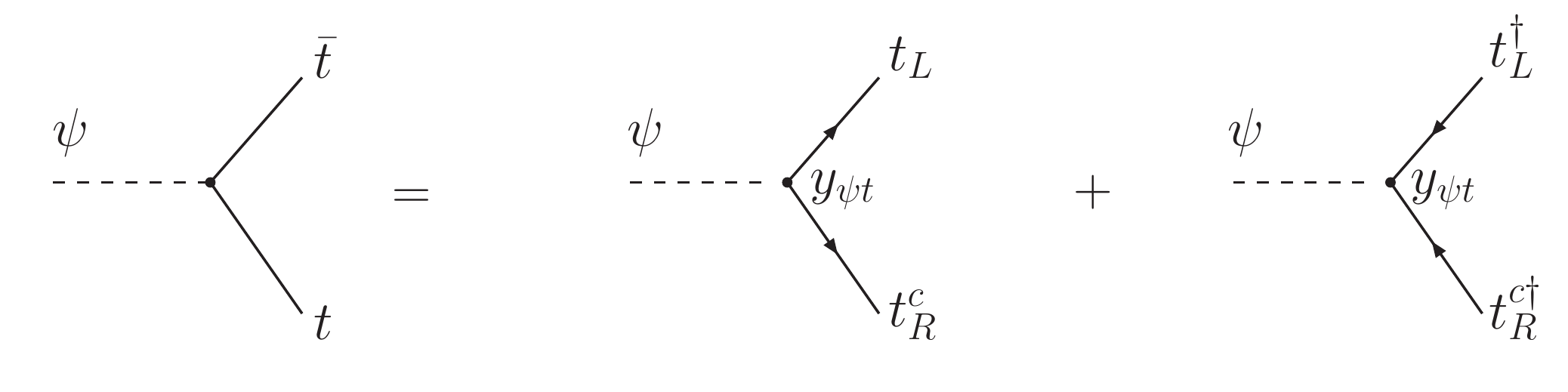}
	\caption{$\psi\rightarrow t\bar{t}$}\label{psi-to-tt}
\end{figure}

Consider the process $\psi\rightarrow t\bar{t}$ as an example. Since
\be {\cal L}\supset -y_{\psi t} \psi(t_L t_R^c+t_L^\dag {t_R^c}^\dag)\,,
\ee
$\psi$ can decay into $t \bar{t}$ 
(see Fig. \ref{psi-to-tt}) where we define the four component Dirac spinors
\be 
t=
\left(                 
\begin{array}{c}   
	t_L   \\  
	{t_R^c}^\dag \\  
\end{array}
\right)\, , 
\qquad
\bar{t}=
\left(                 
\begin{array}{c}   
	t_R^c   \\  
	{t_L}^\dag \\  
\end{array}
\right)\, .
\ee
The decay rates of $\psi$ to $t\bar{t}$, $c\bar{c}$ and $u\bar{u}$ all have the following form. Noting that $m_{F}=m_{\bar{F}}$, we find 
\begin{equation}
\Gamma_d(\psi\rightarrow F \bar{F})
={y_{\psi F}^2m_\psi\over 8\pi}\lf[1-4\lf({m_F\over m_{\psi}} \rt)^2 \rt]^{3\over 2} \, ,
\label{m1}
\end{equation}
where the mass of the fermion is given by
\be m_F= y_{\psi F}\sqrt{\langle \psi^2\rangle}\, 
\label{m2}
\ee 
and $m_{\psi}=m=6.49 \times 10^{-6}$.
It is important to note that the decay can only occur once $2m_F<m_\psi$. Since $m_F$ is determined by $\sqrt{\langle \psi^2\rangle}$, $m_F$ can initially be larger than $m_\psi/2$ \footnote{{The time delay of such decays due to the evolution of the inflaton VEV has been noted elsewhere, see for example \cite{Allahverdi:2005mz,Allahverdi:2008pf}. }}. In this case, $\Gamma_d=0$. 
With the expansion of the Universe, the amplitude of the oscillations of $\psi$ will decrease. When $\sqrt{\langle \psi^2\rangle}$ becomes sufficiently small, the decay $\psi\rightarrow F \bar{F}$ will become non-zero at some specific time, which we denote by $t_{F*}$. For $t > t_{F*}$, $\Gamma_d$ will increase as $\sqrt{\langle \psi^2\rangle}$ continues to get smaller.
Eventually, when $m_F \ll m_{\psi}$, it follows from \eqref{m1} that $\Gamma_d$ will approach a constant. That is,
\be
\Gamma_d \longrightarrow  {y_{\psi F}^2 m_\psi \over 8\pi} \equiv \Gamma_d^{\text{max}} \, ,
\label{m3}
\ee
which is the maximal value of $\Gamma_d$. Using \eqref{f7} and \eqref{yc1}  we find that
\begin{equation}
\Gamma^{\rm max}_{d,u}=1.801 \times 10^{-18}~,~\Gamma^{\rm max}_{d,c}=6.116\times 10^{-13}~,~\Gamma^{\rm max}_{d,t}=1.585 \times 10^{-8} \, .
\label{yc2}
\end{equation}
Therefore, a species with a smaller Yukawa coupling constant will be produced earlier, but with a relatively smaller maximal decay rate than a species with a larger Yukawa coupling.

The equation of state for $F\bar{F}$ is given by $p_{F\bar{F}}=\omega_{F\bar{F}}\rho_{F\bar{F}}$, where
\be
\omega_{F\bar{F}}={1\over3}\lf[1-4\lf({m_F\over m_{\psi}} \rt)^2 \rt]\,
\label{m4}
\ee
for $2m_F\leq m_\psi$. When $2m_F\simeq m_\psi$, the decay products $F$ and $\bar{F}$ are highly non-relativistic and $\omega_{F\bar{F}}\approx0$. However, when $2m_F\ll m_\psi$,  $F$ and $\bar{F}$ are relativistic and, hence,  $\omega_{F\bar{F}}\approx 1/3$. In the regime where $\Gamma_d\simeq \Gamma_d^{\text{max}} \ll H$, it is possible to give an approximate analytic solution for $\rho_{F\bar{F}}$ using equations \eqref{f1}-\eqref{f4}. However, this condition can only be satisfied for the up and charm quarks since their Yukawa parameters are relatively small. For the top quark, its Yukawa parameter is sufficiently large that  $\Gamma_d^{\text{max}} > H$. Therefore, {\it the results in the remainder of this subsection apply to $u$ and $c$ quark decays only}. The density function $\rho_{t{\bar{t}}}$ for the top quark can only be computed numerically. This will be carried out in Section 6.

To lowest order, one can ignore $\rho_{F\bar{F}}$ and, hence, $p_{F\bar{F}}$ in \eqref{f1} and \eqref{f2}, as well as $\Gamma_d$ in \eqref{f3}. It follows that $H$ can still be approximated by
\be
H(t) = {2\over 3(t-c_2)}
\label{m5}
\ee
as in \eqref{H1-02}, where $c_{2} \simeq 9.67 \times 10^{6}$. Putting this back into \eqref{f3} and taking $\Gamma_d=\Gamma_d^{\text{max}}$, one can solve this equation for $\psi$. We find that 
\be 
\psi(t)=\sqrt{8\over3}{1\over \omega (t-c_2)}\cdot \exp\lf[{-{\Gamma_d^{\text{max}}\over2}(t-c_2)}\rt]\sin\lf[ \omega(t-c_1)\rt]\, ,
\label{Gammaconst01}
\ee
where $c_1 \simeq 9.78\times10^6$ and
\be 
\omega=\sqrt{m_\psi^2-\lf({\Gamma_d^{\text{max}}\over 2}\rt)^2}\,.
\label{omega} 
\ee
Note that in the limit $\Gamma_d^{\text{max}}\rightarrow0$, this expression reproduces the result in \eqref{psi1-02}. Putting expressions \eqref{m3}, \eqref{m4}, \eqref{m5} and \eqref{Gammaconst01} into \eqref{f4}, we find that
\be 
\rho_{F\bar{F}}\approx {4 \Gamma_d^{\text{max}}\over 5 (t-c_2)}\lf[1-\lf({t-c_2\over t_{F*}-c_2}\rt)^{-5/3} \rt]\, .
\label{energy-density00}
\ee
Note that as $t \rightarrow t_{F*}$, this expression for $\rho_{F\bar{F}} \rightarrow 0$. That is, although derived in the in the regime where  $\Gamma_d=\Gamma_d^{\text{max}}$, we find that it remains a good approximation to the density for any $t \geq t_{F*}$ since physically one knows that
\be 
\rho_{F\bar{F}}(t \leq t_{F*})=0\,.
\label{rho-ini01}
\ee
After $t_{F*}$, that is,  when $2m_F<m_\psi$, $\rho_{F\bar{F}}$ will initially increase with time and reach a maximum value of
\be
\rho_{F\bar{F}}^{\text{max}}\approx {0.28 \Gamma_d^{\text{max}}\over t_{F*}-c_2 }
\label{x1}
\ee
at $t\approx c_2+ 1.8(t_{F*}-c_2)$. Then, $\rho_{F\bar{F}}$ will decrease with time as $\rho_{F\bar{F}}\sim (t-c_2)^{-1}$.
It follows from \eqref{ff1} and \eqref{energy-density00} that the fraction of energy density of species $F\bar{F}$ is given by
\be 
\Omega_{F\bar{F}}\approx {2\over 5}{\Gamma_d^{\text{max}}\over H}\lf[1-\lf({t-c_2\over t_{F*}-c_2}\rt)^{-5/3} \rt]<{2\over 5}{\Gamma_d^{\text{max}}\over H}\ll1\,.
\label{x2}
\ee
It is interesting to note that for $\Gamma_d^{\text{max}}\ll H$, by using the approximation in Appendix B, we find that
\be t_{F*}=t_{osc}+{2\sqrt{2}y_{HF} \over 3m_{\psi}^2}-{2\over3H(t_{osc})}\,.\label{tF*}
\ee
As we will determine below, the time at which reheating is finalized is given by $t_{R}\simeq 8 \times 10^{9}$. It then follows from \eqref{m5} and \eqref{x2} that
\begin{equation}
\Omega_{u\bar{u}}(t_{R}) < 8.636 \times 10^{-9}~, ~~\Omega_{c\bar{c}}(t_{R}) < 2.932 \times 10^{-3} \ .
\label{yc3}
\end{equation}
Similarly, using $t_{osc} \simeq 1.096 \times 10^{7}$ from Figure \ref{figbg01}, $H(t_{osc}) \simeq 5.16 \times 10^{-7}$ from Table \ref{Table-at-t} and \eqref{f7}, \eqref{yc1} we find that
\begin{equation}
t_{u*}=9.726 \times 10^{6} < t_{osc}~~, ~t_{c*}=4.413 \times 10^{7} > t_{MD} \, ,
\label{yc4}
\end{equation}
with $t_{MD}\simeq1.387 \times 10^{7}$.
We conclude that although up-type fermions with small Yukawa coupling constants, that is, $u$ and $c$, can be produced relatively early, their contribution to the background evolution of $H$ and the final decay products of $\psi$ are actually negligible. Physically, this is true because if $\Gamma_d\ll H$, the decay products will be diluted by the expansion of the Universe, thus barely effecting the evolution of $H$ and $\psi$. As a proof of this, one can compare, for example, $\Omega_{c\bar{c}}(t_{R})< 2.932 \times 10^{-3}$ against the smallest $\Omega(t_{R})$ computed numerically in Section 6. This is found to be $\Omega_{BB}(t_{R})=4 \times 10^{-3}$.  Noting that the value of  $\Omega_{c\bar{c}}(t_{R})$ is actually dramatically reduced relative to its value in \eqref{yc3} by the decay of the inflaton into the other species discussed below, we conclude that reheating into charm quarks, and therefore, into up quarks is negligibly small. They will, henceforth, be ignored. Only when a Yukawa parameter is large enough that the decay rate becomes comparable and then larger than $H$, will that species play an important role in reheating. As we will see below, this will be the case for the top quark.

\subsubsection{Charginos}
As mentioned previously, the non-zero expectation value of the inflaton--more precisely, the RMS value $\sqrt{\braket{\psi^2}}$--gives rise to effective mass terms for fields, as well as to mixing between different particle species. By diagonalizing the mass matrix for such species, one can determine the correct mass eigenstates into which the inflaton decays. We now examine the first class of such mass eigenstates, which we will label ``charginos", in direct analogy with the mass eigenstates associated with dynamical electroweak symmetry breaking in the \BLMSSM \cite{Martin:1997ns,Ovrut:2015uea}.

The mixing terms can arise from two sources; 1) the superpotential and 2) the``super-covariant derivative" of the $H_u$ Higgs doublet superfield. The former set are parameterized by $y_{Hi} \sqrt{\braket{\psi^2}}$, while the latter have the parameters $g_{a} \sqrt{\braket{\psi^2}}$, where $y_{Hi}$, $g_{a} $ denote Higgs coupled Yukawa parameters and gauge couplings respectively. We give an explicit description of where these terms arise from in Appendix C. Since the third family Yukawa coupling parameters are the largest, we will, for simplicity, assume that
\begin{enumerate}
	\item All Yukawa coupling matrices are diagonal.
	\item Only the third family quark and lepton Yukawa coupling parameters need be considered.
	\item Since the third family neutrino Yukawa coupling parameter is also negligible, it can be dropped as well. 
\end{enumerate}
Dropping all terms which have a neutrino Yukawa coupling $y_{\nu}$ and examining the effective mass Lagrangian for the ``charginos", 
we find that one set of fields which are mixed due to the non-zero value of $\sqrt{\braket{\psi^2}}$ are 
\begin{eqnarray}
\tilde{W}^{+}, \tilde{\psi}_u^{+}, \tau_R^c, \tilde{W}^{-}, \psi_d^{-}, \tau_L \,.
\end{eqnarray}
In order to construct the inflaton potential given in Section 2, we have previously taken the supersymmetric $\mu$ parameter 
to be of ${\cal{O}}(10^{10} {\rm GeV})$. This value is much smaller than the soft masses of the ${W}$-gauginos, as well as the initial values of the mixing terms $y_{H\tau} \sqrt{\braket{\psi^2}}$ and $g_2 \sqrt{\braket{\psi^2}}$. 
We can, therefore, simplify this system further by working in the ``small $\mu$" limit, and drop terms involving $\mu$.  
Of course, as the value of $\sqrt{\braket{\psi^2}}$ decreases, the value of $\mu$ will eventually exceed that of other terms we have not dropped. However, this effect is not significant since it will only occur very near the end of the reheating period. Hence we can, to a good approximation, take $\mu$ to be negligible.

In this limit, we are able to decouple the $\tau_R^c$, $\psi_d^-$ states since there is no longer any mixing between $\psi_d^-$ and $\psi_u^+$.
Examining the effective mass Lagrangian in equation \eqref{eq:L-mass-C}, we see that 
\begin{eqnarray}
\La_{mass} & \supset & 
y_{H\tau} \braket{\tilde{\nu}_{3,L}} \tau_R^{c} \psi_d^{-}  + \mathrm{h.c.}\, 
\label{yc5}
\end{eqnarray}
where, using the formalism presented in \cite{Ovrut:2015uea}, we find that the value of $y_{H\tau}$ at $5.8\times10^{13} ~{\rm GeV}$  is given by
\begin{equation}
y_{H\tau}=3.88 \times 10^{-2} \ .
\label{yc6}
\end{equation}
Note that \eqref{yc5} is a mass term for a Dirac mass 
\begin{eqnarray}
\Psi^{\prime}
 = \begin{pmatrix}
\psi_d^- \\
\tau_R
\end{pmatrix} \, 
\end{eqnarray}
with mass
\begin{equation}
m_{\psi_d \tau} =  y_{\psi_d \tau} \sqrt{\braket{\psi^2}}~~ \ , ~~  y_{\psi_d \tau}=\frac{y_{H\tau} }{\sqrt{6}} \, .
\label{sb1}
\end{equation}
The decay rate to the $\psi_d^-$ and $\tau_R^c$ states is, therefore, analogous to the decay of the inflaton to top quarks, and takes the form
\begin{eqnarray}
\Gamma_d(\psi \rightarrow  \psi_d^- \tau_R^c) &=& \frac{y_{{\psi}_{d}\tau}^2 m_{\psi}}{8 \pi } \left[1 - 4\frac{m_{\psi_d\tau}^2}{m_{\psi}^2}\right]^{3/2}
\label{eq:chargino1}
\end{eqnarray}
The equation of state parameter for $\psi_d\tau$ is
	\be \omega_{\psi_d\tau}={1\over3}\left[1 - 4\frac{m_{\psi_d\tau}^2}{m_{\psi}^2}\right]\,.
\ee

The remaining fermions,
\begin{eqnarray}
\tilde{W}^{+}, \tilde{\psi}_u^{+}, \tilde{W}^{-}, \tau_L
\end{eqnarray}
remain mixed and form the new mass eigenstates $\tilde{C}_1^{\pm}$ and $\tilde{C}_2^{\pm}$, where
\begin{eqnarray}
\begin{pmatrix}
\tilde{C}_1^+ \\
\tilde{C}_2^+
\end{pmatrix} 
= V
\begin{pmatrix}
\tilde{W}^+ \\
\tilde{\psi}_u^+
\end{pmatrix}
\, , \qquad
\begin{pmatrix}
\tilde{C}_1^- \\
\tilde{C}_2^-
\end{pmatrix} 
= U
\begin{pmatrix}
\tilde{W}^- \\
\tau_L 
\end{pmatrix} \ .
\end{eqnarray}
Determining the matrices $U$, $V$ is straightforward. The explicit form of both are given in appendix C.1.
The states $\tilde{C}_1^{\pm}$ and $\tilde{C}_2^{\pm}$ form Dirac fermions with masses $m_{\tilde{C}_1}$ and $m_{\tilde{C}_2}$ respectively. These are also presented in appendix C.1.
We find that the large value of $m_{\tilde{C}_2}$ makes the decay of the inflaton to $\tilde{C}_2^{\pm}$ kinematically impossible. Hence, only $\tilde{C}_1^{\pm}$ are produced.
The decay rate of the inflaton to the mass eigenstates $\tilde{C}^{\pm}_1$ is then given by
\begin{eqnarray}
\Gamma_d(\psi \rightarrow \tilde{C}_1^+ \tilde{C}_1^- ) &=& 
\gamma^2 \frac{(m_\psi^2 - 4m_{\tilde{C}_1}^2)^{3/2} }{8\pi m_{\psi}^2}
\nn\\
&=&
{\gamma^2 m_\psi \over 8\pi}\lf[1-4{m_{\tilde{C}_1}^2\over m_\psi^2}\rt]^{3/2} \ ,
\end{eqnarray}
where 
\begin{eqnarray}
\gamma=\frac{g_2}{\sqrt{6}}(U_{1W}V_{iu} + U_{1\tau}V_{1W}) \, ,~ 
m^2_{\tilde{C}_1}=\frac{1}{2} \left((x_1)^2 +2 (x_2)^2-\sqrt{(x_1)^4+4 (x_1)^2 (x_2)^2} \right)~~
\end{eqnarray}
and $x_1 = M_2$, $x_2 = g_2 \sqrt{\langle \psi^{2} \rangle }/ \sqrt{6}$. The elements of $U, V$ are given in appendix 
C.1. 
The equation of state parameter for $\tilde{C}_1^{\pm}$ is
\be \omega_{\tilde{C}_1^{\pm}}={1\over3}\lf[1-4{m_{\tilde{C}_1}^2\over m_\psi^2}\rt]\,.
\ee

It is important to note that the rate and equation of state for the inflaton decay into $\tilde{C}_1^{\pm}$ depend on the soft $SU(2)_{L}$ gaugino mass $M_{2}$. However, this will vary statistically over the interval $[m/f, fm]$, where $m=1.58 \times 10^{13}~\mathrm{GeV}$ and $f=3.3$. Generically, it will be different for each of the 215 valid black points discussed in Section 3. To avoid having to do a separate analysis for each of these 215 black points, we will, instead, note that one expects their average value, denoted by $M$, to be near the center of the interval. Furthermore, for concreteness, we will henceforth assume that 
\begin{equation}
M=m=1.58 \times 10^{13}~\mathrm{GeV} \ .
\label{yc8}
\end{equation}
Looking ahead, we note that inflaton decays into different species will depend on the gaugino soft masses $M_{R}$ and $M_{B-L}$, as well as on $M_{2}$. Therefore, for concreteness, we will henceforth make the generic assumption that
\begin{equation}
M_2 \simeq M_{R} \simeq M_{B-L} = M=m = 1.58 \times 10^{13} \mathrm{GeV} \, .
\label{yc9}
\end{equation}
Secondly, we note that the rate and equation of state for the inflaton decay into $\tilde{C}_1^{\pm}$ also depend on the $SU(2)_{L}$ gauge parameter $g_{2}$. This quantity is evaluated, as are all the other gauge couplings, by running it from its measured value at the electroweak scale up to the scale of reheating at $\sim 5.8\times10^{13} ~{\rm GeV}$. Hence, its value will essentially be the same for all 215 valid black points. Using the formalism developed in \cite{Ovrut:2015uea}, we find that at $5.8\times10^{13} ~{\rm GeV}$
\begin{equation}
g_{2}=0.57 \ .
\label{yc10}
\end{equation}
Again, looking ahead we find that inflaton decays into different species will depend on the gauge couplings $g_{R}$ and $g_{BL}$ as well as on $g_{2}$, all evaluated at the reheating scale of $5.8\times10^{13} ~{\rm GeV}$. Using the formalism developed in \cite{Ovrut:2015uea}, we find that at $5.8\times10^{13} ~{\rm GeV}$
\begin{equation}
g_{3}=0.60~, ~~g_{2}=0.57~,~~g_{R}=0.56~,~~g_{BL}=0.56 
\label{yc11}
\end{equation}
with an average value of $g=0.57$. Therefore, for simplicity of calculation, we will henceforth make the generic assumption that
\begin{equation}
g_{2} \simeq g_{R} \simeq g_{BL} = g= 0.57 \, .
\label{yc12}
\end{equation}

As with the top quark, the expressions for the energy densities $\rho_{\psi\tau}$ and the $\rho_{\tilde{C}\tilde{C}}$ charginos can only be computed numerically. We will carry this out in Section 6.

\subsubsection{Neutralinos}
We now turn to the second set of particles which mix due to the non-zero inflaton VEV. We refer to these as  ``neutralinos", in analogy with states described by the \BLMSSM . Again, we will ignore all  terms multiplied by a neutrino Yukawa coupling parameter $y_{H\nu}$. 
Making the same assumptions about Yukawa couplings as before, the effective mass Lagrangian for the ``neutralinos" is
such that the states that mix are
\begin{eqnarray}
\tilde{B}, \tilde{W}_R, \tilde{W}^0, \psi_d^0, \psi_u^0, \nu_{3L}, \nu_{3R}^c \, .
\end{eqnarray}
Once again, we take $\mu$ to be negligible compared to other terms in the effective mass Lagrangian. 
In this limit, the state $\psi_d^0$ decouples. It follows that the effective mixed state mass matrix, $M_{\tilde{N}}$, is six-by-six. To proceed, this must be diagonalized. We leave the details of this to appendix \ref{appendix-a}, but note that to simplify our expressions we again make the assumptions given in \eqref{yc9} and \eqref{yc12}.
Additionally, we define
\begin{eqnarray}
u \equiv \frac{1}{\sqrt{6}} \sqrt{\braket{\psi^2}} \ .
\label{yc13}
\end{eqnarray}
Diagonalizing the mass matrix $M_{\tilde{N}}$, we find the six mass eigenstates given in Table \ref{table:n-eigenstates}.
\begin{table}
	\begin{center}
		\begin{tabular}{|c|c|c|}
			\hline 
			Mass & Degeneracy & State \\ 
			\hline 
			0 & 1 & $\tilde{N}_1$ \\ 
			\hline 
			$\frac{1}{2}(M - \sqrt{M^2 + 12u^2})$ & 2 & $\tilde{N}_{2a}$ , $\tilde{N}_{2b}$  \\ 
			\hline 
			$M$ & 1 & $\tilde{N}_{3}$ \\ 
			\hline 
			$\frac{1}{2}(M + \sqrt{M^2 + 12u^2})$ & 2 & $\tilde{N}_{4a}$, $\tilde{N}_{4b}$ \\ 
			\hline 
		\end{tabular}
	\end{center}
	\caption{~Mass eigenstates of the neutralino mass matrix $M_{\tilde{N}}$.The masses $M$ and $u$ are defined in \eqref{yc8} and \eqref{yc13} respectively} \label{table:n-eigenstates}
\end{table}
Of the six eigenstates, only $\tilde{N}_1$ and $\tilde{N}_{2a}$, $\tilde{N}_{2b}$ are kinematically accessible to the decay of the inflaton. The decay processes and rates are
\begin{eqnarray}
&\,&\Gamma (\psi \rightarrow \tilde{N}_{2a} \tilde{N}_{2a}) 
=\frac{\gamma_a^2 m_{\psi}}{16 \pi } \left[1 - 4 {m_{\tilde{N}_2}^2\over m_\psi^2}\right]^{3/2} \ ,
\nn \\
&\,&\Gamma (\psi \rightarrow \tilde{N}_{2b} \tilde{N}_{2b}) 
=
\frac{\gamma_b^2 m_{\psi}}{16 \pi }\left[1 - 4 {m_{\tilde{N}_2}^2 \over m_\psi^2} \right]^{3/2} \ ,
\nn \\
&\,&\Gamma (\psi \rightarrow \tilde{N}_{2a} \tilde{N}_{2b} ) 
=
\frac{\gamma_c^2 m_{\psi}}{8 \pi }\left[1 - 4 {m_{\tilde{N}_2}^2\over m_\psi^2}\right]^{3/2} \ ,
\label{eq:neutralino1}
\end{eqnarray}
where $\tilde{N}_{2a}$ and $\tilde{N}_{2b}$ have the same mass presented in Table \ref{table:n-eigenstates} and
\begin{eqnarray}
\gamma_a &=& \left(\frac{7g}{2\sqrt{3}}\right)  
\frac{\frac{1}{2} u (M + \sqrt{M^2 + 12u^2}) }{M^2 + 12 u^2 + M \sqrt{M^2 + 12 u^2}} \ ,
\nn \\
\nn \\
\gamma_b &=&\left( \frac{9g}{2}\right) 
\frac{u \sqrt{u^2 + \frac{1}{6}(M^2 + M \sqrt{M^2 + 12 u^2}) }}{M^2 + 12 u^2 + M \sqrt{M^2 + 12 u^2}} \ , \label{eq:neutralino2} \quad \\
\nn  \\
\gamma_c &=& \left(\frac{\sqrt{3}g}{2}\right)
\frac{u \sqrt{u^2 + \frac{1}{6}(M^2 + M \sqrt{M^2 + 12 u^2}) }}{M^2 + 12 u^2 + M \sqrt{M^2 + 12 u^2}} 
- \left(\frac{g}{4}\right) 
\frac{u (M  + \sqrt{M^2 + 12u^2}) }{M^2 + 12 u^2 + M \sqrt{M^2 + 12 u^2}}  \ . \nn \\
\nn
\end{eqnarray}
Since $\tilde{N}_{2a}$ and $\tilde{N}_{2b}$ have the same mass, their equations of states parameters are given by the same form
\be \omega_{N_2}={1\over3}\left[1 - 4 {m_{\tilde{N}_2}^2\over m_\psi^2}\right]\,.
\ee

As with the top quark and the charginos, the expressions for the energy densities $\rho_{\tilde{N}\tilde{N}}$ for the neutralinos can only be computed numerically. We will carry this out in Section 6.

\subsubsection{Gauge Bosons}
The covariant derivatives of $H_u^0$,  $\tilde{\nu}_{3L}$ and  $\tilde{\nu}_{3R}^c$ couple the inflaton $\psi$ to the associated gauge bosons and, furthermore, give  an effective mass to these bosons. This occurs via the following terms in the Lagrangian
\begin{eqnarray}
\La_{gauge-coupling} &\supset&
-\frac{g_2^2}{4}\left( |H_u^0|^2 + |\tilde{\nu}_{3L}|^2 \right)W^{0\mu}W_{\mu}^0
- \frac{g_2^2}{2} \left( |H_u^0|^2 + |\tilde{\nu}_{3L}|^2 \right)W^{+\mu}W_{\mu}^-
\nn \\
&&
-g_R^2 \left(
q_{R_u}^2|H_u^0|^2 + q_{R_\nu}^2 |\tilde{\nu}_{3R}^c|\right)W_R^{\mu}W_{R\mu} 
-g_R^2 \left( q_{BL_\nu}^2 |\tilde{\nu}_{3R}^c| + q_{BL_L}^2|\tilde{\nu}_{3L}|^2
\right)B^{\mu}B_{\mu} 
\nn \\
&=&
-\frac{g_2^2}{12} \psi^2 W^{0\mu}W_{\mu}^0
-\frac{g_2^2}{6} \psi^2 W^{+\mu}W_{\mu}^-
-\frac{g_R^2}{12} \psi^2 W_R^{\mu}W_{R\mu}
-\frac{g_{BL}^2}{3} \psi^2 B^{\mu}B_{\mu}  \ .
\label{mr1}
\end{eqnarray}
To find the mass for each species of vector boson, as well as to determine their coupling parameter to the inflaton, we expand the inflaton around its root mean squared VEV as in \eqref{rz1}. Inserting this into the final expression in \eqref{mr1} and, as previously discussed, denoting $\delta \psi$ simply as $\psi$, we find that
\begin{eqnarray}
\La_{gauge-coupling} &\supset& -\frac{1}{2} m_{W^{0}}^{2} W^{0\mu}W_{\mu}^0
- m_{W^{\pm}}^{2} W^{+\mu}W_{\mu}^-
-\frac{1}{2} m_{W_{R}}^2 W_R^{\mu}W_{R\mu}
-\frac{1}{2} m_{W_{B}}^2 B^{\mu}B_{\mu} \nn \\
&&-\frac{g_{2}^{2}}{6} \sqrt{\langle \psi^{2} \rangle} \psi W^{0\mu}W_{\mu}^0 - \frac{g_{2}^{2}}{3} \sqrt{\langle \psi^{2} \rangle} \psi W^{+\mu}W_{\mu}^-
\label{mr3} \\
&&-\frac{g_{R}^{2}}{6} \sqrt{\langle \psi^{2} \rangle} \psi W_R^{\mu}W_{R\mu} -\frac{2g_{BL}^{2}}{3}  \sqrt{\langle \psi^{2} \rangle} \psi B^{\mu}B_{\mu} \nn  \ ,
\end{eqnarray}
where the masses are given by
\begin{equation}
\noindent m_{W^{0}}=m_{W^{\pm}}=\frac{g_{2}\sqrt{\langle \psi^{2} \rangle}}{ \sqrt{6}} ,~m_{W_{R}}=\frac{g_{R}\sqrt{\langle \psi^{2} \rangle}}{ \sqrt{6}} ,~m_{W_{B}}=\sqrt{\frac{2}{3}} g_{BL}\sqrt{\langle \psi^{2} \rangle} \ . \label{rz3} \\
\end{equation}

For a generic coupling $G_{i} \psi W_{i}^{\mu} W_{i \mu}$ with identical W-bosons, the decay rate and the equation of state parameter are given by
\begin{eqnarray}
&&\Gamma_d(\psi\rightarrow W_i^\mu W_{i\mu}) =\frac{G_{i}^{2}}{32 \pi m_{\psi}}\lf[1-4{m_{W_i}^2\over m_\psi^2}\rt]^{1/2} \ , \label{wow1}
\\ 
&&  \omega_{W_i}={1\over3}\lf(1-4{m_{W_i}^2\over m_\psi^2}\rt) \
\label{rz4}
\end{eqnarray}
respectively. For decays into two different W-bosons, one multiplies expression \eqref{wow1} by 2.  It then follows from \eqref{mr3} that the decay rates of the inflaton to the four gauge bosons, and the associated equation of state parameters, are 
\ba \Gamma_d(\psi\rightarrow W_0^\mu W_{0\mu})
&=&{g_2^4\langle \psi^2 \rangle\over 1152\pi m_\psi}\lf[1-4{m_{W_0}^2\over m_\psi^2}\rt]^{1/2}\,,
\\
 \omega_{W_0}&=&{1\over3}\lf[1-4{m_{W_0}^2\over m_\psi^2}\rt]\,,
\\
\Gamma_d(\psi\rightarrow W^{-\mu}W^+_\mu)
&=&{g_2^4\langle \psi^2 \rangle\over 144 \pi m_\psi}\lf[1-4{m_{W_{\pm}}^2\over m_\psi^2}\rt]^{1/2}\,,
\\
\omega_{W_{\pm}}&=&{1\over3}\lf[1-4{m_{W_{\pm}}^2\over m_\psi^2}\rt]\,,
\\
\Gamma_d(\psi\rightarrow W_{R\mu}W^\mu_R)
&=&{g_2^4\langle \psi^2 \rangle\over 1152\pi m_\psi}\lf[1-4{m_{W_R}^2\over m_\psi^2}\rt]^{1/2}\,,
\\
\omega_{W_R}&=&{1\over3}\lf[1-4{m_{W_R}^2\over m_\psi^2}\rt]\,,
\\
\Gamma_d(\psi\rightarrow B^\mu B_{\mu})
&=&{g_{BL}^4\langle \psi^2 \rangle \over 72\pi m_\psi}\lf[1-4{m_B^2\over m_\psi^2}\rt]^{1/2}\,,
\\
\omega_{B}&=&{1\over3}\lf[1-4{m_{B}^2\over m_\psi^2}\rt]\,,
\ea
where the gauge boson masses are given in \eqref{rz3}.

As with the top quark, charginos and neutralinos, the expressions for the energy densities $\rho_{WW}$ and $\rho_{BB}$ for the gauge fields can only be computed numerically. We will carry this out in Section 6. To simplicity the calculations, we will again use the approximation for the gauge couplings presented in \eqref{yc12}.

\begin{figure}[htbp]
\includegraphics[scale=.2,width=0.9\textwidth]{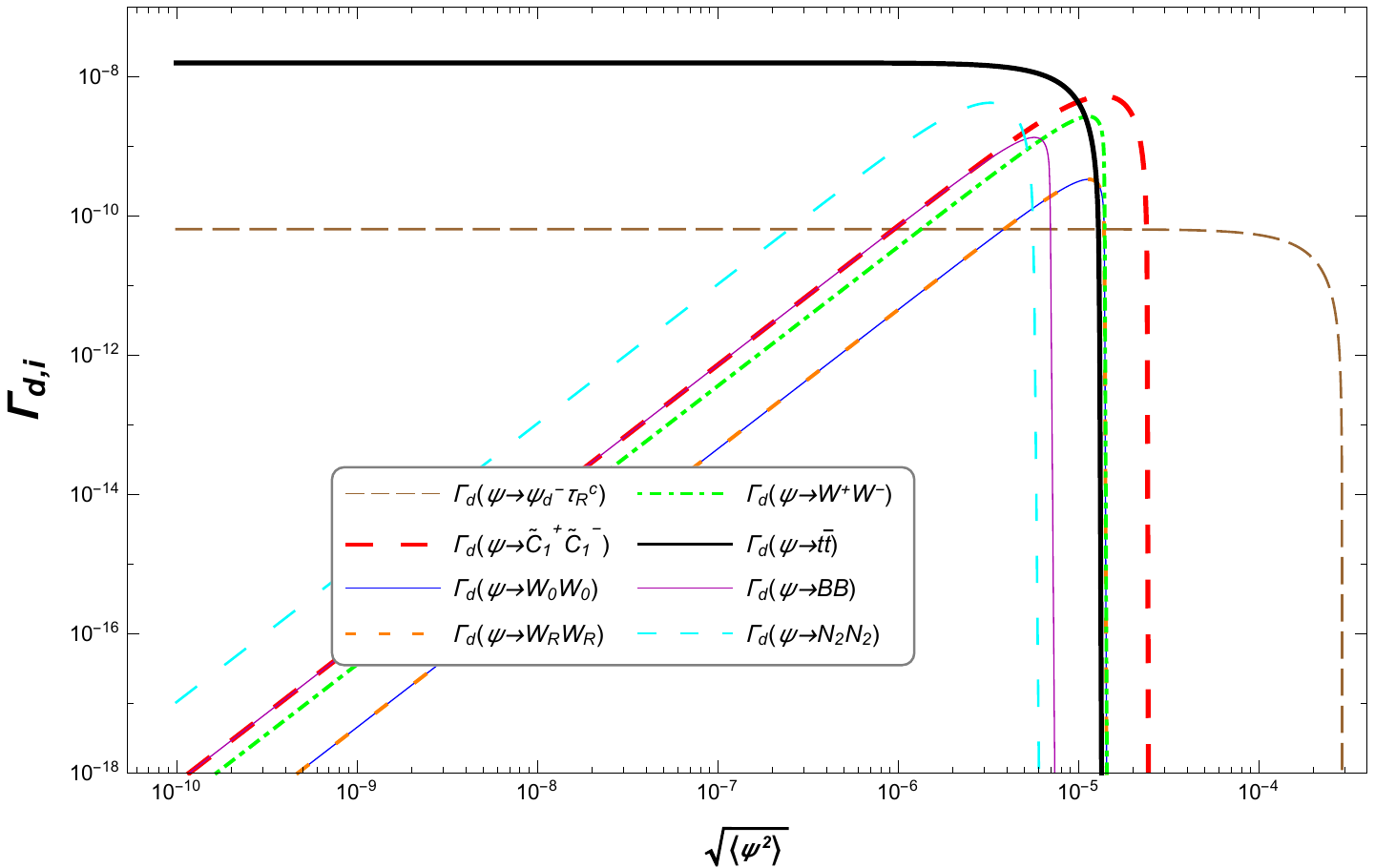}
\caption{The rates $\Gamma_{d,i}$ for different decay processes plotted with respect to $\sqrt{\braket{\psi^2}}$. The Yukawa couplings are all evaluated at the reheating scale of $5.8\times10^{13} ~{\rm GeV}$. Since $\sqrt{\braket{\psi^2}}$ will decrease with time, $\psi\rightarrow \psi_d^-\tau_R^c$ will be the first decay process to become non-zero, whereas the decay $\psi \rightarrow \tilde{N}_{2} \tilde{N}_{2}$ will be turned on last. We have set $M_{P}=1$.}
\label{figGammaA01}
\end{figure}

\subsubsection{Scalars}
The inflaton can couple to other scalar fields via the supersymmetric F-term and D-term potentials, as well as the soft supersymmetry breaking terms. These couplings give rise to a potential three-body decay vertex, as well as mass terms and mixing terms. 
To find out the mass eigenstates into which the inflaton can decay, one must  examine the mass matrix $\mathcal{M}$ whose elements are
\begin{eqnarray}
\mathcal{M}_{ij} = \frac{\partial^2 V}{\partial \phi_i \partial \phi_j^*} \bigg|_{\psi = \sqrt{\braket{\psi^2}}}\,,
\end{eqnarray}
where $i,j$ run over all scalars other than the inflaton and 
\begin{eqnarray}
V = V_F + V_D + V_{soft}\,.
\end{eqnarray}
To simplify our calculations, we assume that all scalars have identical soft masses given by
\begin{eqnarray}
m = 1.58 \times 10^{13} \mathrm{GeV}
\end{eqnarray}
and take the gauge couplings to have their average value at order $5.8\times10^{13} ~{\rm GeV}$, as discussed above. That is,
\begin{eqnarray}
g_{3} \simeq g_2 \simeq  g_R  \simeq  g_{BL} \simeq  g = 0.57\,.
\label{night1}
\end{eqnarray}
Diagonalizing $\mathcal{M}$, we find that the eigenvalues of $\mathcal{M}$ are either $m^2$ or larger. It follows that there are no decays of the inflaton to scalars.

{Additionally, we note that as we are describing a supersymmetric theory of inflation, it is also natural to consider the spin-$\frac{3}{2}$ superpartner of the graviton, the gravitino. The over-production of gravitinos during reheating can potentially be hazardous to nucleosynthesis \cite{Pagels:1981ke,Khlopov:1984pf,Moroi:1993mb}. However, in our present discussion, we expect that the mass of the gravitino is of order $m \sim 10^{13}$ GeV (that is, the same order as the soft supersymmetry breaking scale) and hence its production from the decay of the inflaton is heavily suppressed.}

\section{Numerical calculation} \label{numer}

As discussed in previous sections, the inflaton can decay into different species with different time dependent decay rates. The root mean squared value of $\psi$, that is, $\sqrt{\braket{\psi^2}}$, decreases with the decrease of the oscillatory amplitude of $\psi$ due to the expansion of the Universe and the decay of the inflaton. Thus, different decay processes will begin at different times since the masses of the decay products depend on $\sqrt{\braket{\psi^2}}$.  We plot the decay rates for different processes with respect to $\sqrt{\braket{\psi^2}}$ in Figure \ref{figGammaA01}. We did not plot the decay rates for $u\bar{u}$ and $c\bar{c}$ because, as discussed above, they are too small to have any substantial effect. The values of $\sqrt{\braket{\psi^2}}$ at which  each relevant process is turned on can be found in Table \ref{tabledecay-of-A}. Even though we have simplified the computations by ignoring the $u$ and $c$ quark decays, it remains impossible to find analytical solutions for \eqref{f1}-\eqref{f4} to account for all the relevant decay processes simultaneously. Therefore, in this section, we will numerically solve \eqref{f1}-\eqref{f4} to find the solutions for $H$, $\psi$ and $\rho_i$. From this, one can determine the relative energy densities of the different species at the end of the reheating epoch, as well as the reheating temperature. 

\begin{table*}[htbp!]
	\begin{center}{\footnotesize
		 	\begin{tabular}{|c|c|c|}
			\hline 
	Decay processes &  Value of $\sqrt{\braket{\psi^2}}$  & Value of $\sqrt{\braket{\psi^2}}$   \\  
			& at turn on ($M_p=1$)& at turn on	(GeV)\\		
			\hline
		$\psi\rightarrow \psi_d^-\tau_R^c$ & $2.05\times10^{-4}$ & $4.99\times 10^{14}$ GeV \\ 
			\hline 
		$\psi \rightarrow \tilde{C}_1^+ \tilde{C}_1^-$ & $2.41\times10^{-5}$ & $5.88\times 10^{13}$ GeV  \\ 
			\hline 
        $\psi\rightarrow W_0^\mu W_{0\mu}$, $W_{R\mu}W^\mu_R$,  $W^{-\mu}W^+_\mu$ & $1.39\times10^{-5}$ & $3.39\times 10^{13}$ GeV \\ 
            \hline 
        $\psi\rightarrow t\bar{t}$ & $1.31\times10^{-5}$ & $3.19\times 10^{13}$ GeV \\ 
            \hline 
        $\psi\rightarrow B^\mu B_{\mu}$ & $6.97\times10^{-6}$ & $1.70\times 10^{13}$ GeV\\ 
            \hline
        $\psi \rightarrow \tilde{N}_{2a} \tilde{N}_{2a}$, $\tilde{N}_{2b} \tilde{N}_{2b}$, $\tilde{N}_{2a} \tilde{N}_{2b}$ &$5.69\times10^{-6}$ & $1.39\times 10^{13}$ GeV \\ 
            \hline  
		\end{tabular} }  
	\end{center}
	\caption{Values of $\sqrt{\braket{\psi^2}}$ at which each decay process is turned on. We use the Yukawa couplings evaluated at the reheating scale of order $5.8\times10^{13} ~{\rm GeV}$.}\label{tabledecay-of-A}
\end{table*}

\subsection{Initial conditions}

To find the solutions for $H$, $\psi$ and $\rho_i$ by numerically solving \eqref{f1}-\eqref{f4}, one needs the initial conditions for $H$, $\psi$ and $\rho_i$. In principle, one can solve these equations starting from the beginning of inflation. Thus, the initial conditions would be set by inflation. However, such an approach would take a great deal of computing time due to the severe oscillations of $\psi$ after $t_{osc}$. Furthermore, ignoring the $u$ and $c$ quark decays, it follows from Table \ref{tabledecay-of-A} that the first decay process to turn on is $\psi \rightarrow \psi_d^- \tau_R^c$. The time at which this decay commences can be computed from the expression
\begin{equation}
t_{\psi_d^-\tau_R^c*}=t_{osc}+{2\sqrt{2}y_{H\tau} \over 3m_{\psi}^2}-{2\over3H(t_{osc})}\, \ ,
\label{psi-tau} 
\end{equation}
where $t_{osc} \simeq 1.096 \times 10^{7}$ from Figure \ref{figbg01}, $H(t_{osc}) \simeq 5.16 \times 10^{-7}$ from Table \ref{Table-at-t} and $y_{H\tau}$ was given in \eqref{yc6}.
This expression was first presented in \eqref{tF*} for top-quark decays, but can be used here since both $u$ and $c$ are being neglected. The result is
\begin{equation}
t_{\psi_d^-\tau_R^c*} \simeq 8.78 \times 10^{8}
\label{bird2}
\end{equation}
which is much later than $t_{osc}$, thus exacerbating  the computing time even more. 
Therefore, to save computing time, we will start our calculation from an arbitrarily chosen time $t=t_{\text{I}}$, where $t_{\text{I}}$ is close to, but smaller than, $t_{{\psi_d^-\tau_R^c}*}$. The exact value of $t_{\text{I}}$ is for technical convenience only.
We will set $t_{\text{I}}=8\times10^8$. When $t \leq t_{{\psi_d^-\tau_R^c}*}$, we can neglect all decay effects. Thus, we can approximately set $\rho_i=0$ at $t_{\text{I}}$ for all decay products. Using \eqref{psi1-02} and \eqref{H1-02}, one can obtain the initial conditions for $\psi$ and $H$ at $t_{\text{I}}$, respectively.
Of course, the decay rates for each process are zero until the corresponding process is turned on.

As can be ascertained from Fig. \ref{figGammaA01} and Table \ref{tabledecay-of-A},  the second decay to turn on is $\psi \rightarrow \tilde{C}_1^+ \tilde{C}_1^-$. It is clear from this data that the associated time, $t_{\tilde{C}_1*}$ , will be much later than $t_{\psi_d^-\tau_R^c*}$. As we will see below,
\begin{equation}
t_{\tilde{C}_1*} \simeq 6.45 \times 10^{9}  \ .
\label{bird3}
\end{equation}
Therefore, to further reduce the time for computation, we will divide the numerical calculations into two parts. First, we will numerically compute from $t=t_{\text{I}}$ to some time $t=t_{\text{II}}<t_{\tilde{C}_1 *}$  by using the iterative method described in next subsection. Again, the choice of the value of $t_{\text{II}}$ is for technical convenience only, which will not make any physical difference as long as $t_{\text{II}}<t_{\tilde{C}_1 *}$. We will choose $t_{\text{II}}=5\times 10^9$. 
Second, we numerically compute, also using the iterative method, from $t=t_{\text{II}}$ to the time $t_{R}$ where reheating has been completed. The initial conditions for $\psi$, $H$ and $\rho_i$ at $t=t_{\text{II}}$ will be set by the numerical solutions of the first part; that is, for  $t_{\text{I}}<t<t_{\text{II}}$.

\subsection{Iterative method}

In \eqref{f1}-\eqref{f4}, the equation of state parameters $\omega_i$ and the decay rates $\Gamma_{d,i}$ depend on the root mean square value of $\psi$, that is, $\sqrt{\braket{\psi^2}}$. This makes these background equations very difficult to solve--even numerically--especially when the $\sqrt{\braket{\psi^2}}$-dependence becomes complicate after $t_{\text{I}}$. Hence, we will solve them by iteration. We accomplish this by using Eq. (\ref{psi1-02}), which is the solution for $\psi$ without considering any decay, as the first input $\psi$ for $\sqrt{\braket{\psi^2}}$ in $\omega_i$ and $\Gamma_{d,i}$. Then, we treat  $\omega_i$ and $\Gamma_{d,i}$ simply as some time-dependent functions so that we can find solutions for \eqref{f1}-\eqref{f4} numerically. This gives the first output $\psi(t)$. The first output $\psi(t)$ will be identical with the first input $\psi(t)$ until some observable decay processes are turned on. Once $\sum_i \Gamma_{d,i}$ becomes effectively nonzero, the oscillations of $\psi$ will be damped more quickly. From then on, the output $\sqrt{\braket{\psi^2}}$ will be smaller than that of the input.

Next, we use this first output $\psi(t)$ as a second input $\psi$ for $\sqrt{\braket{\psi^2}}$ in $\omega_i$ and $\Gamma_{d,i}$ and numerically solve \eqref{f1}-\eqref{f4} again. This then leads to the second output $\psi(t)$ that is closer to the final solution. By repeating this method, the output $\psi(t)$ will become closer and closer to the real solution of $\psi$. We repeat this method iteratively multiple times until the output $\psi(t)$ is almost identical with the input $\psi(t)$, which means we have found the correct solution for $\psi$. Using this method also leads to solutions for $H$ and $\rho_i$ to some reasonable accuracy.

Additionally,
since the inflaton will oscillate rapidly around the minimum of its potential during the reheating phase, the oscillations of $\psi$ are too dense to be easily handled in the numerical calculation. The amplitude of the oscillations of $\psi$ will decrease with time. However, the frequency of the oscillations is almost a constant as long as $\sum_i \Gamma_{d,i}\ll  m_\psi$, as we will demonstrate in the numerical results. Hence, as a good approximation we can set
\be \psi(t)=A(t) \sin\lf[m_\psi (t-c_1) \rt]\,,\label{eq:ansatz}
\ee
where  $c_1 \simeq 9.78\times10^6$. This  is very useful in getting rid of the obstacles described above.
Then, by using the discussion in Appendix \ref{Apptrick}, we can simply replace $\sqrt{\braket{\psi^2}}$ in $\omega_i$ and $\Gamma_{d,i}$ with $A(t)/\sqrt{2}$. Therefore, for every input $\psi(t)$ in the iterative calculation, we can focus on $A(t)$ instead of $\psi(t)$.
We apply this iterative method to both the first part of the calculation, where $t_{\text{I}}<t<t_{\text{II}}$, and to the second part, where $t>t_{\text{II}}$, which we respectively denote as part I and part II.
Eventually, when the input $A(t)$ and the output oscillatory amplitude are almost same, we can conclude that we have found the correct solution for $A(t)$ (or equivalently $\psi(t)$), $H(t)$ (or equivalently $a(t)$) and $\rho_i(t)$ (or equivalently $\Omega_i(t)$).

The possible corrections to $\sum_i\Omega_i$ from the accuracy of the solution of $\psi$ can be defined as
\be \Delta\Omega_{\psi}={\rho_{\psi\text{out}}-\rho_{\psi\text{in}}\over 3M_p^2 H^2}\,,\label{eq:Delta-Omega}
\ee
where $\rho_{\psi\text{in}}=1/2\dot{\psi}_{\text{in}}^2+V(\psi_{\text{in}})$, $\rho_{\psi\text{out}}=1/2\dot{\psi}_{\text{out}}^2+V(\psi_{\text{out}})$ and $\psi_{\text{in}}$, $\psi_{\text{out}}$ are the input and the output $\psi$, respectively. Note that Eq. (\ref{eq:ansatz}) should be used when we transform between $\psi$ and $A$.

\subsection{Numerical results}

\begin{figure}[htbp]
	\subfigure[~~$t_{\text{I}}\leq t\leq t_{\text{II}}$]{\includegraphics[width=.49\textwidth]{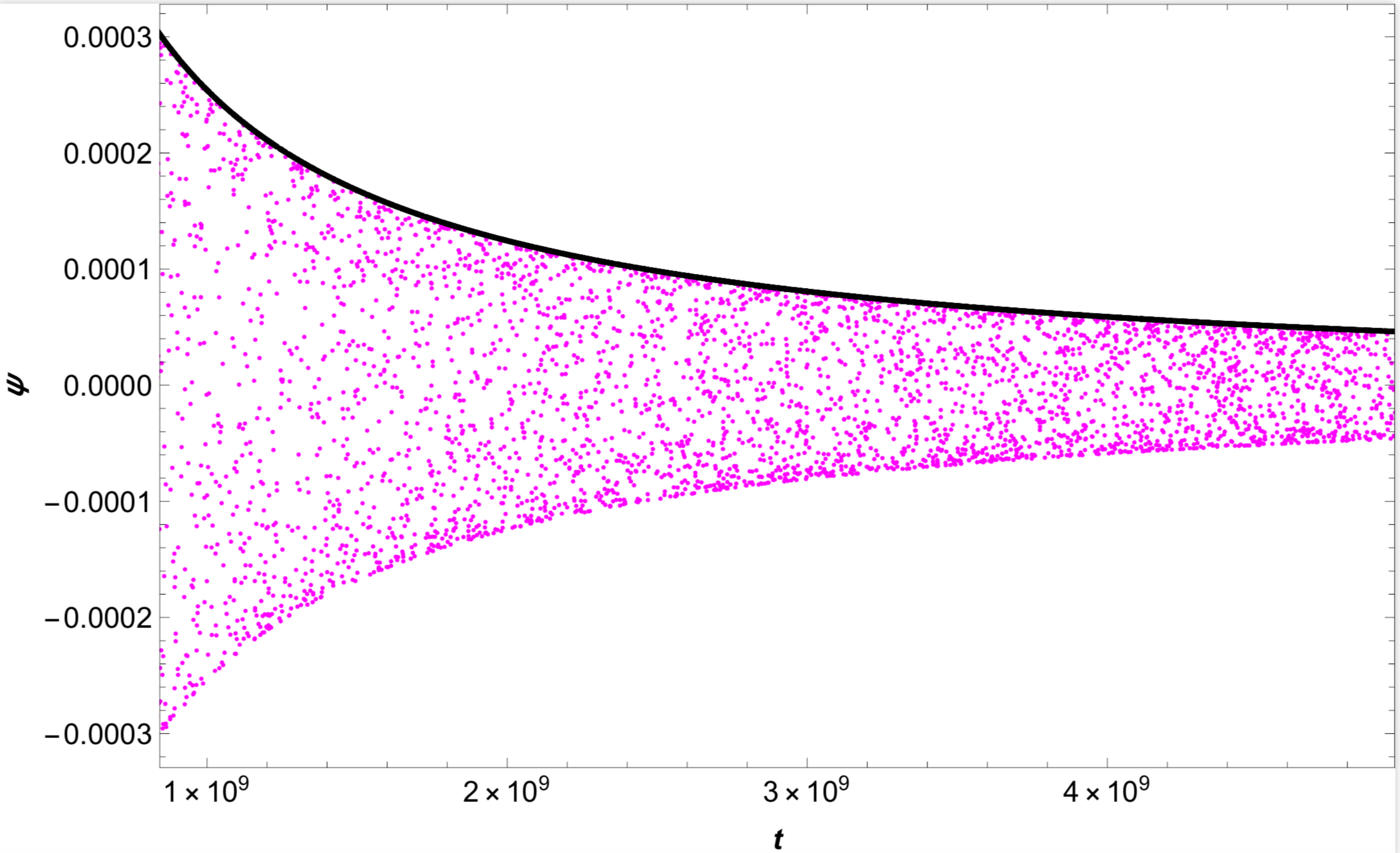} }
	\subfigure[~~$t\geq t_{\text{II}}$]{\includegraphics[width=.49\textwidth]{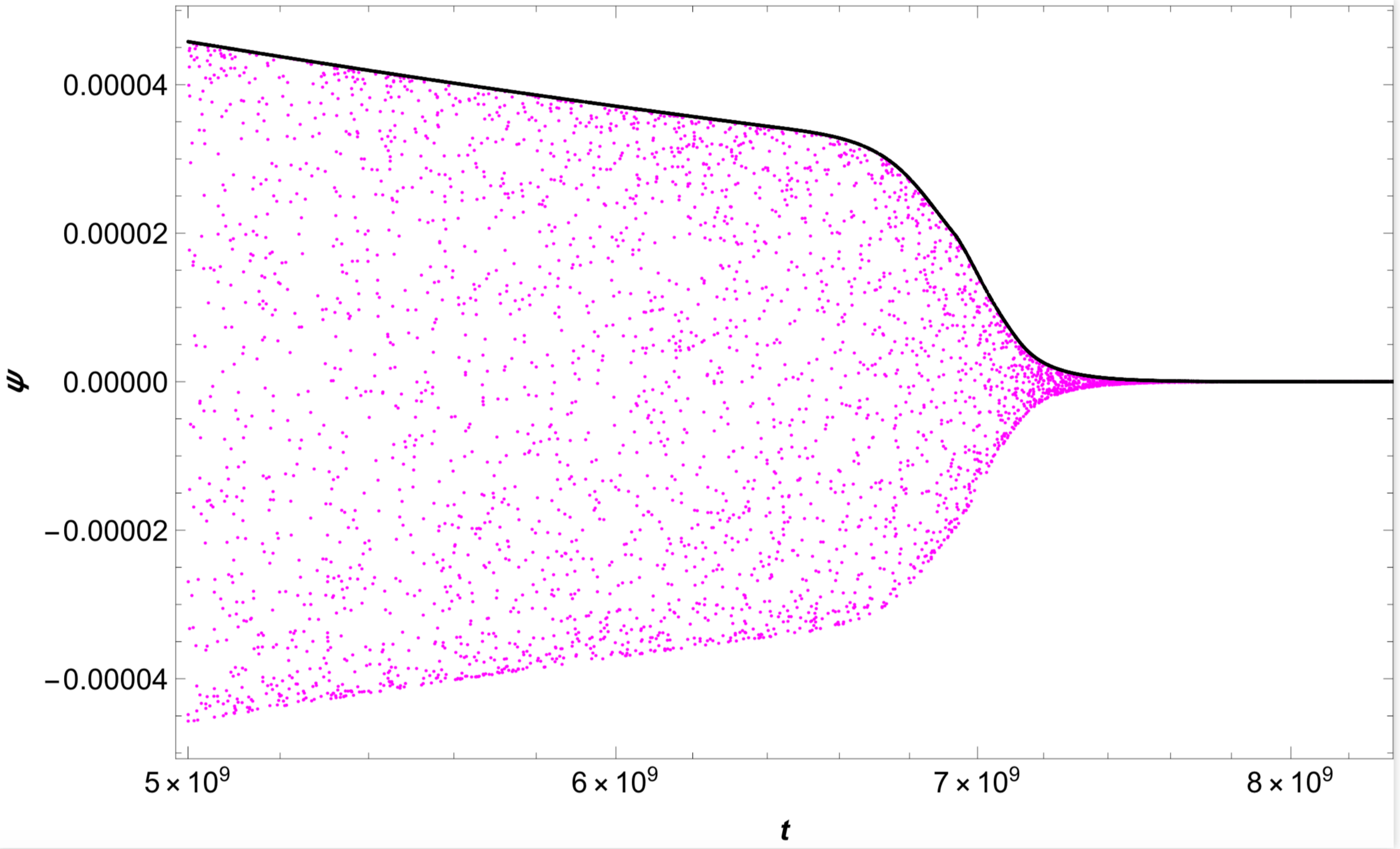} }
	\subfigure[~~$t\geq t_{\text{I}}$]{\includegraphics[width=.49\textwidth]{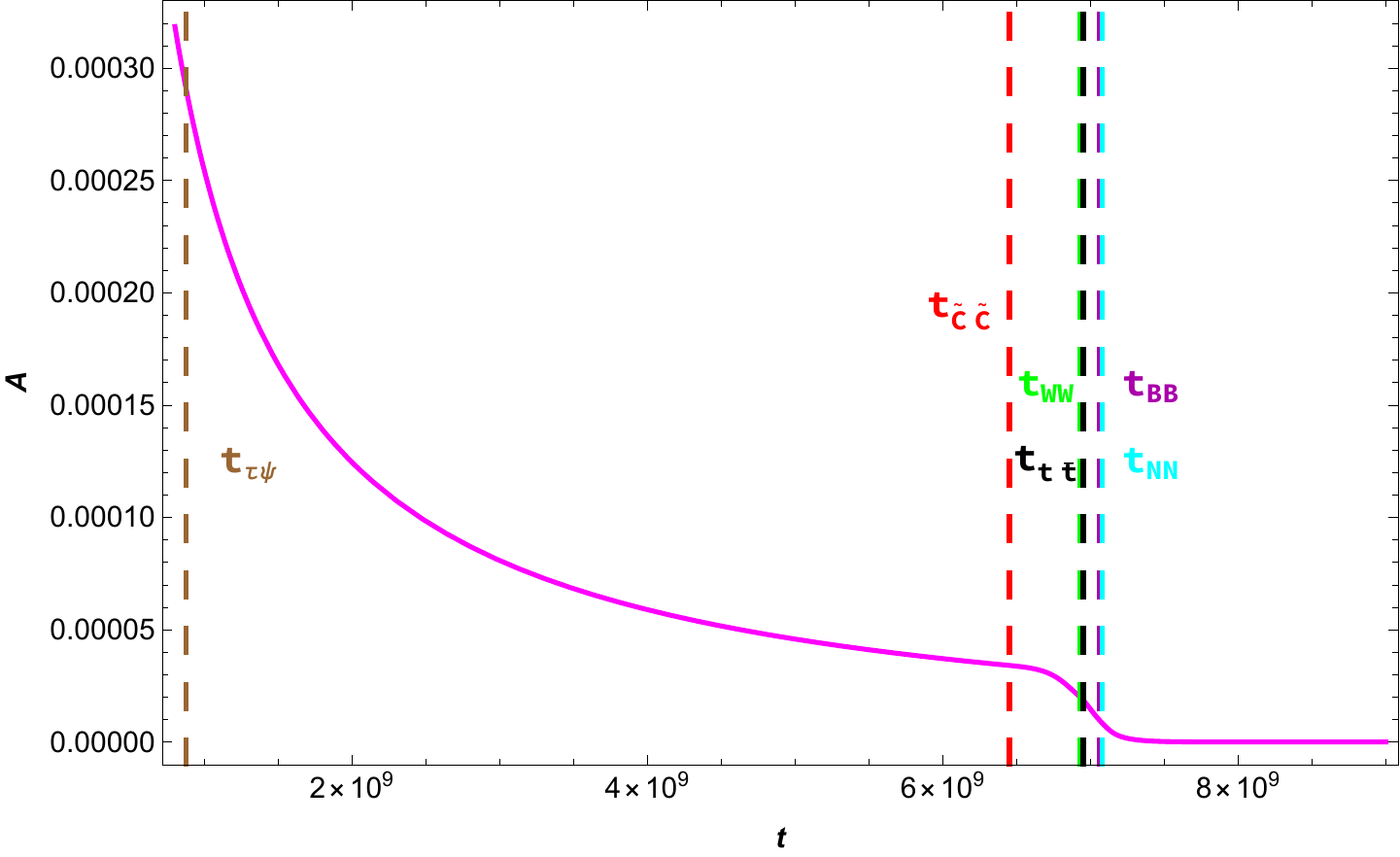} }
	\subfigure[~~$t\geq t_{\text{I}}$]{\includegraphics[width=.49\textwidth]{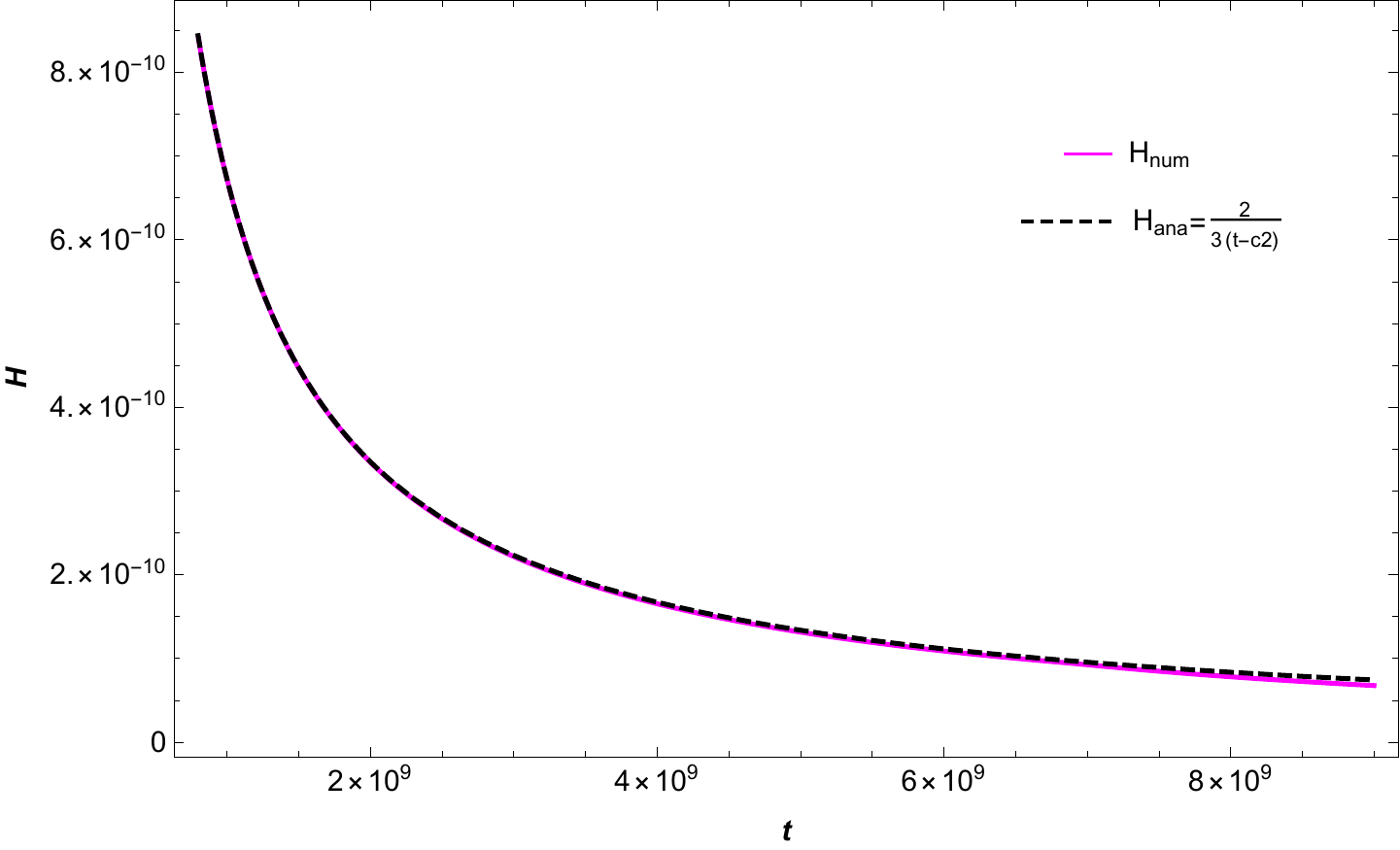} }
	\caption{~In both (a) and (b), the 5000 magenta points are randomly chosen from the curve of the output $\psi(t)$, in the last ($n$th) round of iteration, during the time intervals $t_{\text{I}}\leq t \leq t_{\text{II}}$ and $t\geq t_{\text{II}}$ respectively. Their upper boundaries are identical to the input $A(t)$ in the last round of iteration, that is, $A_{n}$, which is the black curves in (a) and (b). In (c), we plot the upper boundaries of the magenta points, that is, $A_{n+1}$, from $t_{\text{I}}$ to sometime after reheating. In addition,  the time at which each specie is turned on is marked with a vertical line, where $t_{\tau\psi*}=8.78\times10^8$, $t_{\tilde{C}\tilde{C}*}=6.45\times10^9$, $t_{WW*}=6.93\times10^9$, $t_{t\bar{t}*}=6.95\times10^9$, $t_{BB*}=7.06\times10^9$ and $t_{NN*}=7.08\times10^9$. In (d), we plot the solution for $H$ in the last round of iteration, that is, $H_{\text{num}}$, and also (\ref{H1-02}) as a reference,  from $t_{\text{I}}$ to sometime after reheating.  We always set $t_{\text{I}}=8\times10^8$ and $t_{\text{II}}=5\times10^9$ in our numerical calculations for technical convenience and everywhere set $M_{P}=1$.} \label{figpsi-in-out}
\end{figure}

After several iterations, we obtained the final solution for $A(t)$ (or, equivalently, for $\psi(t)$). For the last round of $n$ such iterations, we plot the input $A(t)$ (let us denote it by $A_{n}(t)$), the corresponding output $\psi(t)$ or $A(t)$ (let us denote it by $\psi_{n+1}(t)$ or $A_{n+1}(t)$, respectively) and the solution of $H$ in Fig. \ref{figpsi-in-out}. $A_{n}(t)$ is actually the numerical output $A(t)$ of the $(n-1)$th round of iteration.  Note that the oscillations of $\psi$ are too dense to be plotted completely. Instead, we simply plot $5000$ random points from the curve of $\psi_{n+1}(t)$ for both part I and part II. Thus $A_{n+1}$ corresponds to the upper boundaries of the magenta points in Fig. \ref{figpsi-in-out}, while $A_{n}$ is the black curves in  Fig. \ref{figpsi-in-out}. We can see that $A_{n}$ and $A_{n+1}$ are almost same. We can, therefore, treat $\psi_{n+1}$ as the actual solution of $\psi$.  
We have verified that their deviation is sufficiently small for our interests in this paper.\footnote{In fact, we find that the maximum of $\Delta \Omega_\psi$ in the final iteration is smaller than $10^{-4}$.}

\begin{figure}[htbp]
	\subfigure[~~$t\geq t_{\text{I}}$]{\includegraphics[scale=2,width=0.9\textwidth]{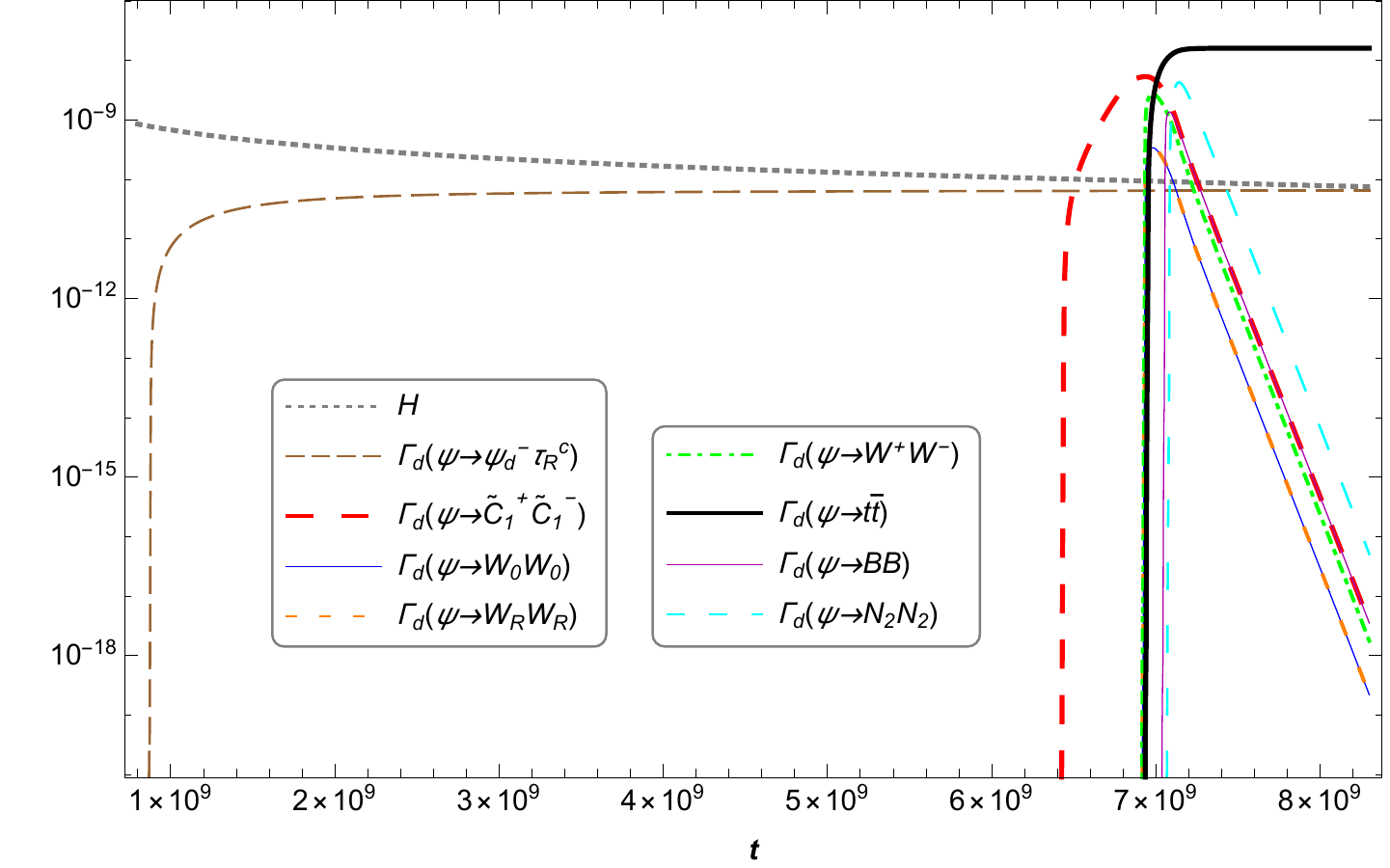}}
	\subfigure[~~$t\geq t_{\text{II}}$]{\includegraphics[scale=2,width=0.9\textwidth]{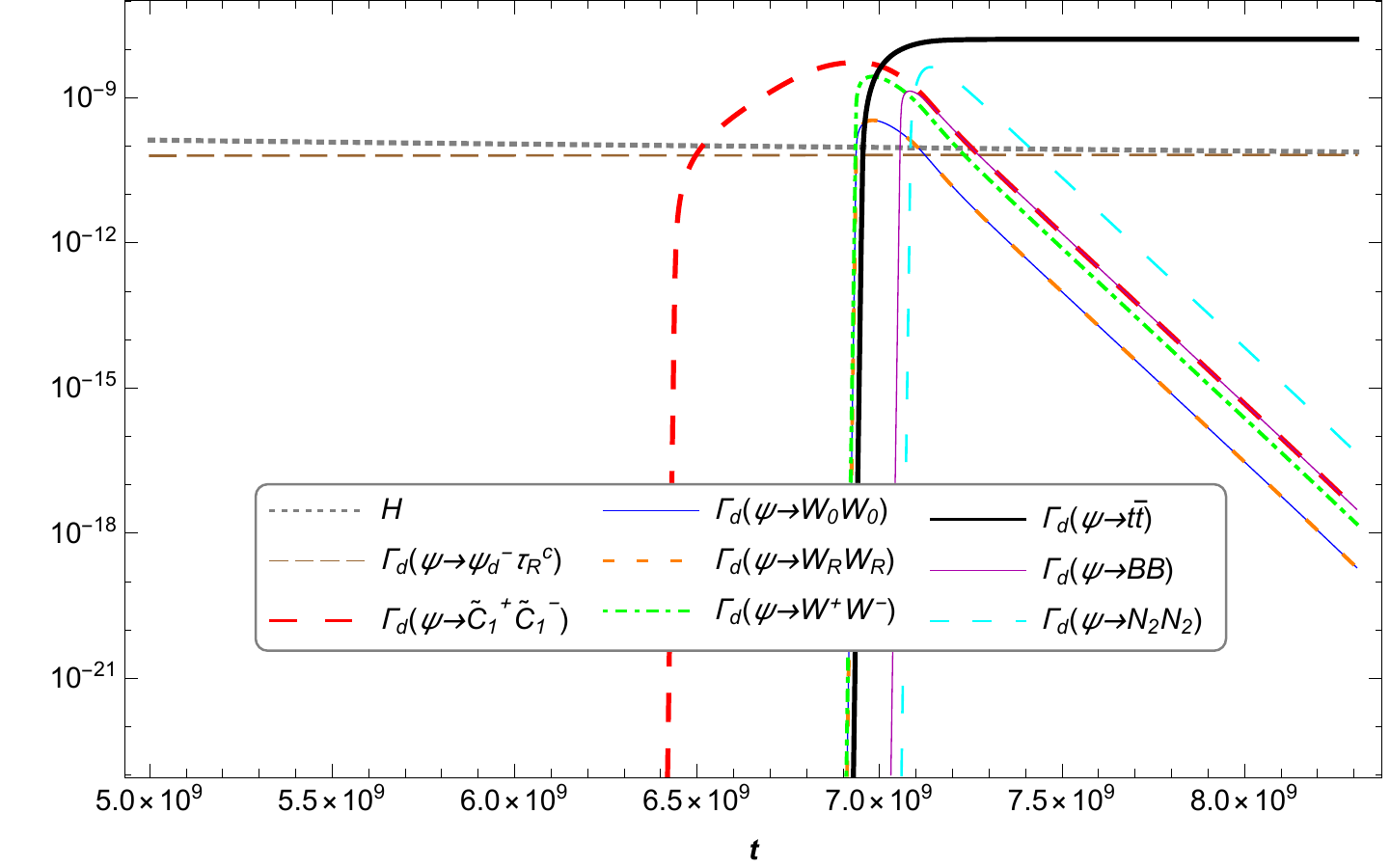}}
	\caption{~In (a), we plot the evolutions of $H$ and the decay rates $\Gamma_{d,i}$ from $t_{\text{I}}$ to sometime after the end of reheating.
	In (b), we plot $H$ and $\Gamma_{d,i}$ from $t_{\text{II}}$ to sometime after the end of reheating. Note that $\Gamma_d(\psi\rightarrow \psi_d^-\tau_R^c)$ is very close to $H$ when $t>t_{\text{II}}$. We set $t_{\text{I}}=8\times10^8$ and $t_{\text{II}}=5\times10^9$. In both (a) and (b),  $\Gamma(\psi\rightarrow \tilde{N}_2\tilde{N}_2)$ is the total decay rate for three processes $\psi \rightarrow \tilde{N}_{2a} \tilde{N}_{2a}$, $\tilde{N}_{2b} \tilde{N}_{2b}$, $\tilde{N}_{2a} \tilde{N}_{2b}$ and we have set $M_{P}=1$.}
	\label{figHGamma}
\end{figure}

We plot the decay rates $\Gamma_{d,i}$ and the Hubble parameter $H$ in Fig. \ref{figHGamma}. Obviously, $\sum_i \Gamma_{d,i}\ll m_\psi=6.49\times 10^{-6} (=1.58\times 10^{13}$ GeV) throughout. Thus the decay of $\psi$ cannot significantly effect its oscillatory frequency. As can be seen in Fig. \ref{figHGamma}, the decay process $\psi\rightarrow \psi_d^-\tau_R^c$ turns on much earlier than the other species, as was quantified above. Its decay rate reaches its maximal value and then becomes a constant. This is comparable to, but smaller than, $H$ prior to the other species in Fig. \ref{figHGamma} turning on. Thus the backreaction from $\psi\rightarrow \psi_d^-\tau_R^c$ cannot be neglected. The decay rate of $\psi\rightarrow t\bar{t}$ is similar to that of $\psi\rightarrow \psi_d^-\tau_R^c$, but turns on later and with a much larger maximal value. The decay rates of other species in Fig. \ref{figHGamma} first increase with time after they are turned on. However, since they are proportional to $\braket{\psi^2}$, when $\braket{\psi^2}$ is small enough they achieve a maximum and then decrease with time. Note that $\Gamma(\psi\rightarrow \tilde{N}_2\tilde{N}_2)$ is the total decay rate for the three processes $\psi \rightarrow \tilde{N}_{2a} \tilde{N}_{2a}$, $\tilde{N}_{2b} \tilde{N}_{2b}$, $\tilde{N}_{2a} \tilde{N}_{2b}$.

\begin{figure}[htbp]
	\subfigure[~~{ $t\geq t_{\text{I}}$}]{\includegraphics[width=.9\textwidth]{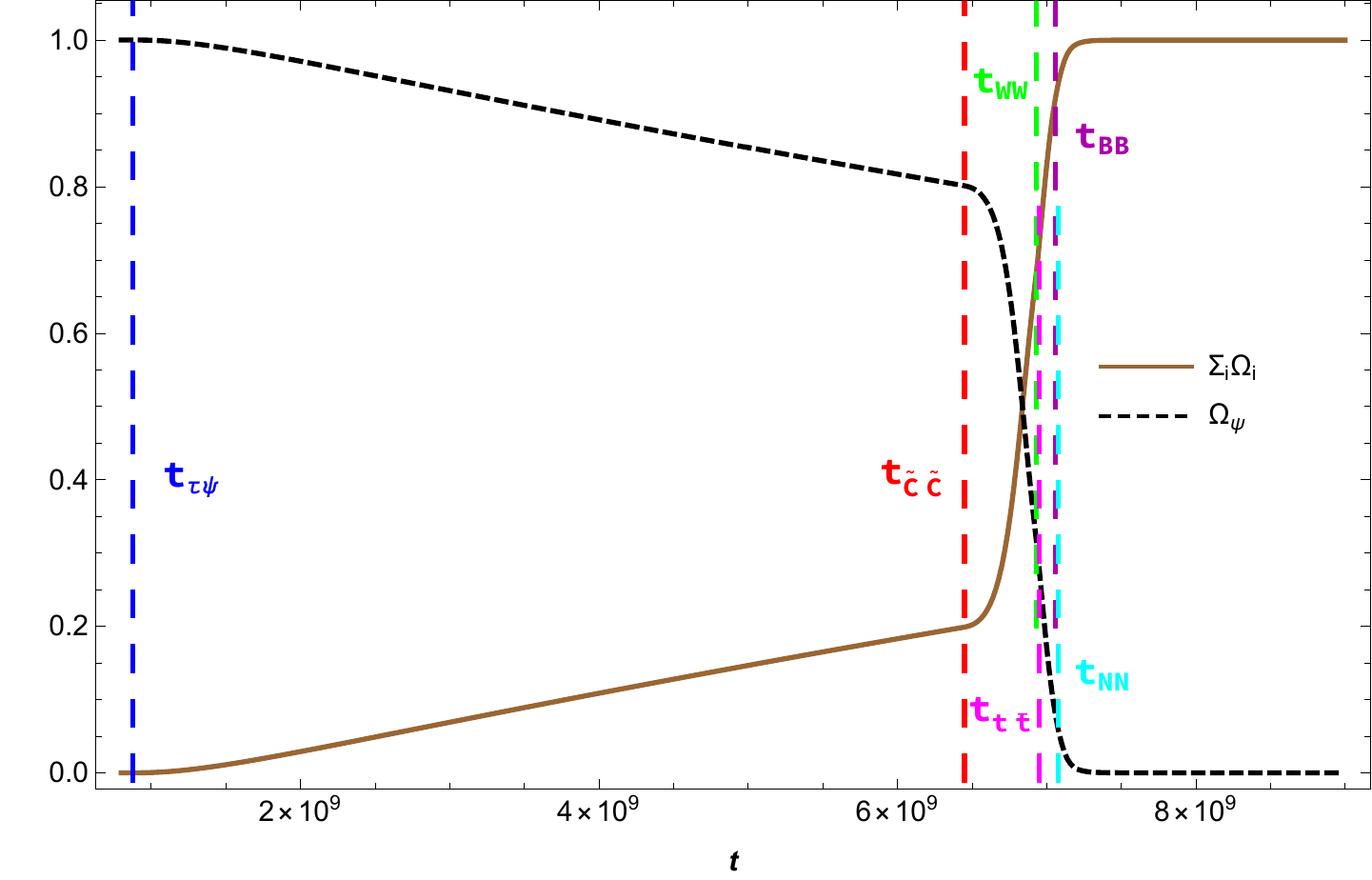} }
	\subfigure[~~{ $t\geq t_{\text{II}}$}]{\includegraphics[width=.9\textwidth]{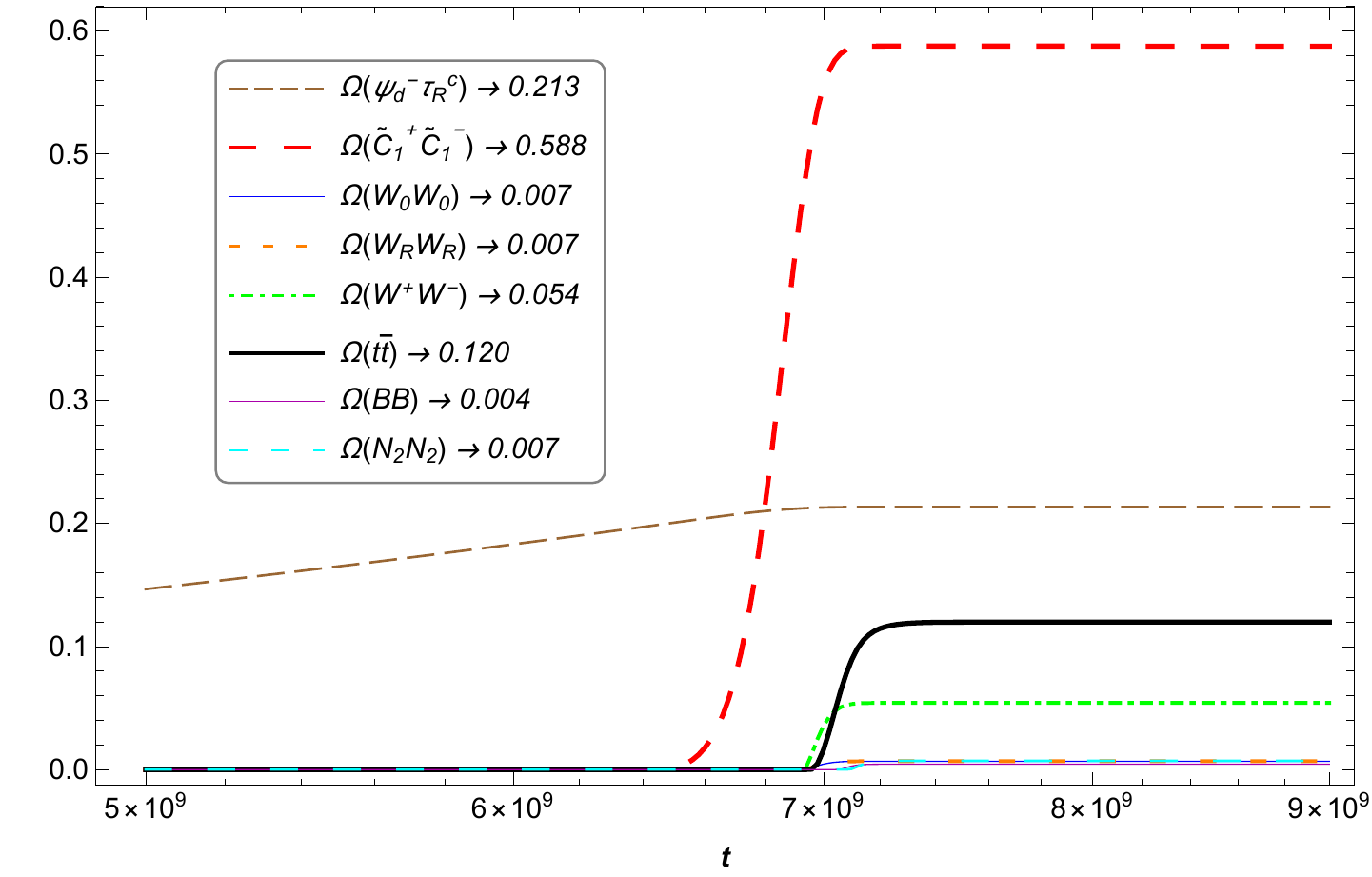} }
	\caption{~In (a), we plot the evolutions of $\Omega_\psi$ and $\sum_i \Omega_i$ from $t_{\text{I}}$ to sometime after the end of reheating.
	The time at which each specie is turned on is marked with a vertical line, where $t_{\tau\psi*}=8.78\times10^8$, $t_{\tilde{C}\tilde{C}*}=6.45\times10^9$, $t_{WW*}=6.93\times10^9$, $t_{t\bar{t}*}=6.95\times10^9$, $t_{BB*}=7.06\times10^9$ and $t_{NN*}=7.08\times10^9$.
	In (b), we plot 
	each  $\Omega_i$ from $t_{\text{II}}$ to sometime after the end of reheating.
	Note that $\Omega(\psi_d^-\tau_R^c)=0.183$ at $t_{\text{II}}$.}
	\label{figOmega}
\end{figure}

The evolutions of $\Omega_i$ and $\Omega_\psi$ are displayed in Fig. \ref{figOmega}.
$\Omega_{N_2N_2}$ includes the decay products for all three processes $\psi \rightarrow \tilde{N}_{2a} \tilde{N}_{2a}$, $\tilde{N}_{2b} \tilde{N}_{2b}$, $\tilde{N}_{2a} \tilde{N}_{2b}$. We did not specify them individually because even their sum is very small.
Note that $\sum_i\Omega_i=1-\Omega_\psi$ by definition. Eventually, $\sum_i\Omega_i\rightarrow 1$, since $\Omega_\psi\rightarrow 0$.
We can define the end of the reheating epoch as the time, $t_{R}$, when $\Omega_\psi\rightarrow 0$, which means that all of the energy of the inflaton has been converted to relativistic species of matter. It is clear from Fig. \ref{figOmega} that
\begin{equation}
t_{R} \simeq 8 \times 10^{9} \ .
\label{bird5}
\end{equation}
However, due to numerical imprecision, we may find that $\sum_i\Omega_i \approx 0.9999$ at $t_{R}$, which is more than sufficient for our purposes. Note that the values of the $\Omega_i$ in Fig. \ref{figOmega}(b) have been rounded to three decimal places. When added together, we find that they sum to
\begin{equation}
\sum_i\Omega_i =1.000  \ .
\label{brd1}
\end{equation}

When $t\simeq t_{R}$, we find $H(t_{R}) \simeq 7.8\times 10^{-11} (\simeq 1.9\times 10^8$ GeV). Since the Universe is now dominated by relativistic particle species, that is, $3M_p^2H^2=\rho_{\rm rel}$, one can, assuming the Universe is in thermal equilibrium, find the reheating temperature from the expression
\be \rho_{\rm rel}={\pi^2\over 30}g_* T_{R}^4\,,
\ee
where $g_*$ is the (effectively) massless degrees of freedom and $T_{R}$ is the temperature. It follows that the reheating temperature for the Sneutrino-Higgs  theory is
\be T_{R}=g_*^{-{1\over4}}\sqrt{\sqrt{90}M_p H(t_{R})\over \pi}\approx {3.74\over g_*^{{1/4}}} \times 10^{13} \, \text{GeV}\ ,
\ee
Since reheating takes place to a mixture of standard model particles (such as the top quark and various $W$-bosons) and lighter supersymmetric sparticles (such as $\tilde{C}^{\pm}_{1}$), the counting of the degrees of freedom is complicated. However, all of these species will eventually decay to the standard model particles with right-handed neutrinos, which has $g_{*}=118$. Hence, it is sufficient for our purposes to make a crude  approximation and take this as the number of degrees of freedom at the reheating temperature. It follows that
\begin{equation} 
T_{R} \simeq 1.13 \times 10^{13}~\text{GeV} \ .
\label{gd1}
\end{equation}

In this paper, we are requiring that the \BL breaking scale be well separated from the scale at which reheating takes place; that is \BL breaking occurs at a scale $\ll 10^{13}~{\text{GeV}}$. This simplifies the reheating calculations and, more importantly, allows reheating to occur prior to the breaking of baryon and lepton number. As discussed in Section 3 and Appendix A, the \BL scale can be made arbitrarily small, albeit at the expense of some fine-tuning. Clearly, the above requirement will be fulfilled for the \BL scales between $10^{10}~{\text{GeV}}$ and $10^{12}~{\text{GeV}}$ discussed in Section 3. As a concrete example, we see from Figure \ref{fig:BLScale2} that of the 215 phenomenologically valid black points, the maximal number ($\approx 20$) occur at a \BL scale of $10^{11}$ GeV.
Henceforth, as an example, let us choose this to be the \BL scale. It follows that the associated energy density is $\rho_{BL}=10^{44}~{\text{GeV}}^{4}$. Hence
\begin{equation}
3M_{P}^{2}H(t_{BL})^{2}=\rho_{BL} \Rightarrow H(t_{BL})\simeq 2.371 \times 10^{3}~{\text{GeV}} \ .
\label{gd2}
\end{equation}
Thus $H(t_{R})\gg H(t_{BL})$, which indicates that \BL breaking will occur much later than the end of the reheating epoch. 

\subsection{The Reheating Interval}

The entire analysis of reheating discussed above depends on inputting the numerical values of specific Yukawa parameters and all of the gauge couplings. However, these quantities all ``run'' with energy scale, changing their values to satisfy the renormalization group equations. In this paper, we have made two important assumptions that drastically simplify, and clarify, the calculations of reheating. 1) As we will see below, the interval of active reheating to matter is less than one order of magnitude in GeV units. Hence, these parameters vary only minimally over this small energy range. It is therefore a good approximation to choose a point in the interior of the reheating interval and to evaluate all Yukawa and gauge couplings at this fixed scale. 2) We find a point in the interior of the reheating interval as follows. We choose an arbitrary energy scale within an order of magnitude of where we expect to find the end of reheating. Using all Yukawa and gauge couplings evaluated at this energy, we numerically compute $t_{\psi_d^-\tau_R^c*}$ and $t_{R}$ and the Hubble parameters associated with them. We then take the average of the Hubble parameters, $H_{avg}$, and convert it to a ``matter'' energy density using $3H^{2}M_{P}^{2}=\rho_{avg}$. The interior energy scale is then chosen to be the fourth root of $\rho_{avg}$. Further iteration shows that this interior point remains a good approximation for characterizing the reheating interval.

Specifically, in this paper we do the following. First consider 2). We begin by choosing the arbitrary scale to be $10^{12}~{\rm GeV}$ and use the RGE's to compute all Yukawa and gauge parameters at this energy. The numerical calculation of $\Gamma_{i}$, $\Omega_{i}$ and $H$, as well as $t_{\psi_d^-\tau_R^c*}$ and $t_{R}$, are carried out as described previously in this Section. We find that the associated Hubble parameters are
\begin{equation}
H(t_{\psi_d^-\tau_R^c*}) \simeq 1.6 \times 10^{9}~{\rm GeV}~, ~H(t_{R}) \simeq 1.6 \times 10^{8}~{\rm GeV} 
\label{end1}
\end{equation}
and, hence,
\begin{equation}
H_{avg} \simeq 8.0 \times 10^{8}~{\rm GeV} ~~\Rightarrow~~ \rho_{avg}^{1/4} \simeq 5.8 \times 10^{13}~{\rm GeV} \ .
\label{end2}
\end{equation}
It then follows from assumption 1) that all Yukawa and coupling parameters used in our reheating calculations will be evaluated at $5.8 \times 10^{13}~{\rm GeV}$.

\section{Attaining equilibrium}

In order to define a reheating temperature for the plasma of decay products, one must determine that they have attained equilibrium \cite{Mazumdar:2013gya}. In this Section, we will show that this is indeed the case prior to $t_{R} \simeq 8 \times 10^{9}$. Thermal equilibrium occurs when the interaction rate of the $i$-th decay product, which we denote by $\Gamma^{i}_{int}$, is sufficiently large for all species $i$. Specifically, one requires that
\begin{eqnarray}
\Gamma^{i}_{int} > H \, ,~  \forall ~i \ .
\label{eq:equilib-cond}
\end{eqnarray}
This implies that the mean interaction length, $1/\Gamma^{i}_{int}$, is within the causal horizon $1/H$.

To demonstrate that \eqref{eq:equilib-cond} is indeed satisfied by the end of reheating, let us consider the elastic scattering of the charginos, $\tilde{C}_1^{\pm}$, mediated by gauge bosons. As we have shown in Figure \ref{figOmega}(b), these comprise the largest fraction of inflaton decay products by the end of reheating. We argue that all other interaction processes involving different species present in the plasma will have similar interaction rates -- since these will also involve gauge boson mediated scattering, all with similarly large values of gauge couplings. Therefore, if condition \eqref{eq:equilib-cond} is satisfied for one species, it is safe to assume that it is satisfied for all of them by the time that reheating is completed.

\begin{figure}[htbp]
\begin{center}
\subfigure[]{\includegraphics[scale=1,width=0.49\textwidth]{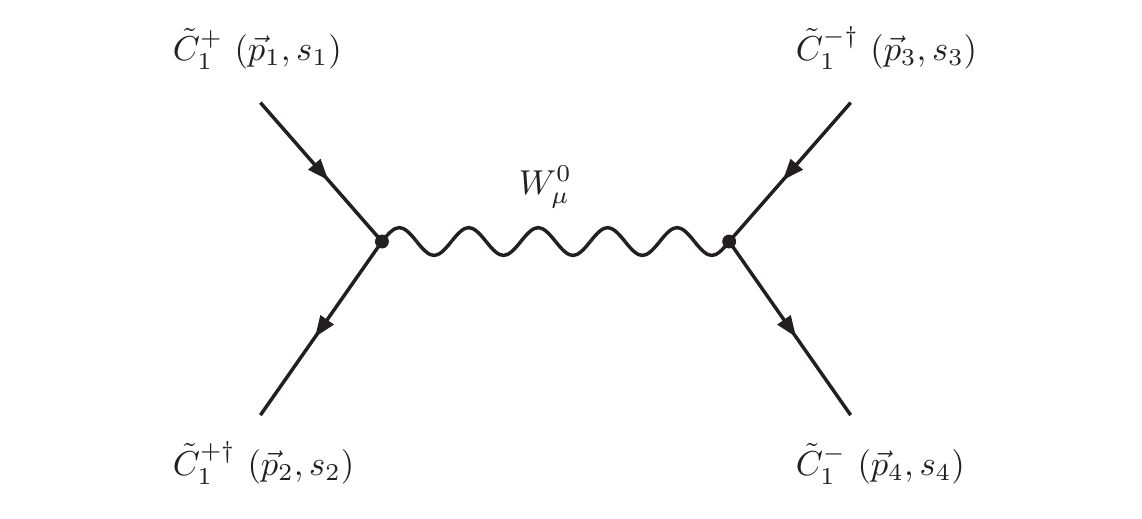}}
\subfigure[]{\includegraphics[scale=1,width=0.49\textwidth]{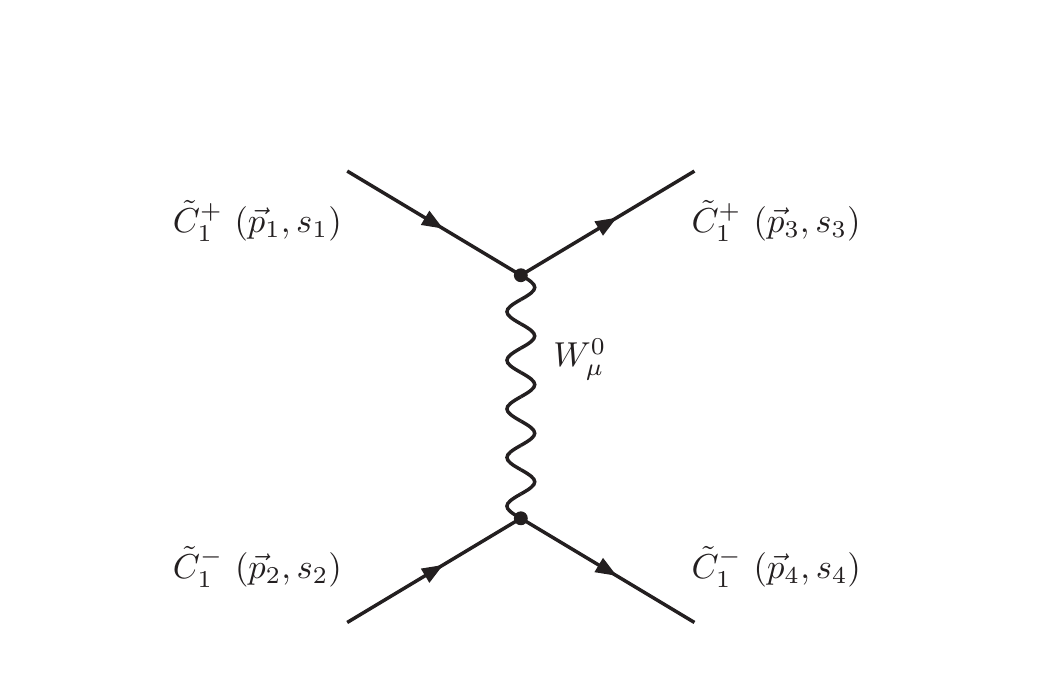}}
\caption{Subfigures (a) and (b) show some of the  $s$- and $t$-channel diagrams respectively contributing to the $\tilde{C}_1^+ \tilde{C}_1^- \rightarrow \tilde{C}_1^+ \tilde{C}_1^-$ scattering process exchanging a $W^{0}_{\mu}$ boson. }
\label{fig:chargino-s-channel}
\end{center}
\end{figure}

\begin{figure}
\begin{center}
\includegraphics[scale=0.2]{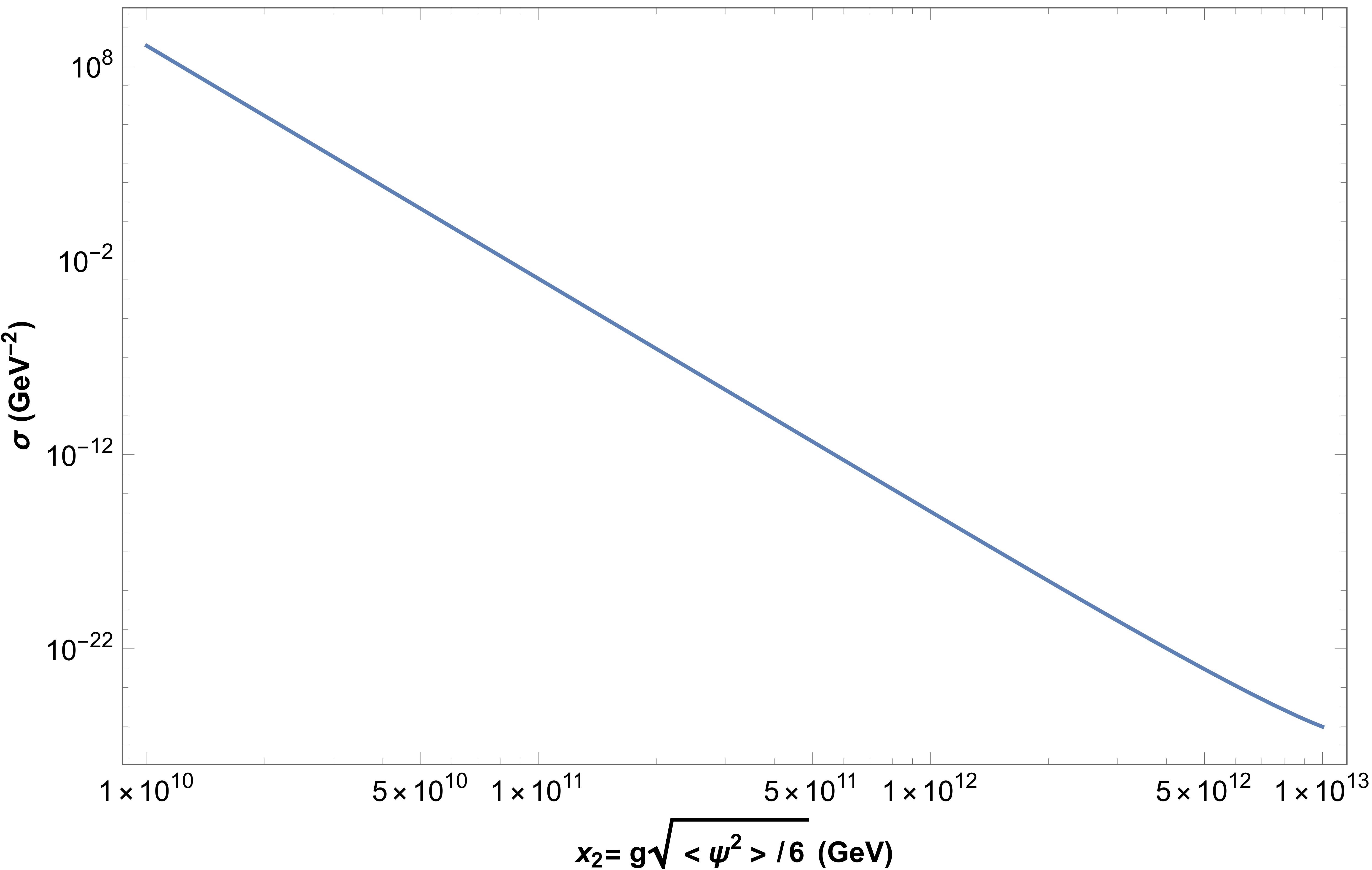}
\caption{A log-log plot of the cross section for the process $\tilde{C}_1^+ \tilde{C}_1^- \rightarrow \tilde{C}_1^+ \tilde{C}_1^-$ plotted against the inflaton expectation value $x_2 = g \sqrt{\braket{\psi^2}}/\sqrt{6}$ , where $g=0.57$ as in the rest of this text.}
\label{fig:chargino-xs}
\end{center}
\end{figure}

\begin{figure}
\begin{center}
\includegraphics[scale=0.8]{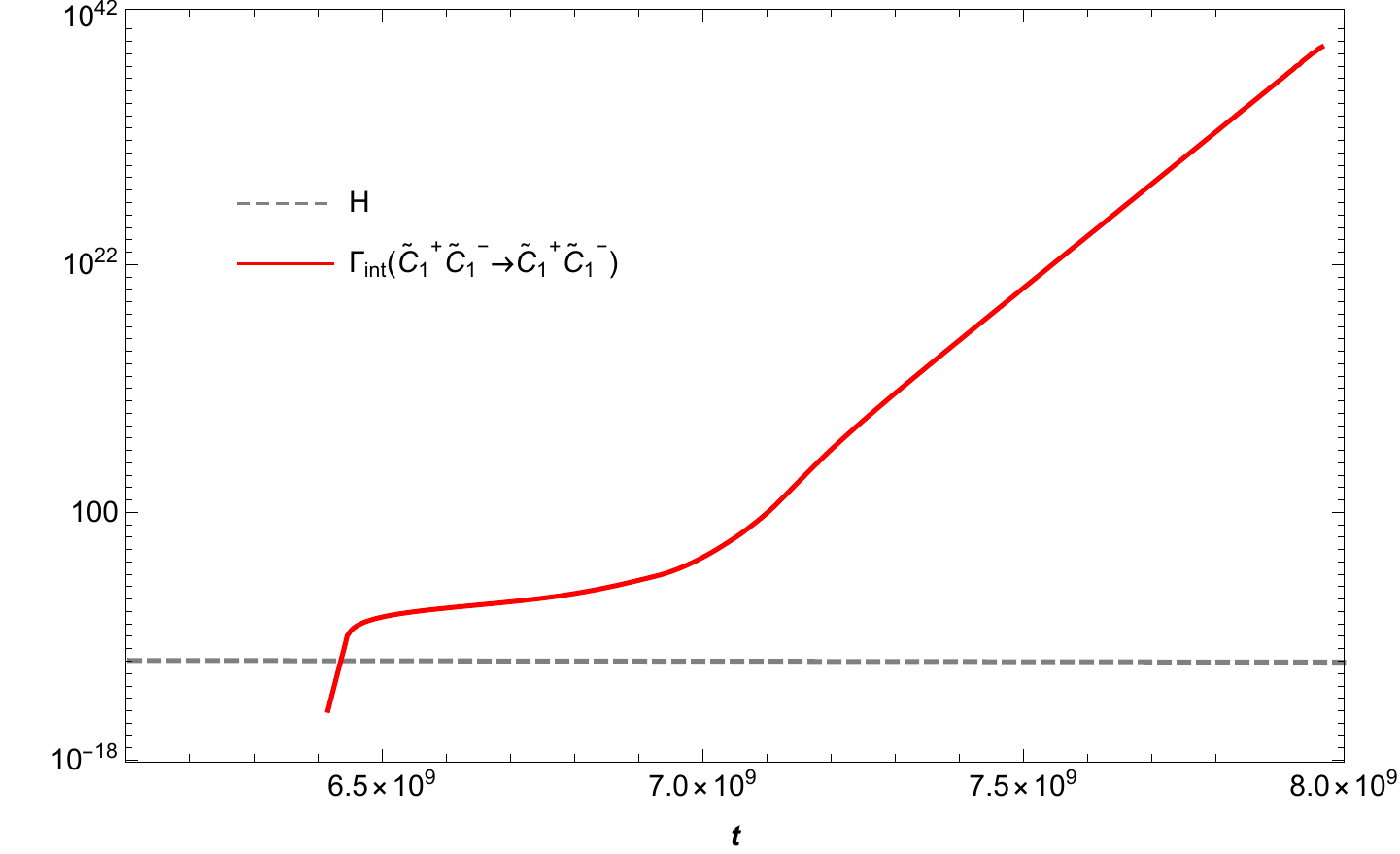}
\caption{The rate $\Gamma_{int}$ for the process $\tilde{C}_1^+ \tilde{C}_1^- \rightarrow \tilde{C}_1^+ \tilde{C}_1^-$ plotted against time (shown in units where $M_P =1$). The rate almost immediately becomes larger than the Hubble parameter $H$, which is approximately 
$7.8 \times 10^{-11}$
at the very end of the plot. The time at the end of reheating, $t_{R} \simeq 8 \times 10^9 $.} 
\label{fig:chargino-Gamma-int}
\end{center}
\end{figure}

For simplicity, let us determine the rate for the process $\tilde{C}_1^+ \tilde{C}_1^- \rightarrow \tilde{C}_1^+ \tilde{C}_1^-$, where, for simplicity, we take the mediating gauge boson to be $W_\mu^{0}$. This process is also mediated by $W_{R \mu}$ and $B_{\mu}$, but since at this energy scale the gauge couplings are all of similar value, see \eqref{night1}, the results will be very similar. Feynman diagrams which contribute to this process are shown in Figure \ref{fig:chargino-s-channel}.
The differential cross-section for this interaction is 
\begin{eqnarray}
\frac{d \sigma}{d t} (\tilde{C}_1^+ \tilde{C}_1^- \rightarrow \tilde{C}_1^+ \tilde{C}_1^-)
&=&
\frac{\mathcal{O}_a^4}{8 \pi s^2}
\left[
\left(\frac{s}{t}\right)^2 + u^2 \left(\frac{1}{s} + \frac{1}{t} \right)^2 + \left( \frac{t}{s} \right)^2
\right]
\label{eq:chargino-dxs}
\end{eqnarray}
where the coupling $\mathcal{O}_a$ is given by
\begin{eqnarray}
\mathcal{O}_a &=& g_2 \left(\frac{1}{2}|V_{1u}|^2 + |V_{1W}|^2  \right)\, .
\end{eqnarray}
The matrix elements $V_{1u}$, $V_{1W}$ are given in Appendix C.1.
We have used the fact that 
\begin{eqnarray}
V_{1u} = U_{1\tau} \, ,  \quad V_{1W} = U_{1W}
\end{eqnarray}
in this calculation. Here, as in the rest of this paper, we follow the conventions and notation outlined in \cite{Dreiner:2008tw}. In the center-of-mass frame, the Mandelstam variables $s$, $t$ and $u$ are
\begin{eqnarray}
s &=& - (p_1 + p_2)^2 = E_{cm}^2 \nn \\
t &=& -(p_3 - p_1)^2 = -\frac{1}{2}(E_{cm}^2 - 4 m_{\tilde{C}_1}^2) (1 - \cos \theta) \nn \\
u &=& -(p_4 - p_1)^2 = -\frac{1}{2}(E_{cm}^2 - 4 m_{\tilde{C}_1}^2) (1 + \cos \theta)\,, 
\label{eq:mandelstam}
\end{eqnarray}
where we have assumed that the incoming states carry energy $E_{cm}/2$, with 
\begin{eqnarray}
E_{cm} = m_{\psi}.
\end{eqnarray}
In deriving equation \eqref{eq:chargino-dxs}, we have ignored the mass of the gauge boson, which quickly becomes negligible as the inflaton VEV decreases. 
The total cross section is then given by integrating 
\begin{eqnarray}
\sigma = \int_{t_-}^{t_+} \frac{d \sigma} {dt} \,, \quad
t_- = t\vert_{\cos \theta = -1} \, , \quad
t_+ = t\vert_{\cos \theta = 1 - \delta} \, ,
\label{eq:chargino-xs}
\end{eqnarray}
where the cutoff $\delta$ must be introduced to remove the collinear divergence. We use the standard result that 
\begin{eqnarray}
\delta = \frac{2m_{\tilde{C}_1}^2}{s}  \, .
\end{eqnarray}
In Figure \ref{fig:chargino-xs}, we plot the resulting cross-section as a function of the parameter $x_2 = g_2 \sqrt{\langle \psi^{2} \rangle }/ \sqrt{6}$, which is proportional to the inflaton expectation value. The interaction rate is given by
\begin{eqnarray}
\Gamma_{int} (\tilde{C}_1^+ \tilde{C}_1^- \rightarrow \tilde{C}_1^+ \tilde{C}_1^-)&=& n \sigma v \ ,
\label{eq:gamma-c}
\end{eqnarray}
where $n$ is the number density of $\tilde{C}_1^{\pm}$ and $v$ is the average particle velocity, which we take to be $v \sim c = 1$. 
For a given particle species with average energy $\braket{E}$, the number density can be approximated by 
the expression
\begin{eqnarray}
n = \frac{\rho}{\braket{E}} \, ,
\end{eqnarray}
The rate $\Gamma_{int} (\tilde{C}_1^+ \tilde{C}_1^- \rightarrow \tilde{C}_1^+ \tilde{C}_1^-)$ is plotted as a function of time along with the Hubble parameter in Figure \ref{fig:chargino-Gamma-int}. It is clear that condition \eqref{eq:equilib-cond} is satisfied well before the completion of reheating. Since the self-scattering interactions of other decay products involve similarly sized gauge couplings, we expect analogous rates for these species--for example, top quarks and  $W$ bosons--to also satisfy condtion \eqref{eq:equilib-cond} by the end of reheating, despite their smaller number density.

\begin{figure}
\begin{center}
\includegraphics[scale=0.8]{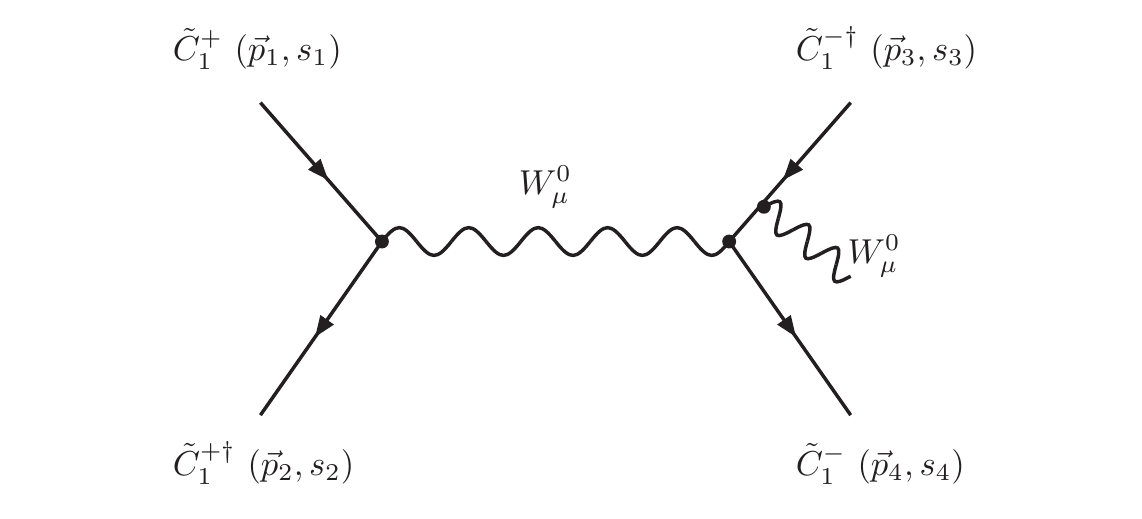}
\caption{Example of a Feynman diagram for $2 \rightarrow 3$ inelastic scattering involving the charginos.}
\label{fig:chargino-2to3}
\end{center}
\end{figure}

Strictly speaking, we have determined that the particles above have attained \textit{kinetic} equilibrium, one of two necessary conditions to ensure that the decay products of the inflaton have thermalized \cite{Enqvist:1993fm,Davidson:2000er,Mazumdar:2013gya}. The second condition, the achievement of \textit{chemical} equilibrium, requires the analysis of number-violating $2 \rightarrow 3$ interactions. An example of such a process is given in Figure \ref{fig:chargino-2to3}. Naively, these interactions may be suppressed by an extra factors of perturbative couplings and hence their rates may be smaller than the number conserving $2 \rightarrow 2$ scattering - although this is not always the case. Without going into the full details of such processes, we will simply argue that their rates are still sufficiently high. That is, a $2 \rightarrow 3$ scattering rate involving the charginos could at worse be
\begin{eqnarray}
\Gamma_{2\rightarrow 3} (\tilde{C}_1^+ \tilde{C}_1^-)
\sim
10^{-2} \,
\Gamma_{int} (\tilde{C}_1^+ \tilde{C}_1^- \rightarrow \tilde{C}_1^+ \tilde{C}_1^-) \, .
\end{eqnarray}
Under this assumption, from figure \ref{fig:chargino-Gamma-int}, we can see that the condition 
\begin{eqnarray}
\Gamma_{2\rightarrow 3} (\tilde{C}_1^+ \tilde{C}_1^-) > H
\end{eqnarray}
is also attained before the end of reheating. We expect that this holds for all other decay products as well.\\

%


\noindent \textbf{Acknowledgments:}
We are most grateful to Mark Trodden, Riccardo Penco and Yun-Song Piao for helpful discussions. 
The work of R.D. and B.A.O. is supported in part by the DOE under contract No. ${\rm DE \mhf SC0007901}$. Y.C. is supported in part by the UCAS Joint PhD Training Program. Y.C. would like to thank the Department of Physics and Astronomy at the University of Pennsylvania for their hospitality during his visit. Y.C. is especially grateful to Yun-Song Piao for his countless support and help.


\section*{Appendix A: ~~Determining the \BL scale}

In this Appendix, we present the details of the renormalization group equations that allow us to specify the desired \BL breaking scale, and then run in energy-momentum up to the unification scale $\langle M_{U} \rangle$ to determine the associated initial values for both $m_{\tilde{\nu}_{R,3}^c}^2 (t_U)$ and $m_{H_{u}}^{2}(t_{U})$. To do this, we rely heavily on the description of the RGE's of the \BLMSSM presented in \cite{Ovrut:2015uea}, as well as on the phenomenological analyses of the low-energy physics discussed in those papers.

\subsection*{A.1.~~~ RG Running of the Sneutrino Mass}
In the upside-down hierarchy, the RGEs describing the running of $m_{\tilde{\nu}_{R,3}^c}^2$ are\footnote{Here, we amend the equation for the running in the region $ t_{BL} < t < t_{SUSY}$ from the expression given in \cite{Ovrut:2015uea}. }
\begin{eqnarray}
&&16\pi^2 \frac{d}{d t} m^2_{\tilde{\nu}_{R,3}^c} (t) = -3 g_{BL}^2 (t) M_{B-L}^2(t) - 2 g_R^2 (t) M_R^2 (t) \qquad \qquad \qquad\ \quad \label{eq:RGE1}\\
&&\qquad \qquad \qquad + \frac{3}{4} g_{BL}^2 (t) S_{B-L}(t) - g_R^2(t) S_R(t) \, , \quad t_{SUSY} < t < t_U \nn 
\end{eqnarray}
\begin{eqnarray}
&& 16 \pi^2 \frac{d}{d t} m^2_{\tilde{\nu}_{R,3}^c} (t)
=\left( \frac{3}{4}g_{BL}^2(t) + \frac{1}{2}g_R^2(t) \right)m^2_{\tilde{\nu}_{R,3}^c} (t) \, , \quad t_{BL} < t < t_{SUSY} \nn \\ \quad
\label{eq:RGE2}
\end{eqnarray}
where we define $t = ln(\frac{M}{M_{Z}})$. Here, $M_{a}$, $g_a$ are the associated gaugino masses and gauge couplings for $a=BL, 3R$, and the $S_{a}$-terms are defined as 
\begin{eqnarray}
&&S_{B-L} = \mathrm{Tr} \left(2 m^2_{\tilde{Q}} - m^2_{\tilde{u}_R^c} - m^2_{\tilde{d}_R^c} - 2m^2_{\tilde{L}} + m^2_{\tilde{\nu}_R^c} + m^2_{\tilde{e}_R^c} \right)
\\
&&S_{R} = m^2_{H_u} - m^2_{H_d} + \mathrm{Tr} \left( -\frac{3}{2}  m^2_{\tilde{u}_R^c} +\frac{3}{2}  m^2_{\tilde{d}_R^c} - \frac{1}{2}  m^2_{\tilde{\nu}_R^c}+ \frac{1}{2}  m^2_{\tilde{e}_R^c}\right) 
\end{eqnarray}
The RGEs for the $M_{a}$ and $S_a$ are given in \cite{Ovrut:2015uea}, and can be solved to yield the following analytic expressions.
\begin{eqnarray}
S_{a}(t) &=& \frac{g_a^2(t)}{g_a^2(t_U)} S_a(t_U) \\
M_{a}(t) &=& \frac{g_a^2(t)}{g_a^2(t_U)}M_a(t_U)
\end{eqnarray}
The running of the gauge couplings $g_a$ can be found by solving the RGEs
\begin{eqnarray}
\frac{d}{dt}\alpha_a^{-1} &=& -\frac{b_{a,S}}{2\pi}\, , \qquad a=BL, R \,, \qquad t_{SUSY} < t < t_U
\nn \\
\frac{d}{dt}\alpha_a^{-1} &=& -\frac{b_a}{2\pi} \, ,\qquad ~~a=BL, R \, , \qquad t_{BL} < t < t_{SUSY} \, 
\nn \\
\frac{d}{dt}\alpha_1^{-1} &=& -\frac{b_1}{2\pi} \, , \qquad \qquad \qquad \quad \quad~~~~~~~ t_{Z} < t < t_{SUSY}
\end{eqnarray}
with the boundary conditions
\begin{eqnarray}
\alpha_{BL} (t_{BL}) = \frac{2}{5} \frac{\alpha_1(t_{BL})}{\cos^2\theta_R} \, , \quad
\alpha_{R} (t_{BL}) = \frac{3}{5} \frac{\alpha_1(t_{BL})}{\sin^2\theta_R} \, , \quad
\alpha_1 (t_Z) = 0.0170 \, .
\end{eqnarray}
The beta functions for the various regimes are $b_{BL, S} = 6, b_{R, S} = 7$ and $b_{BL} = \frac{33}{8}, b_{R} = \frac{53}{12}, b_1 = \frac{41}{10}$.

We can solve the RGEs given in equations \eqref{eq:RGE1} and \eqref{eq:RGE2} to arrive at the following expression for $m^2_{\tilde{\nu}_{R,3}^c}$ at a given scale. We find that
{\small
\begin{eqnarray}
 \noindent m^2_{\tilde{\nu}_{R,3}^c} (t)=m^2_{\tilde{\nu}_{R,3}^c} (t_{U})
+ \frac{3}{4} b_{BL, S}^{-1} M^2_{B-L}(t_U) \left(
\frac{g_{BL}^4(t_U) - g_{BL}^4(t)}{g_{BL}^4(t_U)}
\right)
+ \frac{1}{2} b_{R, S}^{-1} M^2_{R}(t_U) \left(
\frac{g_{R}^4(t_U) - g_{R}^4(t)}{g_{R}^4(t_U)}
\right)
\nn \\
- \frac{3}{8} b_{BL, S}^{-1} S_{B-L} (t_U) \left(
\frac{g_{BL}^2(t_U) - g_{BL}^2(t)}{g_{BL}^2(t_U)}
\right)
+ \frac{1}{2} b_{R, S}^{-1} S_{R} (t_U) \left(
\frac{g_{R}^2(t_U) - g_{R}^2(t)}{g_{R}^2(t_U)}
\right)  \qquad \qquad \quad
\nn \\\label{eq:m-higher}
\end{eqnarray} }
for $t_{SUSY} < t < t_U$, and
\begin{eqnarray}
m^2_{\tilde{\nu}_{R,3}^c} (t) = m^2_{\tilde{\nu}_{R,3}^c} (t_{SUSY})
\left[
\frac{\alpha_{BL}^{-1}(t_U) - \frac{b_{BL}^{SUSY} }{2\pi} (t_{SUSY} - t_U)- \frac{b_{BL} }{2\pi} (t- t_{SUSY})}{\alpha_{BL}^{-1}(t_U) - \frac{b_{BL}^{SUSY} }{2\pi} (t_{SUSY} - t_U)} 
\right]^{\frac{1}{4\pi} \frac{3}{4}(-\frac{b_{BL}}{2\pi})^{-1}}
\nn\\
\cdot
\left[
\frac{\alpha_{R}^{-1}(t_U) - \frac{b_{R}^{SUSY} }{2\pi} (t_{SUSY} - t_U)- \frac{b_{R} }{2\pi} (t- t_{SUSY})}{\alpha_{R}^{-1}(t_U) - \frac{b_{R}^{SUSY} }{2\pi} (t_{SUSY} - t_U)} 
\right]^{\frac{1}{4\pi} \frac{1}{2}(-\frac{b_{R}}{2\pi})^{-1}} \quad
\nn\\
=
m^2_{\tilde{\nu}_{R,3}^c} (t_{SUSY})
\left( 
\frac{g^2_{BL}(t)}{g^2_{BL}(t_{SUSY})}
\right)^{\frac{3}{8b_{BL}}}
\left(
\frac{g^2_{R}(t)}{g^2_{R}(t_{SUSY})}
\right)^{\frac{1}{4b_{R}}} \label{eq:m-lower} \qquad \qquad \qquad \qquad \qquad \quad~
\end{eqnarray}
for $t_{BL} < t < t_{SUSY}$.

Given that we know $m^2_{\tilde{\nu}_{R,3}^c} (t) $ at $t= t_{BL}$, we would like to determine the corresponding value of $m^2_{\tilde{\nu}_{R,3}^c} (t) $ at $t= t_{U}$. Rearranging equations  \eqref{eq:m-higher} and \eqref{eq:m-lower}, we find that
{\small
\begin{eqnarray}
m^2_{\tilde{\nu}_{R,3}^c} (t_{U}) =
- \frac{3}{4} b_{BL, S}^{-1} M^2_{B-L}(t_U) \left(
\frac{g_{BL}^4(t_U) - g_{BL}^4(t_{SUSY})}{g_{BL}^4(t_U)}
\right)
- \frac{1}{2} b_{R, S}^{-1} M^2_{R}(t_U) \left(
\frac{g_{R}^4(t_U) - g_{R}^4(t_{SUSY})}{g_{R}^4(t_U)}
\right)
\nn \\
+ \frac{3}{8} b_{BL, S}^{-1} S_{B-L} (t_U) \left(
\frac{g_{BL}^2(t_U) - g_{BL}^2(t_{SUSY})}{g_{BL}^2(t_U)}
\right)
- \frac{1}{2} b_{R, S}^{-1} S_{R} (t_U) \left(
\frac{g_{R}^2(t_U) - g_{R}^2(t_{SUSY})}{g_{R}^2(t_U)}~~~
\right)
\nn \\
+ m^2_{\tilde{\nu}_{R,3}^c} (t_{BL})
\left( 
\frac{g^2_{BL}(t_{SUSY})}{g^2_{BL}(t_{BL})}
\right)^{\frac{3}{8b_{BL}}}
\left(
\frac{g^2_{R}(t_{SUSY})}{g^2_{R}(t_{BL})}
\right)^{\frac{1}{4b_{R}}} \, .\label{burt1} \qquad \qquad \qquad \qquad \qquad \qquad \qquad \qquad~
\end{eqnarray}}
However, we must now use the fact that $S_{B-L}(t_U)$ and $S_R(t_U)$ contain $m^2_{\tilde{\nu}_{R,3}^c} (t_{U})$. To accomplish this, we redefine the above expressions in terms of two new objects, $S_{B-L}(t_U)^{\prime}$ and $S_R(t_U)^{\prime}$, given by 
\begin{eqnarray}
S_{B-L} &=& \mathrm{Tr} \left(2 m^2_{\tilde{Q}} - m^2_{\tilde{u}_R^c} - m^2_{\tilde{d}_R^c} - 2m^2_{\tilde{L}} + m^2_{\tilde{e}_R^c} \right) +  m^2_{\tilde{\nu}_{R,1}^c} +  m^2_{\tilde{\nu}_{R,2}^c} +  m^2_{\tilde{\nu}_{R,3}^c} 
\nn \\
&=& 
S_{B-L}^{\prime} + m^2_{\tilde{\nu}_{3}^c} \label{burt2}
\\
S_{R} &=& m^2_{H_u} - m^2_{H_d} + \mathrm{Tr} \left( -\frac{3}{2}  m^2_{\tilde{u}_R^c} +\frac{3}{2}  m^2_{\tilde{d}_R^c} + \frac{1}{2}  m^2_{\tilde{e}_R^c}\right) - \frac{1}{2}  m^2_{\tilde{\nu}_{R,1}^c}- \frac{1}{2}  m^2_{\tilde{\nu}_{R,2}^c}- \frac{1}{2}  m^2_{\tilde{\nu}_{R,3}^c}
\nn \\
&=& 
S_{R}^\prime  - \frac{1}{2}  m^2_{\tilde{\nu}_{R,3}^c}
\label{eq:S-prime}
\end{eqnarray}
respectively. For simplicity, we have suppressed the fact that all terms in \eqref{burt2} and  \eqref{eq:S-prime} are evaluated at $t = t_{U}$.
Doing this, and re-arranging terms to extract $m^2_{\tilde{\nu}_{R,3}^c} (t_U) $, we arrive at the expression
{\small
\begin{eqnarray}
m^2_{\tilde{\nu}_{R,3}^c} (t_U) =
\left( 
1 -  \frac{3}{8} b_{BL, S}^{-1}  \left(
\frac{g_{BL}^2(t_U) - g_{BL}^2(t_{SUSY})}{g_{BL}^2(t_U)}
\right)
- \frac{1}{4} b_{R, S}^{-1}   \left(
\frac{g_{R}^2(t_U) - g_{R}^2(t_{SUSY})}{g_{R}^2(t_U)}
\right)
\right)^{-1} \quad
\nn \\
\bigg\{
- \frac{3}{4} b_{BL, S}^{-1} M^2_{B-L}(t_U) \left(
\frac{g_{BL}^4(t_U) - g_{BL}^4(t_{SUSY})}{g_{BL}^4(t_U)}
\right)
- \frac{1}{2} b_{R, S}^{-1} M^2_{R}(t_U) \left(
\frac{g_{R}^4(t_U) - g_{R}^4(t_{SUSY})}{g_{R}^4(t_U)}
\right)
\nn \\
+ \frac{3}{8} b_{BL, S}^{-1} S_{B-L}^\prime (t_U) \left(
\frac{g_{BL}^2(t_U) - g_{BL}^2(t_{SUSY})}{g_{BL}^2(t_U)}
\right)
- \frac{1}{2} b_{R, S}^{-1} S_{R}^\prime (t_U) \left(
\frac{g_{R}^2(t_U) - g_{R}^2(t_{SUSY})}{g_{R}^2(t_U)}
\right)~~~~
\nn \\
+ m^2_{\tilde{\nu}_{R,3}^c} (t_{BL})
\left( 
\frac{g^2_{BL}(t_{SUSY})}{g^2_{BL}(t_{BL})}
\right)^{\frac{3}{8b_{BL}}}
\left(
\frac{g^2_{R}(t_{SUSY})}{g^2_{R}(t_{BL})}
\right)^{\frac{1}{4b_{R}}}
\bigg\} \, . \qquad \qquad \qquad  \qquad \qquad \qquad \qquad \quad~~~
\label{eq:mNusol1}
\end{eqnarray}}

\subsection*{A.2.~~~Cosmological Constraint}
Recall from \eqref{14} and \eqref{25} that, in order to construct our inflaton and match the necessary cosmological data, it is required that
\begin{eqnarray}
m_{H_u}^2(t_U) + m_{\tilde{L}_{3}}^2(t_U) + m_{\tilde{\nu}_{R,3}^c}^2(t_U) &=& 3(1.58 \times 10^{13} \mathrm{GeV})^2 \, .
\label{eq:cosmo}
\end{eqnarray}
We have previously satisfied this condition in our computational framework by not randomly generating $m_{H_u}^2(t_U)$ but, instead, randomly generating the other two masses and using \eqref{eq:cosmo} to calculate the required value of $m_{H_u}^2(t_U)$.
In the context of our present discussion, we must take this expression into account when we enforce a specific \BL scale.
Since this relation involves $m_{\tilde{\nu}_3^c}$, and $m_{H_u}$ enters \eqref{eq:mNusol1} via the $S$-terms, we can see that equations  \eqref{eq:mNusol1} and \eqref{eq:cosmo} are intertwined and must be solved simultaneously. 
To do this, we can re-express these equations in the form
\begin{eqnarray}
m_{\tilde{\nu}_{R,3}^c}^2 (t_U) = A + B m_{H_u}^2(t_U)
\nn \\
m_{\tilde{\nu}_{R,3}^c}^2 (t_U) + m_{H_u}^2(t_U) = C ~ \, , \label{rehan1}
\end{eqnarray}
where
{\small
\begin{eqnarray}
A &=& \left( 
1 - \frac{3}{8 b_{BL, S}} \left(1 - \frac{g_{BL}^2(t_{SUSY})}{g_{BL}^2(t_U)} \right) - \frac{1}{4 b_{R, S}} \left(1 - \frac{g_{R}^2(t_{SUSY})}{g_{R}^2(t_U)} \right)
\right)^{-1}
\nn \\
&& \bigg\{
- \frac{3}{4b_{BL,S}} M_{B-L}^2(t_U)   \left(1 - \frac{g_{BL}^4(t_{SUSY})}{g_{BL}^4(t_U)} \right)
- \frac{1}{2b_{R,S}} M_{R}^2(t_U)   \left(1 - \frac{g_{R}^4(t_{SUSY})}{g_{R}^4(t_U)} \right)
\nn \\
&& 
+ \frac{3}{8b_{BL,S}} S_{B-L}^{\prime}(t_U) \left(1 - \frac{g_{BL}^2(t_{SUSY})}{g_{BL}^2(t_U)} \right)
- \frac{1}{2b_{R,S}} S_{R}^{\prime \prime}(t_U) \left(1 - \frac{g_{R}^2(t_{SUSY})}{g_{R}^2(t_U)} \right)
\nn \\
&&
+ m_{\tilde{\nu}_{R,3}^c} (t_{BL}) \left( \frac{g_{BL}^2(t_{SUSY})}{g_{BL}^2(t_{BL})}\right)^{\frac{3}{8b_{BL}}} \left( \frac{g_{R}^2(t_{SUSY})}{g_{R}^2(t_{BL})}\right)^{\frac{1}{4b_{R}}}
\bigg\}
\nn \\
B &=& \left( 
1 - \frac{3}{8 b_{BL, S}} \left(1 - \frac{g_{BL}^2(t_{SUSY})}{g_{BL}^2(t_U)} \right) - \frac{1}{4 b_{R, S}} \left(1 - \frac{g_{R}^2(t_{SUSY})}{g_{R}^2(t_U)} \right)
\right)^{-1}
\label{burt3} \\
&& \bigg\{
- \frac{1}{2b_{R,S}} \left(1 - \frac{g_{R}^2(t_{SUSY})}{g_{R}^2(t_U)} \right)
\bigg\}
\nn \\
C &=& 3(1.58 \times 10^{13} \mathrm{GeV})^2 - m^2_{\tilde{L}_{3}} (t_U) \nn 
\end{eqnarray}}
with
\begin{eqnarray}
S_R^{\prime \prime} &=& S_R^{\prime} + m^2_{H_u} \ .
\end{eqnarray}
The equations in \eqref{rehan1} are then soluble by inverting the matrix expression
\begin{eqnarray}
\begin{pmatrix}
1 && -B \\
1 && 1
\end{pmatrix}
\begin{pmatrix}
m^2_{\tilde{\nu}_{R,3}^c} (t_U) \\
m_{H_u}^2(t_U)
\end{pmatrix}
&=&
\begin{pmatrix}
A \\
C
\end{pmatrix}
\end{eqnarray}
to give
\begin{eqnarray}
\begin{pmatrix}
m^2_{\tilde{\nu}_{R,3}^c} (t_U) \\
m_{H_u}^2(t_U)
\end{pmatrix}
&=&
\frac{1}{1+ B}
\begin{pmatrix}
1 &&  B\\
-1 && 1
\end{pmatrix}
\begin{pmatrix}
A \\
C
\end{pmatrix}
=
\frac{1}{1+ B}
\begin{pmatrix}
A + BC \\
-A + C
\end{pmatrix} 
\label{eq:simult-matrix}
\end{eqnarray}
and inserting A, B and C in \eqref{burt3}. In particular, for $m^2_{\tilde{\nu}_{R,3}^c} (t_U)$ we find
{\small
\begin{eqnarray}
&&m^2_{\tilde{\nu}_{R,3}^c} (t_U) = \nn \\
&& \frac{1}{ 1 + \left( 1 - \frac{3}{8 b_{BL, S}} \left(1 - \frac{g_{BL}^2(t_{SUSY})}{g_{BL}^2(t_U)} \right) - \frac{1}{4 b_{R, S}} \left(1 - \frac{g_{R}^2(t_{SUSY})}{g_{R}^2(t_U)} \right)
\right)^{-1}
\bigg\{
- \frac{1}{2b_{R,S}} \left(1 - \frac{g_{R}^2(t_{SUSY})}{g_{R}^2(t_U)} \right)
\bigg\}} 
\nn  \\
&&
\Bigg[
\left( 
1 - \frac{3}{8 b_{BL, S}} \left(1 - \frac{g_{BL}^2(t_{SUSY})}{g_{BL}^2(t_U)} \right) - \frac{1}{4 b_{R, S}} \left(1 - \frac{g_{R}^2(t_{SUSY})}{g_{R}^2(t_U)} \right)
\right)^{-1}
\nn \\
&& \bigg\{
- \frac{3}{4b_{BL,S}} M_{B-L}^2(t_U)   \left(1 - \frac{g_{BL}^4(t_{SUSY})}{g_{BL}^4(t_U)} \right)
- \frac{1}{2b_{R,S}} M_{R}^2(t_U)   \left(1 - \frac{g_{R}^4(t_{SUSY})}{g_{R}^4(t_U)} \right)
\nn \\
&& 
+ \frac{3}{8b_{BL,S}} S_{B-L}^{\prime}(t_U) \left(1 - \frac{g_{BL}^2(t_{SUSY})}{g_{BL}^2(t_U)} \right)
- \frac{1}{2b_{R,S}} S_{R}^{\prime \prime}(t_U) \left(1 - \frac{g_{R}^2(t_{SUSY})}{g_{R}^2(t_U)} \right)
\nn \\
&&
+ m_{\tilde{\nu}_{R,3}^c} (t_{BL}) \left( \frac{g_{BL}^2(t_{SUSY})}{g_{BL}^2(t_{BL})}\right)^{\frac{3}{8b_{BL}}} \left( \frac{g_{R}^2(t_{SUSY})}{g_{R}^2(t_{BL})}\right)^{\frac{1}{4b_{R}}}
\bigg\}
\label{eq:mNusol2} \\
&& +
\left( 
1 - \frac{3}{8 b_{BL, S}} \left(1 - \frac{g_{BL}^2(t_{SUSY})}{g_{BL}^2(t_U)} \right) - \frac{1}{4 b_{R, S}} \left(1 - \frac{g_{R}^2(t_{SUSY})}{g_{R}^2(t_U)} \right)
\right)^{-1}
\nn \\
&& 
\times \bigg\{
- \frac{1}{2b_{R,S}} \left(1 - \frac{g_{R}^2(t_{SUSY})}{g_{R}^2(t_U)} \right)
\bigg\} 
\times \left( 3(1.58 \times 10^{13} \mathrm{GeV})^2 - m^2_{\tilde{L}_3} (t_U) \right)
\Bigg] \nn  \,.
\end{eqnarray}}
The solutions to $m^2_{H_u}$ and $m^2_{\tilde{\nu}^c_{R,3}}$ given by equation \eqref{eq:simult-matrix} are implemented in our computational framework as follows: 
\begin{enumerate}
\item The scalar soft mass parameters $\{m^2_{\tilde{Q}_1}, m^2_{\tilde{Q}_2}, m^2_{\tilde{Q}_3}, m^2_{\tilde{u}^c_{R,1}}, m^2_{\tilde{u}^c_{R,2}}, m^2_{\tilde{u}^c_{R,3}}, m^2_{\tilde{d}^c_{R,1}}, m^2_{\tilde{d}^c_{R,2}},\\ 
m^2_{\tilde{d}^c_{R,3}}, m^2_{\tilde{e}^c_{R,1}}, m^2_{\tilde{e}^c_{R,2}}, m^2_{\tilde{e}^c_{R,3}}$ 
$m^2_{\tilde{L}_1}, m^2_{\tilde{L}_2}, m^2_{\tilde{L}_3}, m^2_{\tilde{\nu}^c_{R,1}}, m^2_{\tilde{\nu}^c_{R,2}}, m^2_{H_d}  \}$ and the  soft gaugino masses 
$\{M_{3}, M_{2}, M_{B-L}, M_{R} \}$ are thrown statistically at the unification mass scale $M_U$.
\item The SUSY breaking scale $M_{SUSY}$ is initially approximated by the absolute value of $M_{3}$.
\item The desired value $M_{BL}$ of the \BL scale is inputted by choosing $m_{\tilde{\nu}_{R,3}^c}^2 (M_{BL})$ using \eqref{30}.
\item Using this value of the \BL scale, the initial guess for the $M_{SUSY}$ and the previously generated soft-breaking masses, we determine the values of $m^2_{\tilde{\nu}^c_{R,3}}$ and $m^2_{H_u}$ at $M_U$ using equation \eqref{eq:simult-matrix}.
\item Having a complete set of initial soft mass data, the SUSY scale is iteratively solved for using the relation
\begin{eqnarray}
\sqrt{m_{\tilde{t}_1} (M_{SUSY})m_{\tilde{t}_2} (M_{SUSY}) } = M_{SUSY} \,.
\label{eq:susy}
\end{eqnarray}
In each iteration, the value of $M_{SUSY}$ changes. Hence, one must re-solve for $m^2_{\tilde{\nu}^c_{R,3}}$ and $m^2_{H_u}$ at $M_U$ using equation \eqref{eq:simult-matrix} and the updated value of $M_{SUSY}$. This process continues until the iterative solution for \eqref{eq:susy} converges within an allowed range.
\item We finally check to see if $m^2_{\tilde{\nu}^c_{R,3}} (M_U) < m^2_{\tilde{\nu}^c_{R,1}} (M_U), \, m^2_{\tilde{\nu}^c_{R,2}} (M_U)$, a condition which defines the third family right-handed sneutrino. If this is not satisfied, we discard this set of initial data and then return to the first step.
\item The complete set of soft masses, the final value of $M_{SUSY}$ and the inputted value of the \BL breaking scale are then used to carry out the remaining physical checks on the scale of electroweak breaking, the Higgs mass and the sparticle mass bounds in the manner described in \cite{Ovrut:2015uea}. 
\end{enumerate}
\begin{figure}
\begin{center}
\includegraphics[scale=0.5]{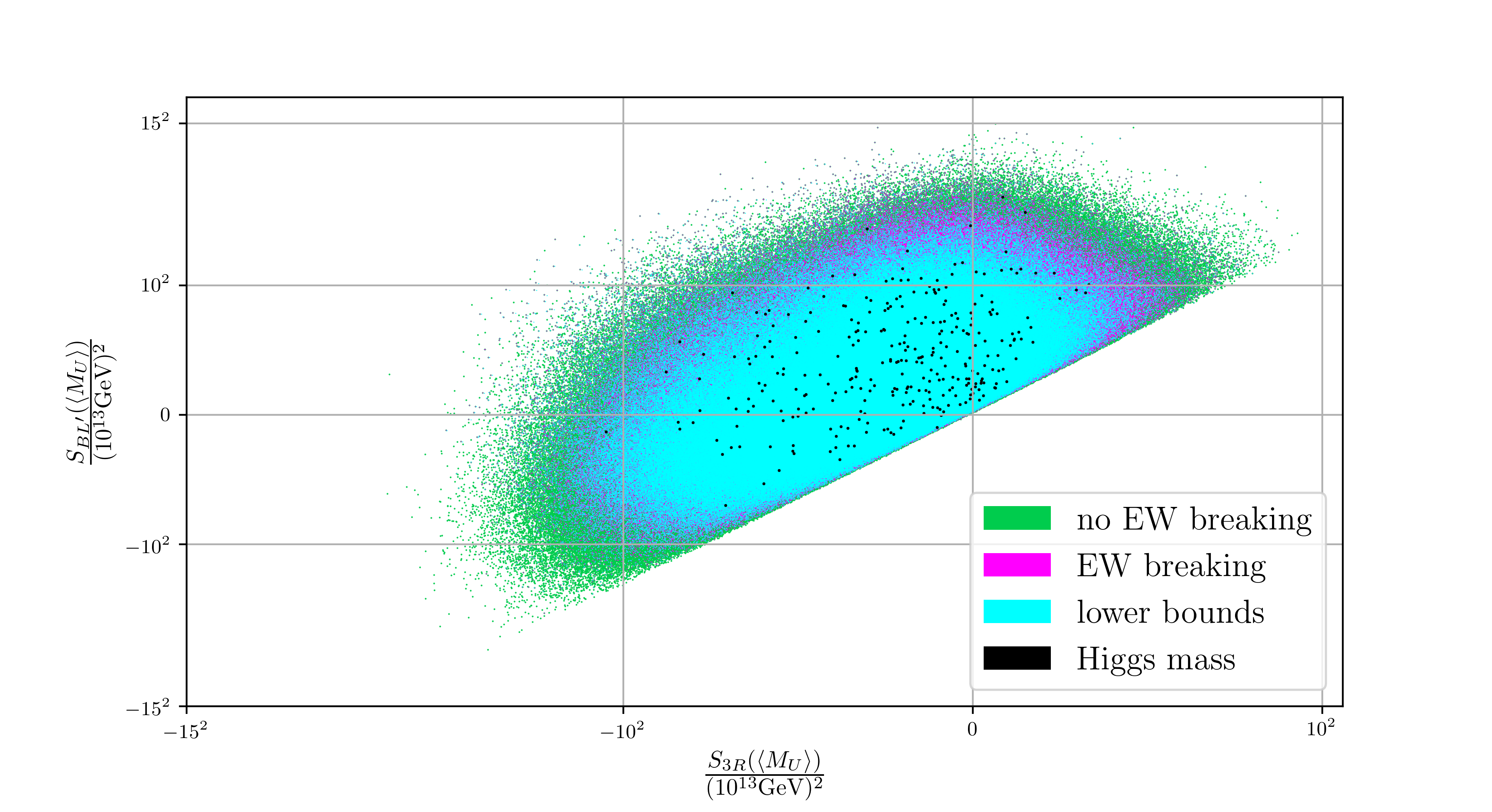}
\caption{Results from generating 50 million sets of initial data where the \BL scale is chosen from a log-uniform distribution between $10^6$ GeV and $10^{14}$ GeV. We find that 6,123,522 points break \BL but not electroweak symmetry, and
1,997,972 points break \BL and electroweak symmetry. Of the latter
1,040,259 points are consistent with current LHC bounds on sparticle searches. Finally, we have 305 points which satisfy all these conditions and are within the $2\sigma$ window of the measured Higgs mass.}
\label{fig:blackpoints4}
\end{center}
\end{figure}

\begin{figure}[h!]
\begin{center}
\includegraphics[scale=0.4]{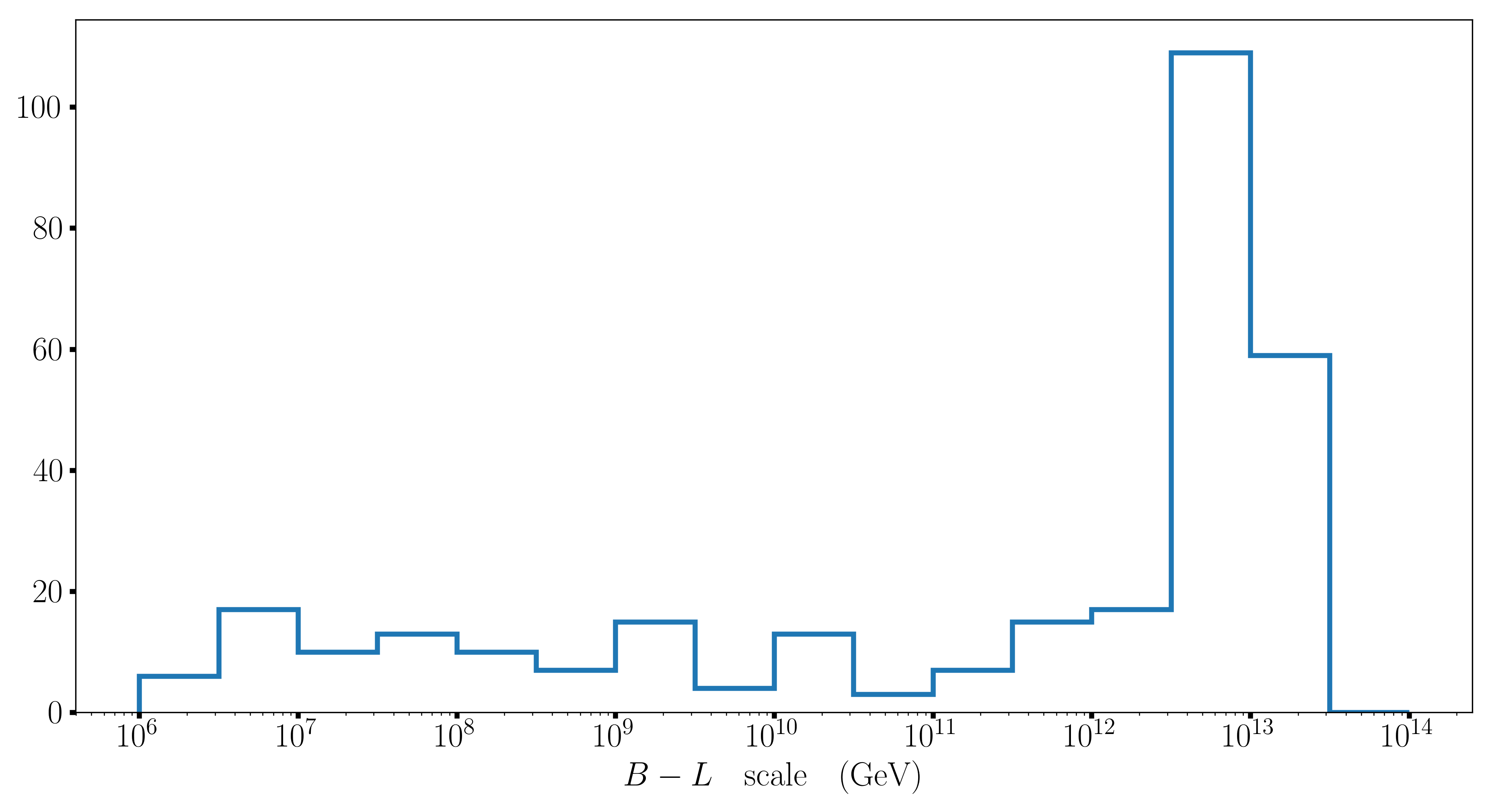}
\caption{Distribution of the \BL breaking scale for the 305 black points displayed in Figure \ref{fig:blackpoints4}.}
\label{fig:BLScale3}
\end{center}
\end{figure}

In the text of this paper, we will use this formalism to generate physically acceptable valid black points whose \BL scale is in the range $10^{10}~ {\rm GeV} \leq M_{BL} \leq 10^{12}~ {\rm GeV}$. This range comfortably accommodates our theory of reheating. Be that as it may, it is of interest to see whether the \BL scale can be substantially reduced to much lower values. With that in mind, in this Appendix we will implement the above procedure and generate 50 million initial throws of the soft masses with the inputted scale of $U(1)_{B-L}$ breaking randomly generated from a log-uniform distribution between $10^6$ GeV and $10^{14}$ GeV. Carrying out our checks, we find that this ultimately leads to 305 sets of initial data which satisfy all phenomenological constraints 1), 2) and 3) presented earlier. These physically valid black points are shown in Figure \ref{fig:blackpoints4}.
The distribution of the \BL breaking scale for the black points in Figure \ref{fig:blackpoints4}
is given in Figure \ref{fig:BLScale3}and verifies that our approach has indeed allowed us to extend the range of the \BL scale to lower values.

\subsection*{A.3.~~~Fine-Tuning}
\begin{figure}
\begin{center}
\includegraphics[scale=0.4]{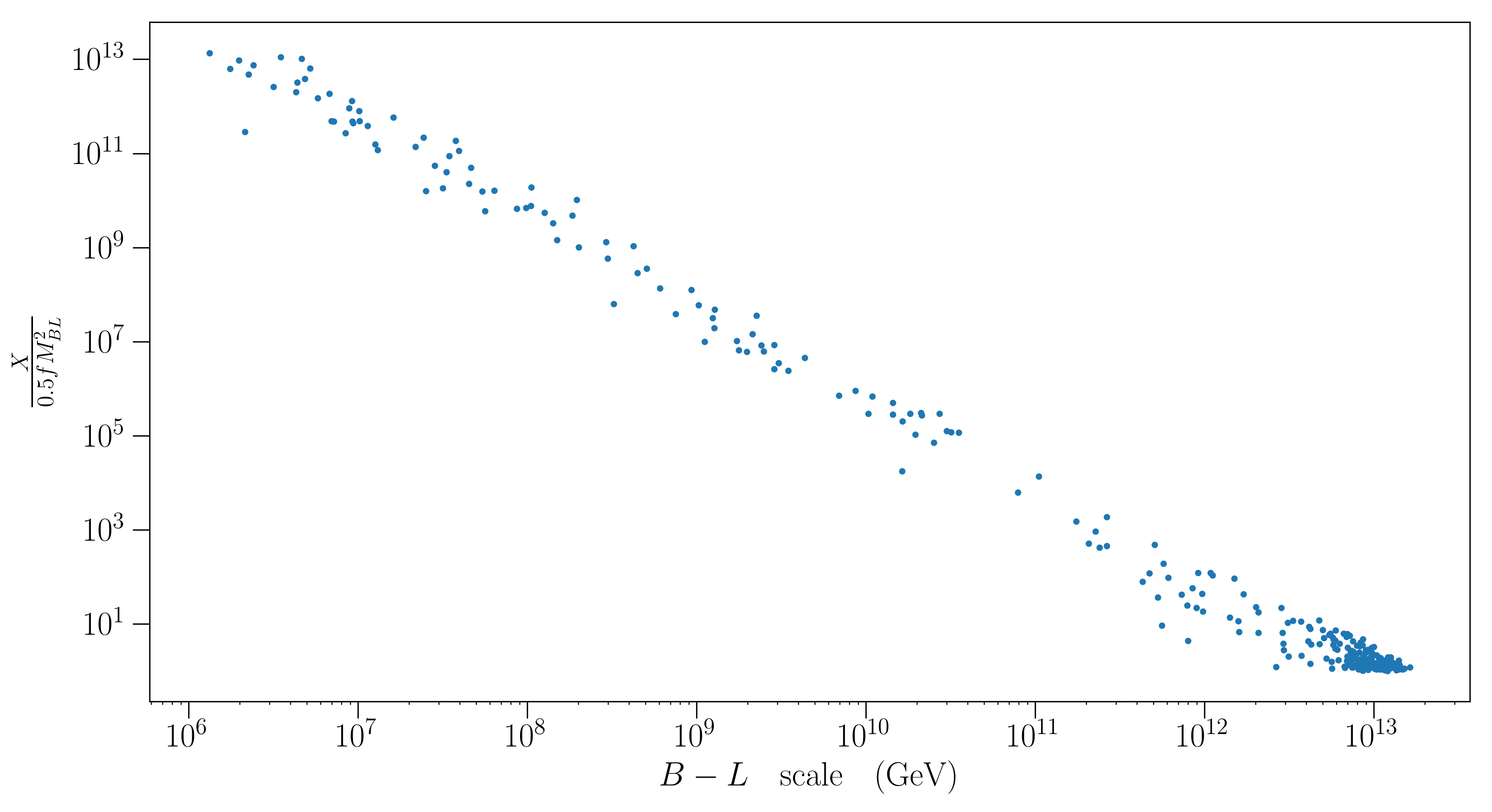}
\end{center}
\caption{Log-log plot of $\frac{X}{\frac{1}{2} f M_{BL}^2}$ against the \BL scale, for the valid black points shown in Figure \ref{fig:blackpoints4} from the scan of 50 million sets of initial conditions. The quantity $\frac{X}{\frac{1}{2} f M_{BL}^2}$ expresses the degree of fine-tuning required to achieve the associated value of the \BL scale.}
\label{fig:fine-tuning-log2}
\end{figure}

As we have seen, in order to ensure that the scale of \BL breaking have a desired value, the mass of the third family right-handed sneutrino
has to be adjusted against the other masses. To quantify the degree of fine-tuning necessary, we re-examine equation \eqref{eq:mNusol1}.
Re-arranging this equation as an expression for $M_{BL}$, we find
\begin{eqnarray}
&&\frac{1}{2}\frac{\left( \frac{g_{BL}^2(t_{SUSY})}{g_{BL}^2(t_{BL})}\right)^{\frac{3}{8b_{BL}}} \left( \frac{g_{R}^2(t_{SUSY})}{g_{R}^2(t_{BL})}\right)^{\frac{1}{4b_{R}}}}{\left( 
1 - \frac{3}{8 b_{BL, S}} \left(1 - \frac{g_{BL}^2(t_{SUSY})}{g_{BL}^2(t_U)} \right) - \frac{1}{4 b_{R, S}} \left(1 - \frac{g_{R}^2(t_{SUSY})}{g_{R}^2(t_U)} \right)
\right) }M_{BL}^2 \nn \\
&=&
\left( 
1 - \frac{3}{8 b_{BL, S}} \left(1 - \frac{g_{BL}^2(t_{SUSY})}{g_{BL}^2(t_U)} \right) - \frac{1}{4 b_{R, S}} \left(1 - \frac{g_{R}^2(t_{SUSY}}{g_{R}^2(t_U)} \right)
\right)^{-1}
\nn \\
&& \bigg\{
- \frac{3}{4b_{BL,S}} M_{B-L}^2(t_U)   \left(1 - \frac{g_{BL}^4(t_{SUSY})}{g_{BL}^4(t_U)} \right)
- \frac{1}{2b_{R,S}} M_{R}^2(t_U)   \left(1 - \frac{g_{R}^4(t_{SUSY})}{g_{R}^4(t_U)} \right)
\nn \\
&& 
+ \frac{3}{8b_{BL,S}} S_{B-L}^{\prime}(t_U) \left(1 - \frac{g_{BL}^2(t_{SUSY})}{g_{BL}^2(t_U)} \right)
- \frac{1}{2b_{R,S}} S_{R}^{\prime}(t_U) \left(1 - \frac{g_{R}^2(t_{SUSY})}{g_{R}^2(t_U)} \right)
\bigg\} 
\nn \\
&&
- m_{\tilde{\nu}_{R,3}^c}^2 (t_U) \, .
\label{eq:fine-tuning1}
\end{eqnarray}
This can be schematically expressed as 
\begin{eqnarray}
\frac{1}{2} f F M_{BL}^2 = F X - m_{\tilde{\nu}_{R,3}^c}^2 (t_U) \, ,
\label{eq:fine-tuning2}
\end{eqnarray}
where
{\small
\begin{eqnarray}
f &=& \left( \frac{g_{BL}^2(t_{SUSY})}{g_{BL}^2(t_{BL})}\right)^{\frac{3}{8b_{BL}}} \left( \frac{g_{R}^2(t_{SUSY})}{g_{R}^2(t_{BL})}\right)^{\frac{1}{4b_{R}}}
\nn \\
F &=& \left( 
1 - \frac{3}{8 b_{BL, S}} \left(1 - \frac{g_{BL}^2(t_{SUSY})}{g_{BL}^2(t_U)} \right) - \frac{1}{4 b_{R, S}} \left(1 - \frac{g_{R}^2(t_{SUSY}}{g_{R}^2(t_U)} \right)
\right)^{-1}
\label{eq:fine-tuning3} \\
X &=& \bigg\{
- \frac{3}{4b_{BL,S}} M_{B-L}^2(t_U)   \left(1 - \frac{g_{BL}^4(t_{SUSY})}{g_{BL}^4(t_U)} \right)
- \frac{1}{2b_{R,S}} M_{R}^2(t_U)   \left(1 - \frac{g_{R}^4(t_{SUSY})}{g_{R}^4(t_U)} \right)
\nn \\
&& 
+ \frac{3}{8b_{BL,S}} S_{B-L}^{\prime}(t_U) \left(1 - \frac{g_{BL}^2(t_{SUSY})}{g_{BL}^2(t_U)} \right)
- \frac{1}{2b_{R,S}} S_{R}^{\prime}(t_U) \left(1 - \frac{g_{R}^2(t_{SUSY})}{g_{R}^2(t_U)} \right)
\bigg\} \, . \nn
\end{eqnarray}}
Equation \eqref{eq:fine-tuning2} shows us precisely where the delicate cancellation necessary to produce a low \BL breaking scale arises. To quantify the degree of fine-tuning, we can plot the ratio
\begin{eqnarray}
\frac{F X}{\frac{1}{2} F f M_{BL}^2} = \frac{X}{\frac{1}{2} f M_{BL}^2}
\label{eq:fine-tuning4}
\end{eqnarray}
against the $M_{BL}$ for those sets of initial data which survive all the phenomenological checks previously outlined. The quantity \eqref{eq:fine-tuning4} is the Barbieri-Giudice (B-G) sensitivity \cite{Barbieri:1987fn} for the \BL breaking scale. 
The results of doing this for the 305 ``black points" in Figure \ref{fig:blackpoints4} are shown in Figure \ref{fig:fine-tuning-log2}.
%


\section*{Appendix B: ~~Useful Approximation}\label{Apptrick}
Consider the quantity
\be 
\langle \psi^2(t)\rangle={1\over 2\delta}\int^{t+\delta}_{t-\delta}\psi^2(\tilde{t})d\tilde{t}\, ,
\ee
where $\delta={2\pi/m_\psi}$ is the period of the oscillations of $\psi$.
In order to remove of the integral in this expression, we use the approximation that $A(t)$ does not change much in the time interval $t-\delta$ to $t+\delta$. This  is true
since $\delta\ll H^{-1}$ as long as $m_\psi\gg H$. It follows that
\ba &\,& \langle \psi^2(t)\rangle={1\over 2\delta}\int^{t+\delta}_{t-\delta}\psi^2(\tilde{t})d\tilde{t}\nn\\
&\,&\qquad\quad\,\approx
{A^2(t)\over 2\delta}\int^{t+\delta}_{t-\delta}\sin^2\lf[m_{\psi}(\tilde{t}-t_{osc}) \rt] d\tilde{t}  \nn\\
&\,&\qquad\quad\,={A^2(t)\over 2}\label{trick00}
\, . 
\ea

\section*{Appendix C: ~~Diagonalization of Mass Matrices}
\label{appendix-a}

In the \BLMSSM we have three families of quark and lepton chiral superfields 
\begin{eqnarray}
Q \sim (\mathbf{3}, \mathbf{2}, 0, 1/3)\, , \,
u_R^c \sim (\mathbf{\bar{3}}, \mathbf{1}, -1/2, -1/3)\, , \,
d_R^c \sim (\mathbf{3}, \mathbf{1}, 1/2, -1/3)\, 
\nn \\
L \sim (\mathbf{1}, \mathbf{2}, 0, -1)\, , \,
e_{R}^c (\mathbf{1}, \mathbf{1}, 1/2, 1)\,  , \,
\nu_{R}^c (\mathbf{1}, \mathbf{1}, -1/2, 1) \, ,\qquad \qquad \qquad ~ \label{t1}
\end{eqnarray}
as well as a pair of Higgs-Higgs conjugate doublet superfields
\begin{eqnarray}
H_u \sim (\mathbf{1}, \mathbf{2}, 1/2, 0)\, , \,
H_d \sim (\mathbf{1}, \mathbf{2}, -1/2, 0)\, ,  \label{t2}
\end{eqnarray}
where we have presented the  transformation properties under the gauge group $SU(3)_C \times SU(2)_L \times U(1)_{3R} \times U(1)_{B-L}$.
The superpotential for the \BLMSSM is given by
\begin{eqnarray}
W = y_u Q H_u u^c_R - y_d Q H_d d^c_R - y_e L H_d e^c_R + y_{\nu} L H_u \nu^c_R + \mu H_u H_d \, ,
\end{eqnarray}
where we assume that the Yukawa parameters are family diagonal and real, and do not display the family index. This gives rise to the Lagrangian 
\begin{eqnarray}
\La_{W} &=& - W \big|_{\theta^2} + \mathrm{h.c} \nn \\
& \supset &
-y_{t} \left(H_u^{0} t_L t_R^c - H_u^{+} b_L t_R^c\right)
-y_{\tau} \left(
\tilde{\tau}_{L} \psi_d^0 \tau_R^c - \tilde{\nu}_{3,L} \psi_d^{-} \tau_R^c 
\right)
\nn \\
&&
- y_{\nu_3} \left(
H_u^{0} \nu_{L,3} \nu_{R,3}^c - H_u^{+} \tau_L \nu_{R,3}^c
+ \tilde{\nu}_{L,3} \psi_u^{0} \nu_{R,3}^c 
+ \tilde{\nu}_{R,3}^c \psi_u^{0} \nu_{L,3}
- \tilde{\nu}_{R,3}^c \psi_u^{+} \tau_L 
\right)
\nn \\
&& 
- \mu \left( \psi_u^{+} \psi_d^{-} - \psi_u^{0} \psi_d^{0} \right) + \mathrm{h.c.}\, ,
\end{eqnarray}
as well as the purely scalar $F$-term potental $V_F = \sum_i\big|\frac{\partial W}{\partial \phi_i} \big|^2$. 

We also have the soft-mass terms for the $SU(2)$, $U(1)_{3R}$ and $U(1)_{B-L}$ gauginos given by
\begin{eqnarray}
\La_{soft} &=&
- \frac{1}{2} M_2 \left( 
\tilde{W}^1 \tilde{W}^1 + \tilde{W}^2 \tilde{W}^2 + \tilde{W}^3 \tilde{W}^3 \right)
- \frac{1}{2} M_{3R} \tilde{W}_{R} \tilde{W}_{R} 
- \frac{1}{2} M_{B-L} \tilde{B} \tilde{B} 
+ \mathrm{h.c.}
\nn \\
&=&
- M_2 \tilde{W}^+ \tilde{W}^- 
- \frac{1}{2} M_2 \tilde{W}^3 \tilde{W}^3
- \frac{1}{2} M_{3R} \tilde{W}_{R} \tilde{W}_{R} 
- \frac{1}{2} M_{B-L} \tilde{B} \tilde{B} 
+ \mathrm{h.c.}\, ,
\end{eqnarray}
where 
\begin{eqnarray}
\tilde{W}^{\pm} &=& \frac{1}{\sqrt{2}} (\tilde{W}^1 \mp \tilde{W}^2)\, .
\end{eqnarray}
Additionally, the following terms arise from the gauge super-covariant derivative:
\begin{eqnarray}
\La_{kinetic} &=&
- \frac{1}{\sqrt{2}}g_2 \left( 
\sqrt{2}H_u^{0*} \tilde{W}^- \psi_u^+ - H_u^{0*} \tilde{W}^3 \psi_u^0 
+ \tilde{\nu}_{L,3}^{*} \tilde{W}^3 \nu_{L,3} 
+ \sqrt{2} \tilde{\nu}_{L,3}^{*} \tilde{W}^{+} \tau_{L}
\right)
\nn \\
&&
- \sqrt{2}g_R \left(
q_{R_u}  H_u^{0*} \tilde{W}_R \psi_u^0  
+ q_{R_\nu} \tilde{\nu}_{R,3}^{c*} \tilde{W}_R \nu_{R,3}^c 
\right)
\nn \\
&&
- \sqrt{2}g_{BL} \left(
q_{BL_L} \tilde{\nu}_{L,3}^{*} \tilde{B} \nu_{L,3} 
+ q_{BL_\nu} \tilde{\nu}_{R,3}^{c*} \tilde{B} \nu_{R,3}^c
\right)
+ \mathrm{h.c.}
\end{eqnarray}
The inflaton, $\phi = (H_u^0 + \tilde{\nu}_{L,3} + \tilde{\nu}_{R,3}^c)/\sqrt{3}$, attains a time-dependent expectation value during both the inflationary and post-inflationary periods. In doing so, it gives rise to fermion mixing terms in the Lagrangian. We now determine the mass eigenstates and eigenvalues due to this mixing.

\subsection*{C.1.  Chargino Mixing}
\label{apx:chargino-mixing}
Dropping terms which have any neutrino Yukawa coupling $y_{\nu}$, the effective mass Lagrangian for the ``charginos" is given by
\begin{eqnarray}
\La_{mass, C} &=& 
- g_2 \braket{ H_u^{0 *} } \tilde{W}^{-} \psi_u^{+} 
- g_2 \braket{ \tilde{\nu}_{L,3}^*} \tilde{W}^+ \tau_L
+  y_{\tau} \braket{\tilde{\nu}_{L,3}} \tau_R^{c} \psi_d^{-} 
\nn\\&\,&
- \mu \psi_u^{+} \psi_d^{-}
- M_2 \tilde{W}^{+} \tilde{W}^{-} 
+ \mathrm{h.c.}
\label{eq:L-mass-C}
\end{eqnarray}
Of course, the expectation values need to be expressed in terms of the RMS value of the inflaton field using
\begin{eqnarray}
\braket{H_u^{0}} = \braket{\tilde{\nu}_{L,3}} =\braket{\tilde{\nu}_{R,3}^c} = \frac{1}{\sqrt{6}}\sqrt{\braket{\psi^2}} \, .
\end{eqnarray}
However, for clarity, we will continue to express the expectation values in terms of the original fields until it becomes necessary to evaluate them.
We can re-express \eqref{eq:L-mass-C} as
\begin{eqnarray}
\La_{mass, C} = -\frac{1}{2} \vec{\Psi}^T M_C \vec{\Psi} + \mathrm{h.c.} \ ,
\end{eqnarray}
where
\begin{eqnarray}
\vec{\Psi}^T = \left(
\tilde{W}^{+}, \tilde{\psi}_u^{+}, \tau_R^c, \tilde{W}^{-}, \psi_d^{-}, \tau_L
\right) 
\label{eq:basis}
\end{eqnarray}
and
\begin{eqnarray}
M_C = 
\begin{pmatrix}
0 && X^T \\
X && 0
\end{pmatrix} \, , ~~
X =
\begin{pmatrix}
M_2 && g_2\braket{ H_u^{0 *} } && 0 \\
0 	&& \mu 						&& -y_{\tau} \braket{\tilde{\nu}_{L,3}} \\
g_2 \braket{ \tilde{\nu}_{L,3}^*} && 0	&& 0 
\end{pmatrix} \, .
\label{eq:mass-matrix}
\end{eqnarray}
The mass eigenstates of this system can be found by diagonalizing the 3-by-3 matrix $X$, using the two 3-by-3 unitary matrices $U$ and $V$ defined by
\begin{eqnarray}
U^* X V^{-1} = 
\begin{pmatrix}
m_{\tilde{C}_1} &&					 	&& \\
&& m_{\tilde{C}_2} 		&& \\
&&						&& m_{\tilde{C}_3}
\end{pmatrix}\, ,
\end{eqnarray}
\begin{eqnarray}
\begin{pmatrix}
\tilde{C}_1^+\\
\tilde{C}_2^+ \\
\tilde{C}_3^+
\end{pmatrix} =
V 
\begin{pmatrix}
\tilde{W}^+ \\
\psi_{u}^+ \\
\tau_R^c
\end{pmatrix} \, ,\quad
\begin{pmatrix}
\tilde{C}_1^- \\
\tilde{C}_2^- \\
\tilde{C}_3^- \\
\end{pmatrix} =
U
\begin{pmatrix}
\tilde{W}^- \\
\psi_d^- \\
\tau_L 
\end{pmatrix}  \ .
\end{eqnarray}
It is easier to find the matrices $U$ 
and $V$ from the expressions
\begin{eqnarray}
U^* X X^\dagger U^T = V X^\dagger X V^{-1} = 
\begin{pmatrix}
m_{\tilde{C}_1}^2 &&					 	&& \\
&& m_{\tilde{C}_2}^2 		&& \\
&&						&& m_{\tilde{C}_3}^2
\end{pmatrix}\, .
\end{eqnarray}
As it stands, the expressions one gets for the mass eigenvalues and mixing matrices $U,V$ involve solving for the roots of a cubic equation and give very cumbersome expressions. We can simplify the situation by working in the limit in which the $\mu$-term is negligible and can be dropped -- a reasonable approximation during the period of reheating, where  $\sqrt{\braket{\psi^2}}$ is initally around  $10^{14}$ GeV and the gaugino mass $M_2$ is always of ${\mathcal{O}}(10^{13} {\rm GeV})$. As $\sqrt{\braket{\psi^2}}$ decreases, the $\mu$-term does become comparable to (and indeed larger than) terms such as $g_2 \sqrt{\braket{\psi^2}} $ that we have kept. However, this will occur after the bulk of reheating has occurred, and thus will have an insignificant effect, so we will simply drop $\mu$ in our subsequent calculations.

\subsubsection*{The Small $\mu$ Limit}
When terms involving $\mu$ are dropped, the system to be diagonalized (presented in \eqref{eq:mass-matrix} ) simplifies tremendously. We are able to decouple the $\tau_R^c$, $\psi_d^-$ states as there is no longer any mixing between $\psi_d^-$ and $\psi_u^+$.
Indeed, examining the effective mass Lagrangian in equation \eqref{eq:L-mass-C}, we see that 
\begin{eqnarray}
\La_{mass} & \supset & 
y_{\tau} \braket{\tilde{\nu}_{L,3}} \tau_R^{c} \psi_d^{-}  + \mathrm{h.c.}\, ,
\end{eqnarray}
which looks like a Dirac mass for the fermion
\begin{eqnarray}
\Psi_{Dirac} = \begin{pmatrix}
\psi_d^- \\
\tau_R
\end{pmatrix} \, ,
\end{eqnarray}
with mass $ y_{\tau} \braket{\tilde{\nu}_{L,3}}$.
We can then simplify the system given in equations \eqref{eq:basis}, \eqref{eq:mass-matrix} to %
\begin{eqnarray}
\vec{\Psi}^T = \left(
\tilde{W}^{+}, \tilde{\psi}_u^{+}, \tilde{W}^{-}, \tau_L 
\right) \ ,
\end{eqnarray}
\begin{eqnarray}
M_C = 
\begin{pmatrix}
0 && X^T \\
X && 0
\end{pmatrix} \, ,\quad
X =
\begin{pmatrix}
M_2 && g_2\braket{ H_u^{0 *} } \\
g_2 \braket{ \tilde{\nu}_{L,3}^*} && 0 
\end{pmatrix}\, .
\end{eqnarray}
Expressing the matrix $X$ schematically as 
\begin{eqnarray}
X = 
\begin{pmatrix}
x_1 && x_2 \\
x_2 && 0
\end{pmatrix}
\end{eqnarray}
where $x_1 = M_2$, $x_2 = g_2 \sqrt{\langle \psi \rangle} / \sqrt{6}$,  it follows that
\begin{eqnarray}
X X^\dagger = X^\dagger X = 
\begin{pmatrix}
(x_1)^2 + (x_2)^2 && x_1 x_2 \\
x_1 x_2 && (x_1)^2 + (x_2)^2
\end{pmatrix} \ .
\label{eq:xSquare}
\end{eqnarray}
The eigenvalues of \eqref{eq:xSquare} are then given by 
\begin{eqnarray}
&\,&m^2_{\tilde{C}_1} = \frac{1}{2} \left((x_1)^2 +2 (x_2)^2-\sqrt{(x_1)^4+4 (x_1)^2 (x_2)^2} \right) \, , \quad
\\&\,&
m^2_{\tilde{C}_2} = \frac{1}{2} \left((x_1)^2 +2 (x_2)^2 +\sqrt{(x_1)^4+4 (x_1)^2 (x_2)^2} \right)\,.
\nn \\
\end{eqnarray}
The matrices $U$ and $V$ become
\begin{eqnarray}
U = V =
\frac{1}{\sqrt{4(x_2)^2 + (x_1 - \sqrt{(x_1)^2 + 4x_2^2})^2}}
\begin{pmatrix}
x_1 - \sqrt{(x_1)^2 + 4x_2^2} && 2x_2 \\
x_1 + \sqrt{(x_1)^2 + 4x_2^2} && 2x_2
\end{pmatrix}\,.
\end{eqnarray}
It follows that the associated mass eigenstates are  
\begin{eqnarray}
\begin{pmatrix}
\tilde{C}_1^+ \\
\tilde{C}_2^+
\end{pmatrix} 
= V
\begin{pmatrix}
\tilde{W}^+ \\
\tilde{\psi}_u^+
\end{pmatrix}
\, , \qquad
\begin{pmatrix}
\tilde{C}_1^- \\
\tilde{C}_2^-
\end{pmatrix} 
= U
\begin{pmatrix}
\tilde{W}^- \\
\tau_L
\end{pmatrix}
\end{eqnarray}
with masses $m_{\tilde{C}_1}$, $m_{\tilde{C}_2}$.

\begin{figure}
\begin{center}
\includegraphics[scale=1.0]{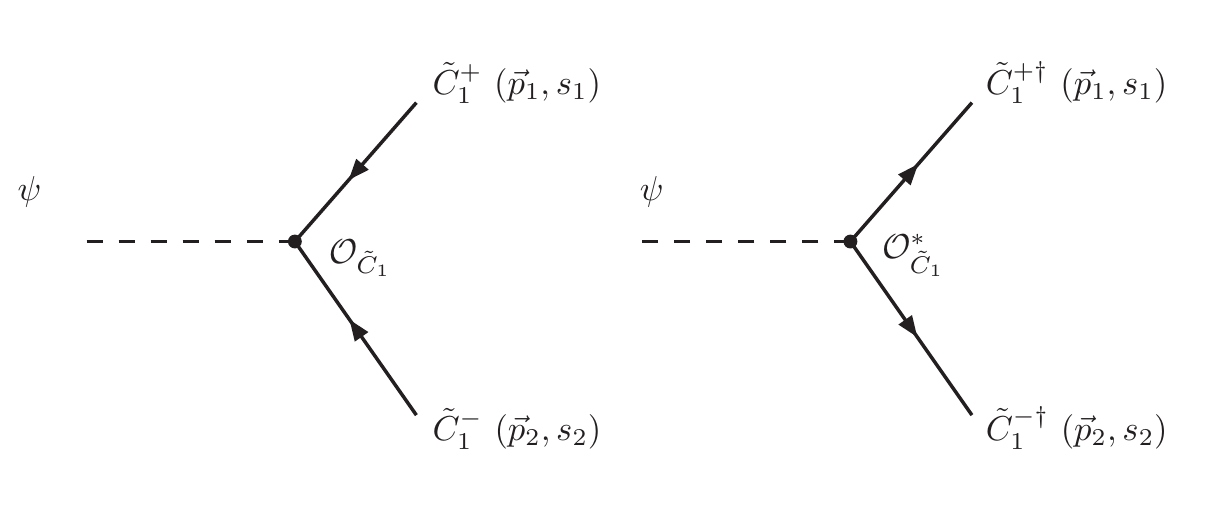}
\end{center}
\caption{Diagrams which contribute to the decay $\psi \rightarrow \tilde{C}_1^+ \tilde{C}_2^-$. Here, the coupling $\mathcal{O}_{\tilde{C}_1} =-\frac{g_2}{\sqrt{6}}(U_{1W}^*V_{1u}^* + U_{1\tau}^*V_{1W}^*) $.}
\label{fig:chargino-decay}
\end{figure}

Decays of the inflaton into the charginos arise from the vertices
\begin{eqnarray}
\La_{decay,C} &=&
-\frac{g_2}{\sqrt{6}} \psi \tilde{W}^- \psi_u^+ 
-\frac{g_2}{\sqrt{6}} \psi \tilde{W}^+ \tau_L
+ \mathrm{h.c.}
\end{eqnarray}
Assuming that the states  with mass $m_{\tilde{C}_2}$ are too heavy to be produced, we will be interested in the decays to $\tilde{C}_1^+$, $\tilde{C}_1^-$ only. Rotating the vertices above, we find that 
\begin{eqnarray}
\La_{decay,C} &=&
-\frac{g_2}{\sqrt{6}}\psi\left(
U_{1W}^* V_{1u}^* + U_{1\tau}^*V_{1W}^*  
\right)\tilde{C}_1^+ \tilde{C}_1^-
\end{eqnarray}
where 
\begin{eqnarray}
U_{1W} = V_{1W} =
\frac{x_1 - \sqrt{(x_1)^2 + 4x_2^2}}{\sqrt{4(x_2)^2 + (x_1 - \sqrt{(x_1)^2 + 4x_2^2})^2}} \, , \nn \\
U_{1\tau} = V_{1u} =
\frac{2x_2}{\sqrt{4(x_2)^2 + (x_1 - \sqrt{(x_1)^2 + 4x_2^2})^2}} 
\, .
\end{eqnarray}
For later use, we note that in the limit that $x_2 \ll x_1$, the above matrix elements have the form
\begin{eqnarray}
&\,&U_{1W} = V_{1W} = - \frac{x_2}{x_1} + \frac{3}{2}\left( \frac{x_2}{x_1} \right)^3 + \mathcal{O}\left( \frac{x_2}{x_1} \right)^5  \, , \nn \\
&\,&U_{1\tau} = V_{1u} =
1 - \frac{1}{2} \left( \frac{x_2}{x_1} \right)^2
+ \frac{11}{8} \left( \frac{x_2}{x_1} \right)^4 
+ \mathcal{O}\left( \frac{x_2}{x_1} \right)^6
\, .
\end{eqnarray}
The decay rate for the process $\psi \rightarrow \tilde{C}_1^+ \tilde{C}_1^- $ can be determined to be 
The decay rate for the process $\psi \rightarrow \tilde{C}_1^+ \tilde{C}_1^- $ (shown in figure \ref{fig:chargino-decay}) can be determined to be
\begin{eqnarray}
\Gamma = 
\Bigg(\frac{(|\alpha|^2 + |\beta|^2)(m_\psi^2 - 2m_{\tilde{C}_1}^2)
	- (\alpha \beta^* + \beta \alpha^*)(2m_{\tilde{C}_1}^2)}{16 \pi m_{\psi}^3}\Bigg)
(m_\psi^2 - 4m_{\tilde{C}_1}^2)^{1/2}
\nn \\
\label{j1}
\end{eqnarray}
where
\begin{eqnarray}
|\alpha|^2 = |\beta|^2
= \frac{g_2^2}{6} |U_{1W}V_{1u} + U_{1\tau}V_{1W}|^2\,.
\end{eqnarray}
Since $\alpha, \beta$ are real, we can simplify this further to get
\begin{eqnarray}
\Gamma = 
\gamma^2 \frac{(m_\psi^2 - 4m_{\tilde{C}_1}^2)^{3/2} }{8\pi m_{\psi}^2}\,,
\end{eqnarray}
where 
\begin{eqnarray}
&\,&\gamma = \frac{g_2}{\sqrt{6}}(U_{1W}V_{1u} + U_{1\tau}V_{1W}) \, ,\qquad\\
&\,&
m^2_{\tilde{C}_1} = \frac{1}{2} \left((x_1)^2 +2 (x_2)^2-\sqrt{(x_1)^4+4 (x_1)^2 (x_2)^2} \right) \ .
\end{eqnarray}

\subsection*{C.2.   Neutralino Mixing}

Again, ignoring terms which come with a factor of a neutrino Yukawa coupling $y_\nu$, the effective mass Lagrangian for the ``neutralinos" is given by
\begin{eqnarray}
\La_{mass, N} &=&
\mu \psi_u^0 \psi_d^0 - \frac{1}{2} M_2 \tilde{W}^0 \tilde{W}^0 - \frac{1}{2} \tilde{W}_R \tilde{W}_R - \frac{1}{2} M_{B-L} \tilde{B} \tilde{B} 
\nn \\
&&
+ \frac{1}{\sqrt{2}} g_2\braket{H_u^{0*}} \tilde{W}^0 \psi_u^0 - \frac{1}{\sqrt{2}} g_2 \braket{\tilde{\nu}_{L,3}^*} \tilde{W}^0 \nu_{L,3} 
\label{j2} \\
&&
- \sqrt{2} g_R q_{R_u}\braket{H_u^{0*}} \tilde{W}_R \psi_u^0 - \sqrt{2} g_R q_{R_\nu} \braket{\tilde{\nu}_{R,3}^{c*}} \tilde{W}_R \nu_{R,3}^{c} 
\nn \\
&&
- \sqrt{2} g_{BL} q_{BL_L} \braket{\tilde{\nu}_{L,3}^{*}} \tilde{B} \nu_{L,3} - \sqrt{2} g_{BL} q_{BL_\nu} \braket{\tilde{\nu}_{R,3}^{c^*}} \tilde{B} \nu_{R,3}^{c}
\nn \\
&&
+ \mathrm{h.c.}
\nn
\end{eqnarray}
Here, $q_{G_x}$ denotes the $U(1)_G$ charge of the field $x=H_u,\nu_R^c,L$, where $G= R, B-L$.
We can express this as
\begin{eqnarray}
\La_{mass, N} &=& -\frac{1}{2} \vec{\psi}^{~T} M_N \vec{\psi} + \mathrm{h.c.} \, ,
\end{eqnarray}
where
\begin{eqnarray}
\vec{\psi}^{~T }= \left( \tilde{B}, \tilde{W}_R, \tilde{W}^0, \psi_d^0, \psi_u^0, \nu_{L,3}, \nu_{R,3}^c \right)
\end{eqnarray}
and
{\footnotesize
\begin{eqnarray}
 M_N 
=
\begin{pmatrix}
M_{B-L} & 0 & 0 & 0 & 0 & -\sqrt{2}g_{BL}  \braket{\tilde{\nu}_{L,3}^*} & \sqrt{2}g_{BL}  \braket{\tilde{\nu}_{R,3}^{c*}} \\ 
0 & M_R & 0 & 0 & \frac{g_R}{\sqrt{2}}  \braket{H_u^{0*}} & 0 & -\frac{g_R}{\sqrt{2}}  \braket{\tilde{\nu}_{R,3}^{c*}} \\ 
0 & 0 & M_2 & 0 & -\frac{g_2}{\sqrt{2}} \braket{H_u^{0*}} & \frac{g_2}{\sqrt{2}} \braket{\tilde{\nu}_{L,3}^*} & 0 \\ 
0 & 0 & 0 & 0 & -\mu & 0 & 0 \\ 
0 & \frac{g_R}{\sqrt{2}}  \braket{H_u^{0*}} & -\frac{g_2}{\sqrt{2}} \braket{H_u^{0*}} & -\mu & 0 & 0 & 0 \\ 
-\sqrt{2}g_{BL}  \braket{\tilde{\nu}_{L,3}^*} & 0 & \frac{g_2}{\sqrt{2}} \braket{\tilde{\nu}_{L,3}^*} & 0 & 0 & 0 & 0 \\ 
\sqrt{2}g_{BL}  \braket{\tilde{\nu}_{R,3}^{c*}} & -\frac{g_R}{\sqrt{2}}  \braket{\tilde{\nu}_{R,3}^{c*}} & 0 & 0 & 0 & 0 & 0
\end{pmatrix} .
\nn \\
\end{eqnarray}}
For simplicity, let us schematically re-express this to give
\begin{eqnarray}
M_N =
\begin{pmatrix}
x_1 & 0 & 0 & 0 & 0 & x_2 & x_3 \\ 
0 & x_4 & 0 & 0 & x_5 & 0 & x_6 \\ 
0 & 0 & x_7 & 0 & x_8 & x_9 & 0 \\ 
0 & 0 & 0 & 0 & x_{10} & 0 & 0 \\ 
0 & x_5 & x_8 & x_{10} & 0 & 0 & 0 \\ 
x_2 & 0 & x_9 & 0 & 0 & 0 & 0 \\ 
x_3 & x_6 & 0 & 0 & 0 & 0 & 0
\end{pmatrix}  \, .
\end{eqnarray}

\subsubsection*{The Small $\mu$ Limit}

\begin{figure}[h!]
\begin{center}
\includegraphics[scale=1.0]{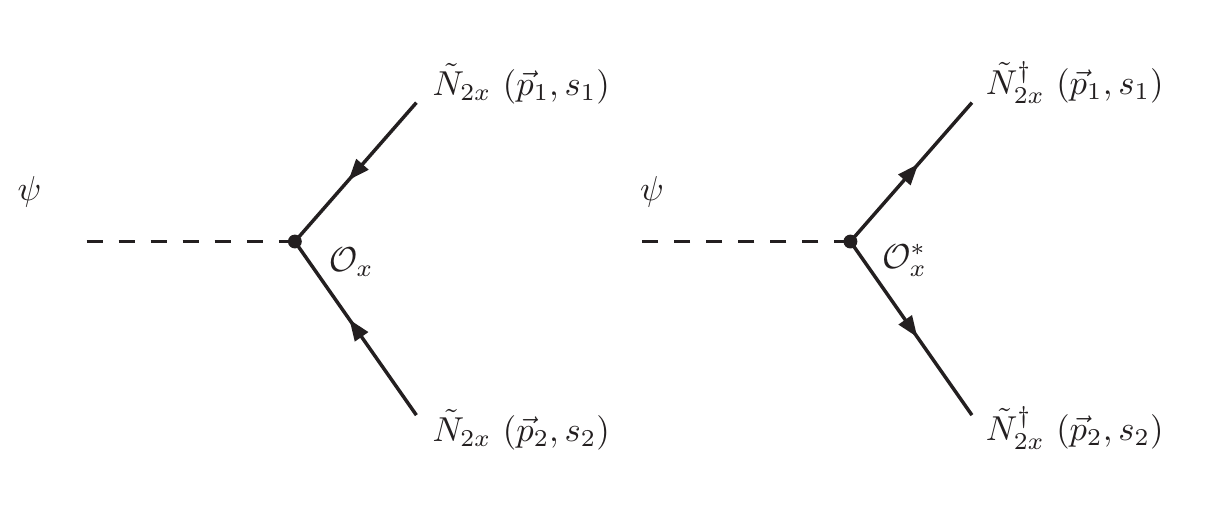}
\end{center}
\caption{Feynman diagrams which contribute to the processes $\psi \rightarrow \tilde{N}_{2a} \tilde{N}_{2a}$ and $\psi \rightarrow \tilde{N}_{2b} \tilde{N}_{2b}$.}
\label{fig:neutralino-decay1}
\end{figure}

\begin{figure}[h!]
\begin{center}
\includegraphics[scale=1.0]{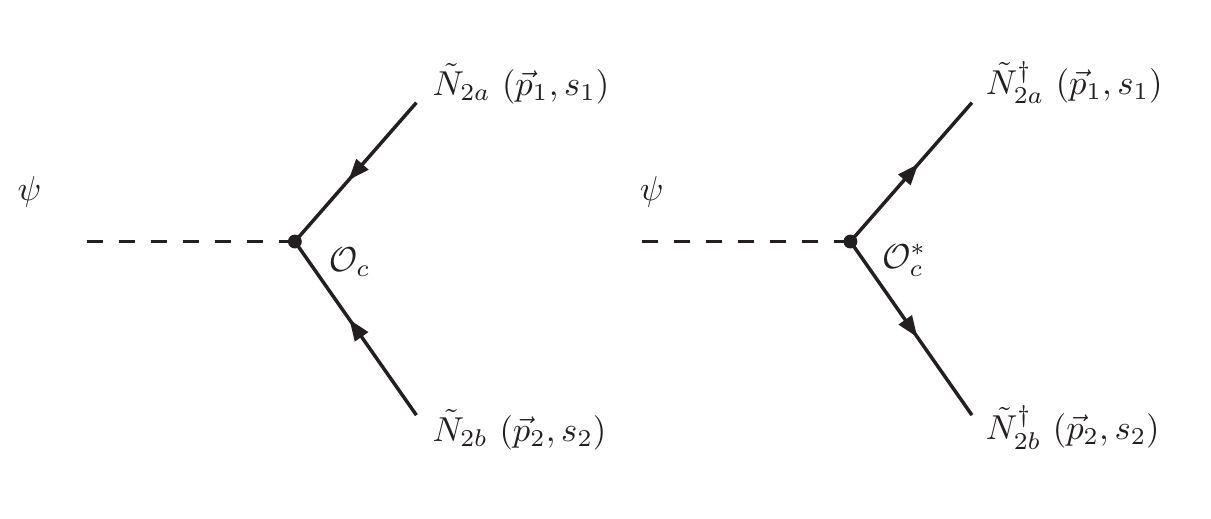}
\end{center}
\caption{Feynman diagrams which contribute to the processes $\psi \rightarrow \tilde{N}_{2a} \tilde{N}_{2b}$.}
\label{fig:neutralino-decay2}
\end{figure}

In this limit, the $\psi_d^0$ state decouples from the mixing. Hence, we only have a six-dimensional system to analyze. This is given by
\begin{eqnarray}
\vec{\psi}^{~T} = \left( \tilde{B}, \tilde{W}_R, \tilde{W}^0, \psi_u^0, \nu_{L,3}, \nu_{R,3}^c \right) \ ,
\label{eq:nu-basis}
\end{eqnarray}
\begin{eqnarray}
M_N =
\begin{pmatrix}
x_1 & 0 & 0  & 0 & x_2 & x_3 \\ 
0 & x_4 & 0  & x_5 & 0 & x_6 \\ 
0 & 0 & x_7  & x_8 & x_9 & 0 \\ 
0 & x_5 & x_8 & 0 & 0 & 0 \\ 
x_2 & 0 & x_9 & 0 & 0 & 0 \\ 
x_3 & x_6 & 0 & 0 & 0 & 0
\end{pmatrix} \ .
\end{eqnarray}
Let us make some further simplifying assumptions in order to find the eigenvalues and eigenvectors of this system. 
\begin{itemize}
	\item Take $x_1 = x_4 = x_7 = M \sim 1.58 \times 10^{13}~$GeV.
	\item Take $-x_2 = x_3 = x_5 = -x_6 = - x_8 = x_9 = u \sim g \sqrt{\braket{\psi^2}}$, where $g\sim 0.57$
\end{itemize}
The mass matrix $N$ then takes the form
\begin{eqnarray}
M_N = 
\begin{pmatrix}
M & 0 & 0 & 0 & -u & u \\ 
0 & M & 0 & u & 0 & -u \\ 
0 & 0 & M & -u & u & 0 \\ 
0 & u & -u & 0 & 0 & 0 \\ 
-u & 0 & u & 0 & 0 & 0 \\ 
u & -u & 0 & 0 & 0 & 0
\end{pmatrix}  \ .
\end{eqnarray}
The eigenvalues and eigenvectors of this system can now be evaluated and are given in table \ref{table:n-eigenstates} in the main body of this paper.
We find that the lightest eigenstates, that is, those into which the inflaton can decay,
are given in terms of the states in \eqref{eq:nu-basis} by
{\footnotesize
\begin{eqnarray}
&\,&\tilde{N}_{1} = 
\frac{1}{\sqrt{3}} \left( 0, 0, 0, 1, 1, 1 \right)^T \ ,
\nn \\
&\,&\tilde{N}_{2a} =
\left(
-u, 2u, -u, -\frac{1}{2} (M + \sqrt{M^2 + 12u^2}), 0, \frac{1}{2} (M + \sqrt{M^2 + 12u^2})
\right)^T 
\bigg/\sqrt{M^2 + 12u^2 + M \sqrt{M^2 + 12 u^2}} \ ,
\nn \\
&\,&\tilde{N}_{2b} =
\bigg(
\sqrt{3}u, 0, -\sqrt{3}u, 
-\sqrt{ u^2 + \frac{1}{6}(M^2 +  M \sqrt{M^2 + 12 u^2})}, 
2\sqrt{ u^2 + \frac{1}{6}(M^2 +  M \sqrt{M^2 + 12 u^2})}, 
\nn \\
&& \qquad
-\sqrt{ u^2 + \frac{1}{6}(M^2 +  M \sqrt{M^2 + 12 u^2})}
\bigg)^T 
\bigg/\sqrt{M^2 + 12u^2 + M \sqrt{M^2 + 12 u^2}} \ .
\end{eqnarray}
}
For completeness, the heavier eigenstates are 
{\footnotesize
\begin{eqnarray}
&\,&\tilde{N}_{3} = 
\frac{1}{\sqrt{3}} \left(1, 1, 1, 0, 0, 0 \right)^T \ ,
\nn \\
&\,&\tilde{N}_{4a} =
\bigg(
\sqrt{ u^2 + \frac{1}{6}(M^2 +  M \sqrt{M^2 + 12 u^2})}, 
-2\sqrt{ u^2 + \frac{1}{6}(M^2 +  M \sqrt{M^2 + 12 u^2})} 
\nn \\
&& \qquad\qquad
\sqrt{ u^2 + \frac{1}{6}(M^2 +  M \sqrt{M^2 + 12 u^2})} \ ,
- \sqrt{3}u,
0,
\sqrt{3}u
\bigg)^T 
\bigg/\sqrt{M^2 + 12u^2 + M \sqrt{M^2 + 12 u^2}} \ ,
\nn \\
&\,&\tilde{N}_{4b} =
\bigg(
-\sqrt{3}\sqrt{ u^2 + \frac{1}{6}(M^2 +  M \sqrt{M^2 + 12 u^2})}, 
0,
\sqrt{3}\sqrt{ u^2 + \frac{1}{6}(M^2 +  M \sqrt{M^2 + 12 u^2})},
-u,
2u, 
-u
\bigg)^T
\nn \\
&& \qquad\qquad
\bigg/\sqrt{M^2 + 12u^2 + M \sqrt{M^2 + 12 u^2}}  \,.
\nn \\
\end{eqnarray}
}
The mass matrix can be diagonalized using the matrix $N$, where
\begin{eqnarray}
N &=& \begin{pmatrix}
\tilde{N}_1,
\tilde{N}_{2a}, 
\tilde{N}_{2b}, 
\tilde{N}_{3},
\tilde{N}_{4a}, 
\tilde{N}_{4b}, 
\end{pmatrix}^T \,  
\end{eqnarray}
such that
{\footnotesize
\begin{eqnarray}
&&
N^* M_N N^{-1} =
\nn\\
&&
\begin{pmatrix}
0 &  &  &  &  &  \\ 
& \frac{1}{2}(M - \sqrt{M^2 + 12u^2})&  &  &  &  \\ 
&  & \frac{1}{2}(M - \sqrt{M^2 + 12u^2}) &  &  &  \\ 
&  &  & M &  &  \\ 
&  &  &  & \frac{1}{2}(M + \sqrt{M^2 + 12u^2}) &  \\ 
&  &  &  &  & \frac{1}{2}(M + \sqrt{M^2 + 12u^2})
\end{pmatrix} \ .
\nn \\
\end{eqnarray}
}
Without the VEVs, the last six terms in \eqref{j2} are
\begin{eqnarray}
\La_{decay, N} &=&
\frac{1}{\sqrt{2}} g_2{H_u^{0*}} \tilde{W}^0 \psi_u^0 - \frac{1}{\sqrt{2}} g_2 {\tilde{\nu}_{L,3}^*} \tilde{W}^0 \nu_{L,3} 
- \sqrt{2} g_R q_{R_u}{H_u^{0*}} \tilde{W}_R \psi_u^0 - \sqrt{2} g_R q_{R_\nu} {\tilde{\nu}_{R,3}^{c*}} \tilde{W}_R \nu_{R,3}^{c} 
\nn \\&&\qquad
- \sqrt{2} g_{BL} q_{BL_L} {\tilde{\nu}_{L,3}^{*}} \tilde{B} \nu_{L,3} - \sqrt{2} g_{BL} q_{BL_\nu} {\tilde{\nu}_{R,3}^{c^*}} \tilde{B} \nu_{R,3}^{c}
+ \mathrm{h.c.}
\nn \\
&=& 
\frac{1}{\sqrt{6}} \psi \bigg(
\frac{1}{\sqrt{2}} g_2 \tilde{W}^0 \psi_u^0 - \frac{1}{\sqrt{2}} g_2  \tilde{W}^0 \nu_{L,3} 
- \sqrt{2} g_R q_{R_u}\tilde{W}_R \psi_u^0 - \sqrt{2} g_R q_{R_\nu}  \tilde{W}_R \nu_{R,3}^{c} 
\nn \\&& \qquad
- \sqrt{2} g_{BL} q_{BL_L}  \tilde{B} \nu_{L,3} - \sqrt{2} g_{BL} q_{BL_\nu}  \tilde{B} \nu_{R,3}^{c}
\bigg) + \mathrm{h.c.}
\end{eqnarray}
Rotating the Lagrangian to the lightest mass eigenstates $\tilde{N}_1$, $\tilde{N}_{2a}$ and $\tilde{N}_{2b}$, we find that the term $\psi \tilde{N}_1 \tilde{N}_1$ is absent, and hence the decay $\psi \rightarrow \tilde{N}_1 \tilde{N}_1$ is not allowed. Furthermore, we find that
\begin{eqnarray}
\La_{decay,N}
&\supset& \frac{1}{\sqrt{6}}
\bigg(
\frac{g_2}{\sqrt{2}}
(N_{(2a)0}^*N_{1u}^* - N_{(2a)0}^* N_{1L}^*)
- 
\frac{g_R}{\sqrt{2}}
(N_{(2a)R}^*N_{1u}^* - N_{(2a)R}^* N_{1\nu}^*)
\nn \\
&& \qquad \qquad
-
\sqrt{2}g_2
(N_{(2a)B}^*N_{1\nu}^* - N_{(2a)B}^* N_{1L}^*)
\bigg) \psi \tilde{N}_{1} \tilde{N}_{2a}
\nn \\ &&
+
\frac{1}{\sqrt{6}}
\bigg(
\frac{g_2}{\sqrt{2}}
(N_{(2b)0}^*N_{1u}^* - N_{(2b)0}^* N_{1L}^*)
- 
\frac{g_R}{\sqrt{2}}
(N_{(2b)R}^*N_{1u}^* - N_{(2b)R}^* N_{1\nu}^*)
\nn \\
&& \qquad \qquad
-
\sqrt{2}g_2
(N_{(2b)B}^*N_{1\nu}^* - N_{(2b)B}^* N_{1L}^*)
\bigg) \psi \tilde{N}_{1} \tilde{N}_{2b}
\nn \\
&&
+ \mathrm{h.c.}
\nn \\
&=& 0 \, ,
\end{eqnarray}
since $N_{iu} = N_{iL} = N_{i\nu} = \frac{1}{\sqrt{3}}$. It follows that there is no decay of $\psi$ to $\tilde{N}_1 \tilde{N}_{2a}$ or to $\tilde{N}_1 \tilde{N}_{2b}$.
Searching the Lagrangian for the remaining kinematically allowed decay terms, we find  
{
\begin{eqnarray}
\La_{decay, N}
&\supset&
\frac{1}{\sqrt{6}} \Big(
\frac{g_2}{\sqrt{2}} N^*_{(2a)0} \left( N^{*}_{(2a)u} - N^{*}_{(2a)L} \right)
-\frac{g_R}{\sqrt{2}} N^*_{(2a)R} \left(  N^{*}_{(2a)u} - N^{*}_{(2a)\nu}\right)
\nn\\
&\,&- \sqrt{2} g_{BL} N^*_{(2a)B} \left(N^{*}_{(2a)\nu} - N^{*}_{(2a)L} \right) 
\Big)\psi \tilde{N}_{2a} \tilde{N}_{2a}
+
\frac{1}{\sqrt{6}} \Big( (2a) \rightarrow (2b) \Big) \psi \tilde{N}_{2b} \tilde{N}_{2b}
\nn \\
&\,&
+\frac{1}{\sqrt{6}}\Big(
\frac{g_2}{\sqrt{2}} N^*_{(2a)0} \left( N^{*}_{(2b)u} - N^{*}_{(2b)L} \right)
-\frac{g_R}{\sqrt{2}} N^*_{(2a)R} \left(  N^{*}_{(2b)u} - N^{*}_{(2b)\nu}\right)
\nn\\
&\,&- \sqrt{2} g_{BL} N^*_{(2a)B} \left(N^{*}_{(2b)\nu} - N^{*}_{(2b)L} \right) 
\Big)\psi \tilde{N}_{2a} \tilde{N}_{2b}
+
\frac{1}{\sqrt{6}} \Big( (2a) \leftrightarrow (2b) \Big) \psi \tilde{N}_{2a} \tilde{N}_{2b}
\nn \\
&=& \mathcal{O}_a\psi \tilde{N}_{2a} \tilde{N}_{2a} 
+ \mathcal{O}_b \psi \tilde{N}_{2b} \tilde{N}_{2b} 
+ \mathcal{O}_c \psi \tilde{N}_{2a} \tilde{N}_{2b} \, .
\end{eqnarray}
}
This allows the inflaton decays $\psi \rightarrow \tilde{N}_{2a} \tilde{N}_{2a}$, $\psi \rightarrow \tilde{N}_{2b} \tilde{N}_{2b}$ and $\psi \rightarrow \tilde{N}_{2a} \tilde{N}_{2b}$. 
Feynman diagrams for these processes are shown in figures \ref{fig:neutralino-decay1} and \ref{fig:neutralino-decay2}.
The first two decay rates take the same form
\begin{eqnarray}
&\,& \Gamma_d(\psi \rightarrow \tilde{N}_{2x} \tilde{N}_{2x}) 
\nn \\
&=&
\frac{1}{32 \pi m_{\psi}^2} \bigg(
(|\alpha|^2 + |\beta|^2) (m_{\psi}^2 - 2 m_{\tilde{N}_2}^2)
- 2(\alpha \beta^* + \beta \alpha^*) m_{\tilde{N}_2}^2
\bigg) (m_\psi^2 - 4 m_{\tilde{N}_2}^2)^{1/2}\ \nn \\
\label{j3}
\end{eqnarray}
where $\alpha = \beta = i\mathcal{O}_a$  and $i\mathcal{O}_b$ for $x= a, b$ respectively.
The decay rate to $\tilde{N}_{2a}\tilde{N}_{2b}$ takes the form
\begin{eqnarray}
&\,&\Gamma_d(\psi \rightarrow \tilde{N}_{2a} \tilde{N}_{2b}) 
\nn\\
&=&
\frac{1}{16\pi m_{\psi}^2} \bigg[
(|\alpha|^2 + |\beta|^2)(m_{\psi}^2 - 2 m_{\tilde{N}_2}^2)
- 2(\alpha \beta^* + \beta \alpha^*) m_{\tilde{N}_2}^2
\bigg] (m_\psi^2 - 4 m_{\tilde{N}_2}^2)^{1/2} \nn \\
\label{j4}
\end{eqnarray}
with $\alpha = \beta = i\mathcal{O}_c$.
The above expressions simplify to give equations \eqref{eq:neutralino1} and \eqref{eq:neutralino2}.

\bibliographystyle{elsarticle-num}
\bibliography{references-1}

\end{document}